\documentclass[%
 aps,
 jmp,%
 amsmath,amssymb,
 reprint,%
]{revtex4-1}
\usepackage{graphicx}
\usepackage{subfigure}
\usepackage{mathrsfs}
\usepackage[bookmarks=true,colorlinks=true,citecolor=blue,linkcolor=red,urlcolor=magenta]{hyperref}

\usepackage{multirow}
\usepackage{dcolumn}
\usepackage{bm}

\graphicspath{{figures_part1/}}

\begin{document}

\preprint{APS/123-QED}

\title[]{Mechanisms of proton-proton inelastic cross-section growth in multi-peripheral model within the framework of perturbation theory. Part 1}

\author{I.V. Sharf}
\affiliation{ Odessa National Polytechnic University, Shevchenko av. 1, Odessa, 65044, Ukraine.}%

\author{G.O. Sokhrannyi}%
\affiliation{ Odessa National Polytechnic University, Shevchenko av. 1, Odessa, 65044, Ukraine.}%

\author{A.V. Tykhonov}%
\affiliation{ Odessa National Polytechnic University, Shevchenko av. 1, Odessa, 65044, Ukraine.}%
\affiliation{ Department of Experimental Particle Physics, Jozef Stefan Institute,
     Jamova 39, SI-1000 Ljubljana, Slovenia.}%

\author{K.V. Yatkin}%
\affiliation{ Odessa National Polytechnic University, Shevchenko av. 1, Odessa, 65044, Ukraine.}%

\author{N.A. Podolyan}%
\affiliation{ Odessa National Polytechnic University, Shevchenko av. 1, Odessa, 65044, Ukraine.}%

\author{M.A. Deliyergiyev}%
\affiliation{ Odessa National Polytechnic University, Shevchenko av. 1, Odessa, 65044, Ukraine.}%
\affiliation{ Department of Experimental Particle Physics, Jozef Stefan Institute,
     Jamova 39, SI-1000 Ljubljana, Slovenia.}%

\author{V.D. Rusov}%
 \email{siiis@te.net.ua}
\affiliation{ Odessa National Polytechnic University, Shevchenko av. 1, Odessa, 65044, Ukraine.}%
\affiliation{Department of Mathematics, Bielefeld University, 
      Universitatsstrasse 25, 33615 Bielefeld, Germany.}%


\date{\today}

\begin{abstract}
We demonstrate a possibility of computation of inelastic scattering cross-section in a multi-peripheral model by application of the Laplace method to multidimensional integral over the domain of physical process. The constrained maximum point of scattering cross-section integral under condition of the energy-momentum conservation has been found. In the vicinity of this point the integrand is substituted for an expression of Gaussian type, which made possible to compute this integral numerically. The paper has two parts. The hunting procedure of the constrained maximum point is considered and the properties of this maximum point are discussed in the given part of the paper. It is shown that virtuality of all internal lines of the ``comb" diagram reduced at the constrained maximum point with energy growth. In the second part of the paper we give some arguments in favor of consideration of the mechanism of virtuality reduction as the mechanism of the total hadron scattering cross-section growth, which is not taken into account within the framework of Regge theory.
\end{abstract}

\maketitle

%


\section{Introduction}
\label{INTRO}
\begin{figure}
\begin{center}
  \includegraphics{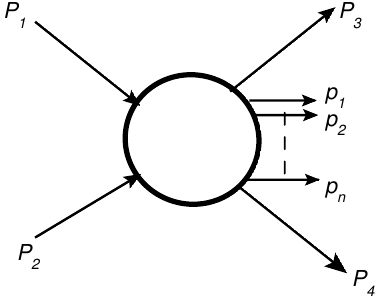}
\end{center}
\caption{A general view of an inelastic scattering diagram.}
  \label{fig:part1_fig01}
\end{figure}

The multi-peripheral model [\onlinecite{springerlink:10.1007/BF02781901}] has been used to describe
hadron scattering for a long time. However, in our opinion,
formal difficulties which appear in the calculation of inelastic
scattering cross-sections have not been overcome.
These difficulties are caused by the fact that inelastic
scattering cross-section with production of a given number
of secondary particles in the final state Fig.\ref{fig:part1_fig01} is described by the multidimensional integral of scattering amplitude squared modulus over the phase volume of final state:
\begin{eqnarray}%
&&{\sigma _n} = \frac{1}{4n!I}\int {\frac{{d{{\vec P}_3}}}{{2{P_{30}}{{(2\pi )}^3}}}} \frac{{d{{\vec P}_4}}}{{2{P_{40}}{{(2\pi )}^3}}}\prod\limits_{k = 1}^n {\frac{{d{{\vec p}_k}}}{{2{p_{0k}}{{(2\pi )}^3}}}   } \nonumber\\
&&\times \Phi{\delta ^{\left( 4 \right)}}\left (  {{P_3} + {P_4} + \sum\limits_{k = 1}^n {{p_k}}  - {P_1} - {P_2}} \right) 
\label{eq1}
\end{eqnarray}%
with
\begin{subequations}
\begin{eqnarray}
&&I = \sqrt {{{({P_1}{P_2})}^2} - {{({M_1}{M_2})}^2}}
\\
&&\Phi = {\left| {T(n,{p_1},{p_2},...,{p_n},{P_1},{P_2},{P_3},{P_4})} \right|^2}
\end{eqnarray}
\end{subequations}
where $M_{1}$ and $M_{2}$ are the masses of colliding particles with four-momenta $P_{1}$ and $P_{2}$;  $T(n, p_1, p_2, \ldots, p_n, P_1, P_2, P_3, P_4)$ is scattering amplitude corresponding to inelastic process shown in Fig.\ref {fig:part1_fig01}; $\delta^{(4)}$ is a four-dimensional delta function describing the conservation laws of energy and three momentum components in this process. Here it is also assumed, that particles with four-momentums $P_3$ and $P_4$ are the same sorts as $P_1$ and $P_2$, respectively, and $n$ secondary particles with four-momentums $p_1, p_2,\ldots, p_n$ are identical.
Since scattering amplitude is, in general, not a product of functions of some variables, and also due to the complexity of integration domain, the multidimensional integral in Eq.\ref{eq1} is not a product of smaller-dimensional ones. In considered inelastic process this domain of phase space of final state particles is determined by the energy-momentum conservation law. As a result, the integration limits for one variable depend on the values of others. In order to overcome these difficulties one usually deals with the multi-Regge kinematics [\onlinecite{Byckling:100542, Collins:111502, bfkl_1976,  Nikitin:113716, levin, levin_2, Ter-Martirosyan, KozlovNSU_2007, Lipatov:2008}]. 
\begin{figure}
\begin{center}
  \includegraphics[scale=0.85]{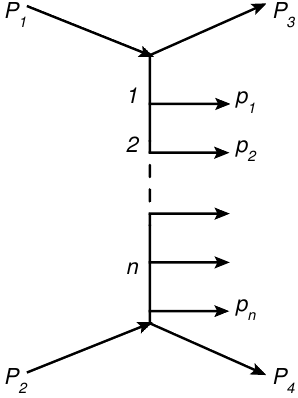}
\end{center}
\caption{An elementary inelastic scattering diagram in the multi-peripheral model (``the comb").}
  \label{fig:part1_fig02}
\end{figure}
The ultimate goal of this paper is to develop a new approach,
based upon well-known Laplace’s method [\onlinecite{DeBruijn:225131}], for the case when scattering amplitude can be represented
as a set of multi-peripheral diagrams Fig.\ref{fig:part1_fig02} within the
framework of perturbation theory.

The essence of this method consists in finding the constrained maximum point of scattering amplitude squared modulus in Eq.\ref{eq1} under four conditions imposed by $\delta^{(4)}$-function of Eq.\ref{eq1}. By expressing the scattering amplitude squared modulus as $|T|^2=exp(ln(|T|^2))$, it is possible to expand the exponent of the exponential function as a Taylor series about a point of the the constrained maximum, coming to nothing more than quadratic terms. Subsequently, we obtain a Gaussian integral whose calculation is reduced to computing the matrix determinant of second derivatives with respect to $ln(|T|^2)$. Now let us consider solution of the listed above problems step by step.

\section{Consideration of the scattering amplitude symmetry properties}
\label{SECTION_2}
First, before we turn to the constrained maximization problem, let’s examine the following simplifications. According to the Feynman diagram technique, the expression for scattering amplitude, which corresponds to a diagram in Fig.\ref{fig:part1_fig02} has a form:
\begin{eqnarray}
&&T\left( {n,{P_3},{P_4},{p_1},{p_2},...,{p_n},{P_1},{P_2}} \right) = {\left( { - ig{{\left( {2\pi } \right)}^4}} \right)^2} \nonumber\\
&&\times {\left( { - i\lambda {{\left( {2\pi } \right)}^4}} \right)^n}{\left( {\frac{{ - i}}{{{{\left( {2\pi } \right)}^4}}}} \right)^{n + 1}} \nonumber\\
&&\times A\left( {n,{P_3},{P_4},{p_1},{p_2}, \cdots ,{p_n},{P_1},{P_2}} \right)
\label{eq2}
\end{eqnarray}
with
\begin{eqnarray}
&&A\left( {n,{P_3},{P_4},{p_1},{p_2}, \cdots ,{p_n},{P_1},{P_2}} \right) =  \nonumber\\
&&= \frac{1}{{{m^2} - {{\left( {{P_1} - {P_3}} \right)}^2} - i\varepsilon }} \nonumber\\
&& \times \frac{1}{{{m^2} - {{\left( {{P_1} - {P_3} - {p_1}} \right)}^2} - i\varepsilon }} \nonumber\\
&&\times \frac{1}{{{m^2} - {{\left( {{P_1} - {P_3} - {p_1} - {p_2}} \right)}^2} - i\varepsilon }} \cdots  \cdots  \cdots \nonumber\\
&&\times \frac{1}{{{m^2} - {{\left( {{P_1} - {P_3} - {p_1} - {p_2} -  \cdots  - {p_{n - 1}}} \right)}^2} - i\varepsilon }}  \nonumber\\
&&\times \frac{1}{{{m^2} - {{\left( {{P_1} - {P_3} - {p_1} - {p_2} -  \cdots  - {p_{n - 1}} - {p_n}} \right)}^2} - i\varepsilon }} \nonumber\\
\label{eq3}
\end{eqnarray}
where $g$ is a coupling constant in the outermost vertices of the diagram; $\lambda$ is a coupling constant in all other vertices; $m$ is the mass of virtual particle field and also secondary particles. As in the original version of multi-peripheral model [\onlinecite{springerlink:10.1007/BF02781901}], pions are taken both as virtual and secondary particles. It was assumed that the particle masses with four-momentums $P_1$, $P_2$, $P_3$, $P_4$ are equal, i.e., $M_1=M_2=M_3=M_4=M$. The proton mass was taken as $M$. Note that the concrete choice of numerical value of mass $M$ has no importance for the results presented in the paper.

As it was noticed in [\onlinecite{Byckling:100542}], for the most ratios of particle masses in the initial and final state, the
virtual particles four-momentums on the diagram of Fig.\ref{fig:part1_fig02} are space-like, i.e., their scalar squares are negative in Minkowski space. The negativity of scalar squares of virtual four-momentums at the given mass configuration $M_1=M_2=M_3=M_4=M$ is easy to prove (see Appendix.\ref{Appendix_A}).

Since the virtual particle four-momenta ${\left( {{P_1} - {P_3}} \right)^2}, {\left( {{P_1} - {P_3} - {p_1}} \right)^2}$, \\
${\left( {{P_1} - {P_3} - {p_1} - {p_2} - ... - {p_n}} \right)^2}$ squares are negative at the physical values of four-momenta of final state particles, the denominators in Eq.\ref{eq1} are equal to zero nowhere in the physical region. Therefore it is possible to reduce $i\varepsilon $ to zero before all calculations.

Due to negativity of the virtual particles four-momenta squares, the magnitude Eq.\ref{eq3} is real and positive. Therefore the search of the constrained maximum point of scattering amplitude squared modulus reduces to the search of the constrained maximum point of function Eq.\ref{eq3}. Hereinafter the expressions Eq.\ref{eq3} and Eq.\ref{eq2}, which differ by constant factor, will be referred to as scattering amplitude.

Let us examine Eq.\ref{eq3} in c.m.s. of colliding particles ${P_1}$ and ${P_2}$. In such a frame of reference the initial and final states have some symmetry, making it possible to use for solving the constrained maximum problem. In particular, the consideration of symmetries makes it possible to reduce the search of the constrained maximum of scattering amplitude to the search of the maximum of its restriction on a certain subset of physical process domain shown in Fig.\ref{fig:part1_fig02}. This restriction is the function of substantially smaller number of independent variables than the initial amplitude.

For the further discussion of these symmetries and related simplifications, it would be convenient to take into account the conservation laws and express the scattering amplitude as a function of independent variables only. After decomposition of the three-dimensional particle momentums in c.m.s. frame to components, which are parallel $p_{k\parallel}$ and orthogonal $\vec{k}_{k\perp}$ to collision axis. These will be labeled the longitudinal and transverse momentum, respectively.

The energy of each particle in the final state can be expressed by their momenta using the mass shell conditions. Having $n+2$ particles in final state Fig.\ref{fig:part1_fig02} give us $3(n+2)$ momentum components of these particles. Since we are looking for a constrained extremum, it is necessary to take into account four relations, which express an energy-momentum conservation law. 
These relations allow us to represent the amplitude in Eq.\ref{eq3} 
as a function of $3n+2$ independent variables. The first $3n$ variables we choose are longitudinal and transverse components of momentums $\vec{p}_1,\vec{p}_2,\ldots,\vec{p}_n$ of particles produced along the ``comb" in Fig.\ref{fig:part1_fig02}. The other two variables are the transverse components of momentum $\vec{P}_{3\perp}$.

Assuming that the $z$-axis coincides with momentum direction $\vec{P}_1$ in c.m.s. and that $x$ and $y$ are the coordinate axes in the plane of transverse momenta, the conservation laws can be represented as:
\begin{eqnarray}%
&&{P_{30}} + {P_{40}} = \sqrt s  - \left( {{p_{10}} + {p_{20}} + ... + {p_{n0}}} \right) \cr
&&{P_{3\parallel }} + {P_{4\parallel }} =  - \left( {{p_{1\parallel }} + {p_{2\parallel }} + ... + {p_{n\parallel }}} \right)  \cr
&&{P_{4 \bot x}} =  - \left( {{p_{1 \bot x}} + {p_{2 \bot x}} + ... + {p_{n \bot x}} + {P_{3 \bot x}}} \right) \cr
&&{P_{4 \bot y}} =  - \left( {{p_{1 \bot y}} + {p_{2 \bot y}} + ... + {p_{n \bot y}} + {P_{4 \bot y}}} \right)
\label{eq4}
\end{eqnarray}%
where
\begin{eqnarray}%
&&  s = {\left( {{P_1} + {P_2}} \right)^2} \cr
&&  {p_{k0}} = \sqrt {{m^2} + {{\left( {{p_{k\parallel }}} \right)}^2} + {{\left( {{p_{k \bot x}}} \right)}^2} + {{\left( {{p_{k \bot y}}} \right)}^2}} \cr
&& {P_{30}} = \sqrt {{M^2} + {{\left( {{P_{3\parallel }}} \right)}^2} + {{\left( {{P_{3 \bot x}}} \right)}^2} + {{\left( {{P_{3 \bot y}}} \right)}^2}}  \cr
&&   {P_{40}} = \sqrt {{M^2} + {{\left( {{P_{4\parallel }}} \right)}^2} + {{\left( {{P_{4 \bot x}}} \right)}^2} + {{\left( {{P_{4 \bot y}}} \right)}^2}}
\label{eq5}
\end{eqnarray}%
Next we substitute in the following definitions:
\begin{eqnarray}%
&& E{}_p \equiv \sqrt s  - \left( {{p_{10}} + {p_{20}} + ... + {p_{n0}}} \right)  \cr
&& {P_{\parallel p}} \equiv  - \left( {{p_{1\parallel }} + {p_{2\parallel }} + ... + {p_{n\parallel }}} \right) \cr
&& {P_{px}} \equiv  - \left( {{p_{1 \bot x}} + {p_{2 \bot x}} + ... + {p_{n \bot x}}} \right) \cr
&& {P_{py}} \equiv  - \left( {{p_{1 \bot y}} + {p_{2 \bot y}} + ... + {p_{n \bot y}}} \right)
\label{eq6}
\end{eqnarray}%
Then, solving the system Eq.\ref{eq4} for the unknown ${P_{3\parallel }}, {P_{4\parallel }},{P_{4 \bot x}},{P_{4 \bot y}}$ gives:
\begin{eqnarray}
  \mbox{\fontsize{10}{14}\selectfont $ {P_{3\parallel }} = \frac{{{E_a}}}{{2 E_b}}\left( { - {P_{\parallel p}} \pm {E_p}\sqrt {1 - \frac{{4\left( {{M^2} + {{\left( {{{\vec P}_{3 \bot }}} \right)}^2}} \right) E_b}}{{{{\left( {E_a} \right)}^2}}}} } \right) $} \cr
  \mbox{\fontsize{10}{14}\selectfont $ {P_{4\parallel }} = \frac{{{E_a}}}{{2 E_b}} \left( { - {P_{\parallel p}} \mp {E_p}\sqrt {1 - \frac{{4\left( {{M^2} + {{\left( {{{\vec P}_{3 \bot }}} \right)}^2}} \right) E_b}}{{{{\left( {E_a} \right)}^2}}}} } \right) $}\cr
\label{eq7}
\end{eqnarray}
where
\begin{subequations}
\begin{eqnarray}
&&{\vec P_{p \bot }} = \left( {{P_{px}},\;{P_{py}}} \right) 
\\
&&{E_a} = E_p^2 - {P_{\parallel p}}^2 - \vec P_{p \bot }^2 - 2\left( {{{\vec P}_{p \bot }} \cdot {{\vec P}_{3 \bot }}} \right) 
\\
&&{E_b} = \left( {E_p^2 - {P_{\parallel p}}^2} \right)
\end{eqnarray}
\end{subequations}
We will discuss the choice of sign in Eq.\ref{eq7} further in the paper. Note that the value of ${P_{3\parallel }}$ will not change, if we change the signs of all transversal momentums in Eq.\ref{eq7} simultaneously.

Substituting ${P_{3\parallel }}$ from Eq.\ref{eq7} to Eq.\ref{eq5}, and resulting $P_{30}$ from Eq.\ref{eq5} into Eq.\ref{eq3}, give us the scattering amplitude as a function of independent variables, for which the conversation laws of all components of energy-momentum four-vector are taken into account. Below, referring to Eq.\ref{eq3}, we assume that these substitutions have already been done. Taking into account this fact, we can define the amplitude Eq.\ref{eq3} as
\begin{eqnarray}%
  \mbox{\fontsize{10}{14}\selectfont $ A\left( {n,{{\vec P}_{3 \bot }},{{\vec p}_{1 \bot }},{{\vec p}_{2 \bot }},...,{{\vec p}_{n \bot }},{p_{1\parallel }},{p_{2\parallel }},...,{p_{n\parallel }}} \right)$ }
\end{eqnarray}%
enumerating only independent variables in the argument list. When the scattering amplitude will be expressed in terms of independent variables only, it is possible to find the ordinary extremum, but not constrained extremum.

In what follows,
let us examine symmetries of the scattering amplitude in c.m.s. Obviously, the system has a symmetry under rotations around the collision axis, it means that there is no preferred direction in the plane orthogonal to collision axis. Hence, if the scattering amplitude has the constrained maximum point, it must be achieved at zero values of the particle momentums components in final state transversal to collision axis. Otherwise these momentums must be somehow directed in the plane of transversal momentums, while all directions are equivalent in this plane.

To show more formally previous conclusion we use an explicit form of amplitude Eq.\ref{eq3}, which transforms to itself, when the signs in front of all transversal momentums are changed:
\begin{eqnarray}%
&& A\left( {n,{{\vec P}_{3 \bot }},{{\vec p}_{1 \bot }},{{\vec p}_{2 \bot }},...,{{\vec p}_{n \bot }},{p_{1\parallel }},{p_{2\parallel }},...,{p_{n\parallel }}} \right) \nonumber\\
&&= A\left( {n, - {{\vec P}_{3 \bot }}, - {{\vec p}_{1 \bot }}, - {{\vec p}_{2 \bot }},..., - {{\vec p}_{n \bot }},{p_{1\parallel }},{p_{2\parallel }},...,{p_{n\parallel }}} \right) \nonumber\\
\label{eq9}
\end{eqnarray}
Take derivative from Eq.\ref{eq9} with respect to any of transversal momentum arguments and after that assuming all transversal momentums equal to zero, we obtain that all partial derivatives with respect to transverse momentum arguments are equal to zero if this arguments are equal to zero.

Thus, for the further search of the constrained maximum we can limit ourselves to reduction of the scattering amplitude on a subset of the values of its independent arguments, which corresponds to zero values of transversal momentums of all particles in final state. This reduction is a function of longitudinal components of momentum ${p_{1\parallel },p_{2\parallel },...,p_{n\parallel } }$, which we designate as ${A_\parallel }\left( {n,{p_{1\parallel }},{p_{2\parallel }},...,{p_{n\parallel }}} \right)$ and from Eq.\ref{eq3} we get:
\begin{eqnarray}%
&& A{\,_\parallel }\left( {n,{p_{1\parallel }},{p_{2\parallel }}, \cdots ,{p_{n\parallel }}} \right) =  \nonumber \\
 && = {\left( {{m^2} - {{\left( {{P_{10}} - {P_{30}}} \right)}^2} + {{\left( {{P_{1\parallel }} - {P_{3\parallel }}} \right)}^2}} \right)^{ - 1}}  \nonumber \\
&& \times \prod\limits_{l = 1}^n {{{\left( {{m^2} - {\alpha_l^2} + {\beta _l^2}} \right)}^{ - 1}}}  
\label{eq:eq_10}
\end{eqnarray}
where
\begin{subequations}
\begin{eqnarray}
&& \alpha_l  = { {{P_{10}} - {P_{30}} - \sum\limits_{k = 1}^l {{p_{k0}}} } }  \\
&& \beta_l  = { {{P_{1\parallel }} - {P_{3\parallel }} - \sum\limits_{k = 1}^l {{p_{k\parallel }}} } }  \\
&& p_{k0}=\sqrt{m^2+(p_{k\parallel})^2}  \\
&& P_{30}=\sqrt{M^2+(P_{3\parallel})^2} 
\end{eqnarray}
\end{subequations}
At the same time, assuming that all transversal momentums equal to zero, we have from Eq.\ref{eq7}:
\begin{eqnarray}%
  \mbox{\fontsize{10}{10}\selectfont ${P_{3\parallel }} = \frac{1}{2} \left( {{P_{\parallel p}} \pm {E_p}\sqrt {1 - \frac{{4{M^2}}}{{{{\left( {{E_p}} \right)}^2} - {{\left( {{P_{\parallel p}}} \right)}^2}}}} } \right) $} \nonumber \\
  \mbox{\fontsize{10}{10}\selectfont ${P_{4\parallel }} = \frac{1}{2} \left( {{P_{\parallel p}} \mp {E_p}\sqrt {1 - \frac{{4{M^2}}}{{{{\left( {{E_p}} \right)}^2} - {{\left( {{P_{\parallel p}}} \right)}^2}}}} }\right) $}
\label{eq11}
\end{eqnarray}
Moreover, we have in c.m.s. ${P_{10}} = {\sqrt s }/2$ and ${P_{1\parallel }} = \sqrt {s/4 - {M^2}}$, where $s$ is determined by Eq.\ref{eq5}.

\begin{figure*}
\begin{center}
\includegraphics[scale=0.8]{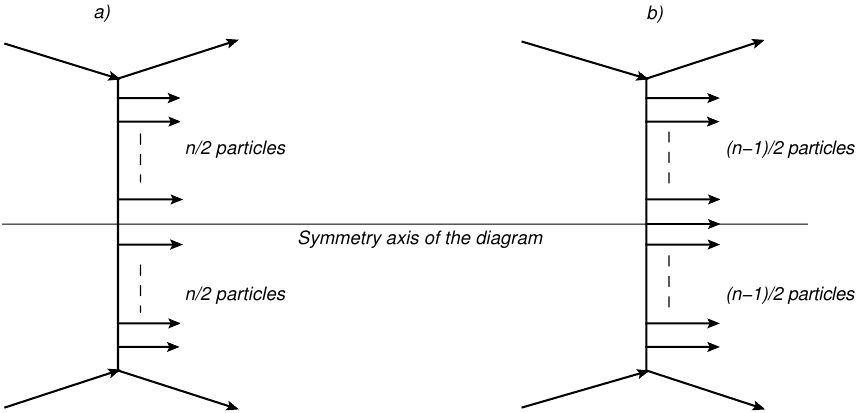}
\end{center}
\caption { An elementary inelastic scattering diagram in the multi-peripheral model with even (a) and with odd (b) number of particles on the ``comb" and it symmetry axis.}
\label{fig:part1_fig03}
\end{figure*}

For the further analysis it is convenient to switch from longitudinal momenta of secondary particles to rapidities ${y_k}$, defined by following relation:
\begin{eqnarray}%
  {p_{k\parallel }} = m \cdot sh\left( {{y_k}} \right),\quad k = 1,\;2,...,n
\label{eq12}
\end{eqnarray}
The function ${A_\parallel }$ can be written as ${A_\parallel } = {A_\parallel }\left( {n,{y_1},{y_2},...,{y_n}} \right)$. The initial state in c.m.s is symmetric with respect to changes in positive direction of collision axis. In addition, those type of diagrams of presented in Fig.\ref{fig:part1_fig02} have an axis of symmetry shown in Fig.\ref{fig:part1_fig03} for the case of even number Fig.\ref{fig:part1_fig03}(a) and for the case of odd number Fig.\ref{fig:part1_fig03}(b) of secondary particles.

From the explicit expression for amplitude Eq.\ref{eq:eq_10} can be shown that this expression will transform to itself, if we place  instead the rapidity of every particle, the rapidity of particle symmetrically arrangement along the axis of symmetry Fig.\ref{fig:part1_fig03} and at the same time change the rapidity sign. Other words, if we change the variables for even number of particles
\begin{eqnarray}%
&&  {y_1} \to  - {y_n},{y_2} \to  - {y_{n - 1}},...,{y_{\frac{n}{2}}} \to  - {y_{\frac{n}{2} + 1}}, \nonumber\\
&&  {y_{\frac{n}{2} + 1}} \to  - {y_{\frac{n}{2}}},...,{y_n} \to  - {y_1}
\label{eq13}
\end{eqnarray}
the expression of restriction amplitude ${A_\parallel }\left( {n,\;{y_{1\parallel }},\;{y_{2\parallel }},...,{y_{n\parallel }}} \right)$ will transform to itself. In case of odd number of particles the transformation similar to Eq.\ref{eq13}, but instead of the rapidity ${y_{\frac{{n - 1}}{2} + 1}}$ which forms the axis of symmetry in Fig.\ref{fig:part1_fig03}, must be substituted  $ - {y_{\frac{{n - 1}}{2} + 1}}$ into the expression for the amplitude. The examined features of function ${A_\parallel }\left( {n,\;{y_{1\parallel }},\;{y_{2\parallel }},...,{y_{n\parallel }}} \right)$ can be expressed  in the following symmetry relations for even number $n$ of secondary particles :
\begin{eqnarray}%
&&  {A_\parallel }\left( {n,{y_1},{y_2},...,{y_{\frac{n}{2}}},{y_{\frac{n}{2} + 1}},...,{y_n}} \right) \nonumber\\
&& = {A_\parallel }\left( {n, - {y_n}, - {y_{n - 1}},..., - {y_{\frac{n}{2} + 1}}, - {y_{\frac{n}{2}}},..., - {y_1}} \right)
\label{eq14}
\end{eqnarray}
and for odd number $n$ of secondary particles, respectively:
\begin{eqnarray}%
&& {A_\parallel }\left( {n,{y_1},{y_2},...,{y_{\frac{{n - 1}}{2}}},{y_{\frac{{n - 1}}{2} + 1}},{y_{\frac{{n - 1}}{2} + 2}},...,{y_{n - 1}},{y_n}} \right) \nonumber\\
&&= {A_\parallel }\left( {n, - {y_n}, - {y_{n - 1}},..., - {y_{\frac{{n - 1}}{2} + 1}}, - {y_{\frac{{n - 1}}{2}}},..., - {y_2}, - {y_1}} \right)  \nonumber\\
\label{eq15}
\end{eqnarray}
Proof of symmetry relations Eq.\ref{eq14} is given in Appendix.\ref{Appendix_B}.

Now let us switch to new variables for even number of $n$
\begin{eqnarray}%
 \mbox{\fontsize{10}{10}\selectfont $ y_1^ +  = \frac{{{y_1} + {y_n}}}{2},y_2^ +  = \frac{{{y_2} + {y_{n - 1}}}}{2},...,y_{\frac{n}{2}}^ +  = \frac{{{y_{\frac{n}{2}}} + {y_{\frac{n}{2} + 1}}}}{2}$ } \cr
  \mbox{\fontsize{10}{10}\selectfont $ y_1^ -  = \frac{{{y_1} - {y_n}}}{2},y_2^ -  = \frac{{{y_2} - {y_{n - 1}}}}{2},...,y_{\frac{n}{2}}^ -  = \frac{{{y_{\frac{n}{2}}} - {y_{\frac{n}{2} + 1}}}}{2}$ }
\end{eqnarray}
and for odd number of $n$
\begin{eqnarray}%
&& y_1^ +  = \frac{{{y_1} + {y_n}}}{2},y_2^ +  = \frac{{{y_2} + {y_{n - 1}}}}{2},...,y_{\frac{{n - 1}}{2}}^ +  = \frac{{{y_{\frac{{n - 1}}{2}}} + {y_{\frac{{n - 1}}{2} + 2}}}}{2}, \nonumber\\
&&y_{\frac{{n - 1}}{2} + 1}^ +  = {y_{\frac{{n - 1}}{2} + 1}}  \nonumber\\
&&y_1^ -  = \frac{{{y_1} - {y_n}}}{2},y_2^ -  = \frac{{{y_2} - {y_{n - 1}}}}{2},...,y_{\frac{{n - 1}}{2}}^ -  = \frac{{{y_{\frac{{n - 1}}{2}}} - {y_{\frac{{n - 1}}{2} + 2}}}}{2}\nonumber\\
\label{eq17}
\end{eqnarray}

In these variables, the symmetry relation Eq.\ref{eq14} becomes
\begin{eqnarray}%
&&  {A_\parallel }\left( {n,y_1^ + ,y_2^ + ,...,y_{\frac{n}{2}}^ + ,y_1^ - ,y_2^ - ,...,y_{\frac{n}{2}}^ - } \right) \nonumber\\
&&   = {A_\parallel }\left( {n, - y_1^ + , - y_2^ + ,..., - y_{\frac{n}{2}}^ + ,y_1^ - ,y_2^ - ,...,y_{\frac{n}{2}}^ - } \right) 
\label{eq18}
\end{eqnarray}
and relation Eq.\ref{eq15} becomes
\begin{eqnarray}%
&&{A_\parallel }\left( {n,y_1^ + ,y_2^ + ,...,y_{\frac{{n - 1}}{2} + 1}^ + ,y_1^ - ,y_2^ - ,...,y_{\frac{{n - 1}}{2}}^ - } \right) \nonumber\\
&&= {A_\parallel }\left( {n, - y_1^ + , - y_2^ + ,..., - y_{\frac{{n - 1}}{2} + 1}^ + ,y_1^ - ,y_2^ - ,...,y_{\frac{{n - 1}}{2}}^ - } \right) \nonumber\\
\label{eq19}
\end{eqnarray}

Computing step by step the partial derivative of function Eq.\ref{eq18} with respect to variables $y_1^ + ,y_2^ + ,...,y_{n/2}^ + $, we get
\begin{eqnarray}%
&& \frac{\partial }{{\partial y_k^ + }}{A_\parallel }\left( {n,y_1^ + ,y_2^ + ,...,y_{\frac{n}{2}}^ + ,y_1^ - ,y_2^ - ,...,y_{\frac{n}{2}}^ - } \right) \nonumber\\
&& =  - \frac{\partial }{{\partial y_k^ + }}{A_\parallel }\left( {n, - y_1^ + , - y_2^ + ,..., - y_{\frac{n}{2}}^ + ,y_1^ - ,y_2^ - ,...,y_{\frac{n}{2}}^ - }  \right)
\label{eq20}
\end{eqnarray}
where $k = 1,\;2,...,\frac{n}{2}$.

It follows from relation Eq.\ref{eq20} for zero values of $y_k^ + $ the derivatives of the function \mbox{\fontsize{10}{10}\selectfont ${A_\parallel }\left( {n,y_1^ + ,y_2^ + , \cdots ,y_{\frac{n}{2}}^ + ,y_1^ - ,y_2^ - ,} \right.\left. { \cdots ,y_{\frac{n}{2} - 1}^ - ,y_{\frac{n}{2}}^ - } \right)$ } vanish. In the same way it can be shown that in case of odd number $n$ of particles that for zero values of $y_k^ + $  the derivatives of scattering amplitude \mbox{\fontsize{10}{10}\selectfont ${A_\parallel }\left( {n,y_1^ + ,y_2^ + ,...,y_{\frac{{n - 1}}{2} + 1}^ + ,y_1^ - ,y_2^ - ,...,y_{\frac{{n - 1}}{2}}^ - } \right)$ } also vanishes.

This shows that for the extremum search we can consider a further reduction of the scattering amplitude on a subset of zero values of variables $y_k^ + $. By virtue of Eqs.\ref{eq17} - \ref{eq18} on this subset we have $y_k^ -  = {y_k},\;k = 1,\;2,...,\;\frac{{n - 1}}{2}$ at even $n$ and $y_k^ -  = {y_k},\;k = 1,\;2,...,\;\frac{n}{2}$ at odd $n$. Designating this reduction as $A_0$ we obtain at even $n$:
\begin{eqnarray}%
&&  {A_0}\left( {n,{y_1},{y_2},...,{y_{\frac{n}{2}}}} \right) \nonumber\\
&&  = {A_\parallel }\left( {n,{y_1},{y_2},...,{y_{\frac{n}{2}}}, - {y_{\frac{n}{2} - 1}},..., - {y_2}, - {y_1}} \right) \nonumber\\
\label{eq21}
\end{eqnarray}
and at odd $n$
\begin{eqnarray}%
&&  {A_0}\left( {n,{y_1},{y_2},...,{y_{\frac{{n - 1}}{2}}}} \right) \nonumber\\
&& = {A_\parallel }\left( {n,{y_1},{y_2},...,{y_{\frac{{n - 1}}{2} + 1}} = 0, - {y_{\frac{{n - 1}}{2} + 1}}, - {y_{\frac{{n - 1}}{2}}},..., - {y_1}} \right) \nonumber\\
\label{eq22}
\end{eqnarray}

If we now examine the formula Eq.\ref{eq:eq_10} on subset, where reduction $A_0$ is considered, we will have ${P_p} =0$ by virtue of Eq.\ref{eq6}. And therefore instead of Eq.\ref{eq11} we have the following expressions:
\begin{eqnarray}%
  \mbox{\fontsize{10}{10}\selectfont $ {P_{3\parallel }} = \frac{{{E_p}}}{2}\sqrt {1 - \frac{{4{M^2}}}{{{{\left( {{E_p}} \right)}^2}}}}; {P_{4\parallel }} = \frac{{-{E_p}}}{2}\sqrt {1 - \frac{{4{M^2}}}{{{{\left( {{E_p}} \right)}^2}}}} $ } 
\label{eq23}
\end{eqnarray}
or
\begin{eqnarray}%
  \mbox{\fontsize{10}{10}\selectfont $ {P_{3\parallel }} = \frac{{{-E_p}}}{2}\sqrt {1 - \frac{{4{M^2}}}{{{{\left( {{E_p}} \right)}^2}}}}; {P_{4\parallel }} = \frac{{{E_p}}}{2}\sqrt {1 - \frac{{4{M^2}}}{{{{\left( {{E_p}} \right)}^2}}}} $ } 
\label{eq24}
\end{eqnarray}

Let us take into account that if we decompose all scalar square terms in denominators of Eq.\ref{eq3} they will include the following difference $P_{1\parallel}-P_{3\parallel}=\sqrt{{s/4}-M^2}-P_{3\parallel}$ and negative Eq.\ref{eq24}, chosen as the $P_{3\parallel}$ in the end will give us greater value in the denominator than in the choice of Eq.\ref{eq23}. Therefore it is naturally to suppose that the main contribution to cross-section Eq.\ref{eq1} gives the range of constrained maximum point  determined by the scattering amplitude, where $P_{3\parallel}$ and $P_{4\parallel}$ are given by Eq.\ref{eq23}, but not by Eq.\ref{eq24}. Hence, considering the expression Eq.\ref{eq21} and Eq.\ref{eq22} for the reduction of the scattering amplitude at zero transverse momentum region, and performing further transformations, we assume that $P_{3\parallel}$ and related with it $P_{30}=\sqrt{M^2+(P_{3\parallel})^2}$ are expressed in terms of the longitudinal momenta of secondary particles by the relation Eq.\ref{eq23}.

After we move on to the appropriate amplitude reduction $A_0$, as in case of diagrams with even and in the case of an odd number of particles we obtain function of particles rapidities located above the axis of symmetry in the diagram. The considered features of symmetry make it possible to simplify the energy parametrization of virtual particles, to which correspond the transversal lines located above axis of symmetry of diagrams in Fig \ref{fig:part1_fig03}.

Applying symmetry relation and conversation of energy, it follows that on subset, on which the considered amplitude reduction $A_0$ is defined, the energy corresponding to the line connecting $n/2$ and $n/2+1$ vertices of the diagram in Fig.\ref{fig:part1_fig02} is equal to zero in case of even number of particles at any values of independent variables (on which $A_0$ depends). Similarly, for an odd number of particles the energy transferred along the line, which joins $(n-1)/2$ and $(n-1)/2+1$ vertices, is equal to $m/2$. The corresponding proof is given in Appendix.\ref{Appendix_C}.

Taking into account these results, reduction of $A_0$ for the diagram in Fig.\ref{fig:part1_fig02} with even number of particles can be written in the form, which is convenient for the further numerical and analytical calculations:
\begin{eqnarray}%
&& {A_0}\left( {n,{y_1},{y_2},...,{y_{\frac{n}{2}}}} \right) = {\left( {{m^2} - {{\left( {\sum\limits_{k = 1}^{\frac{n}{2}} {{E_k}} } \right)}^2} + {{\left( {{S_M}} \right)}^2}} \right)^{ - 2}} \nonumber\\
&&\times \prod\limits_{j = 2}^{\frac{n}{2}} {{{\left( {{m^2} - {{\left( {\sum\limits_{k = j}^{\frac{n}{2}} {{E_k}} } \right)}^2} + {{\left( {{S_M} - \sum\limits_{k = 1}^{j - 1} {{p_{k \parallel }}} } \right)}^2}} \right)}^{ - 2}}} \nonumber\\
&&\times {\left( {{m^2} + {{\left( {{S_M} - \sum\limits_{k = 1}^{\frac{n}{2}} {{p_{k \parallel}}} } \right)}^2}} \right)^{ - 1}}
\label{eq25}
\end{eqnarray}
where
\begin{subequations}
\begin{eqnarray}
&&  {S_M} = \sqrt {s/4 - {M^2}}  - {P_{3\parallel }} \\
&&  {E_k} = m \cdot ch\left( {{y_k}} \right) \\
&&  {p_{k \parallel}} = m \cdot sh\left( {{y_k}} \right)
\end{eqnarray}
\end{subequations}
The similar expression in case of odd number of particles in comb looks like:
\begin{eqnarray}%
&&{A_0} \left( {n,{y_1},{y_2},...,{y_{\frac{n-1}{2}}}} \right) \nonumber\\
&&= {\left( {{m^2} - {{\left( {\frac{m}{2} + \sum\limits_{k = 1}^{\frac{{n - 1}}{2}} {{E_k}} } \right)}^2} + {{\left( {{S_M}} \right)}^2}} \right)^{ - 2}} \nonumber\\
&& \times \prod\limits_{j = 2}^{\frac{{n - 1}}{2}} {{{\left( {{m^2} - {{\left( {\frac{m}{2} + \sum\limits_{k = j}^{\frac{{n - 1}}{2}} {{E_k}} } \right)}^2} + {{\left( {{S_M} - \sum\limits_{k = 1}^{j - 1} {{p_{k \parallel}}} } \right)}^2}} \right)}^{ - 2}}} \nonumber\\
&& \times {\left( {{m^2} - {{\left( {\frac{m}{2}} \right)}^2} + {{\left( {{S_M} - \sum\limits_{k = 1}^{\frac{{n - 1}}{2}} {{p_{k \parallel}}} } \right)}^2}} \right)^{ - 2}}
\label{eq26}
\end{eqnarray}

As it follows from Eqs.\ref{eq25} - \ref{eq26}, it is convenient for the further calculations to make all quantities dimensionless by mass $m$. In dimensionless form, these relations were used for numerical and analytical solution of the extremum for the reduction of the scattering amplitude $A_0$.

\section{Numerical solution for the constrained maximum problem for multi-peripheral scattering amplitude squared module}
\label{SECTION_3}

Numerical solution of the constrained extremum problem was done using Mathcad 2001 [\onlinecite{Brent2009161, Mathcad_OfficialCite}]. As it was shown in the previous section,
since scattering amplitude corresponding to Fig.\ref{fig:part1_fig02} is real and positive, we can search for amplitude maximum instead the maximum of amplitude squared module.
For cases of low values of secondary particles $n$ we search not for the maximum of reduction $A_0$, but for the maximum of total amplitude $A$ defined by Eq.\ref{eq3} with allowance for Eq.\ref{eq11} 
with the choice of positive sign in the front of ${P_{3\parallel }}$ (in order to transform this expression to Eq.\ref{eq23}, when the symmetry properties will be taken into account). These calculations are numerical verification of validity of the discussed above simplifications related to the symmetry properties.

\begin{table}
\caption{
\label{fig:part1_table01}
A typical output of numerical computation of the maximum point of scattering amplitude corresponding to the diagram presented in Fig.\ref{fig:part1_fig02}}
\begin{ruledtabular}
\begin{tabular}{cccccccc}
      index
& \multicolumn{1}{c}{$y$}
&\multicolumn{1}{c}{$p_x$}
&\multicolumn{1}{c}{$p_y$}
&\multicolumn{1}{c}{$P_{3x}$}
&\multicolumn{1}{c}{$P_{3y}$} \\
      \hline
1 	& 3.33917	& 0	& 0	& 0	& 0   \\
2	& 2.75454	& 0	& 0	& -	& -   \\
3	& 2.15274	& 0	& 0	& -	& -   \\
4	& 1.54160	& 0	& 0	& -	& -   \\
5	& 0.92590	& 0	& 0	& -	& -   \\
6	& 0.30901	& 0	& 0	& -	& -   \\
7	& -0.30901		& 0	& 0	& -	& -   \\
8	& -0.92590		& 0	& 0	& -	& -   \\
9	& -1.54160		& 0	& 0	& -	& -   \\
10	& -2.15274		& 0	& 0	& -	& -   \\
11	& -2.75454		& 0	& 0	& -	& -   \\
12	& -3.33917		& 0	& 0	& -     	& -   \\
\end{tabular}
\end{ruledtabular}
\end{table}

A typical result of such calculation for the case of $n = 12$ and energy $\sqrt s  = 55$ GeV using Maximize function of Mathcad 2001 is shown in Table.\ref{fig:part1_table01}. As it follows from Table.\ref{fig:part1_table01}, the numerical computation confirms above conclusion that all the transverse momentums must vanish at the maximum point. Note, also that the set of rapidities corresponding to the maximum point in Table.\ref{fig:part1_table01}, which are determined by numerical computation, confirms conclusion that the diagrams centered at the axis of symmetry have mutually opposite values in the point of rapidity extremum. Similar results were obtained for different numbers of particles $n$ and energies $\sqrt s$.

\begin{figure}
\begin{center}
  \centering
  \subfigure[]{
  \includegraphics[scale=0.45]{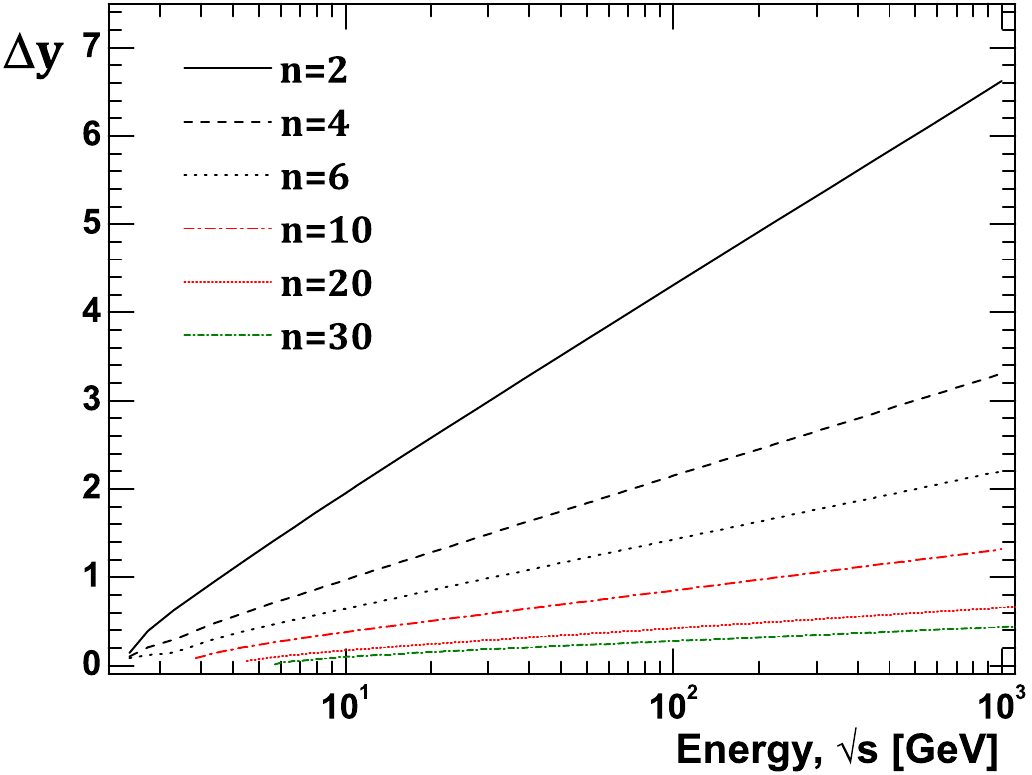}
  \label{fig:part1_fig05a}
  }
  \subfigure[]{
  \includegraphics[scale=0.45]{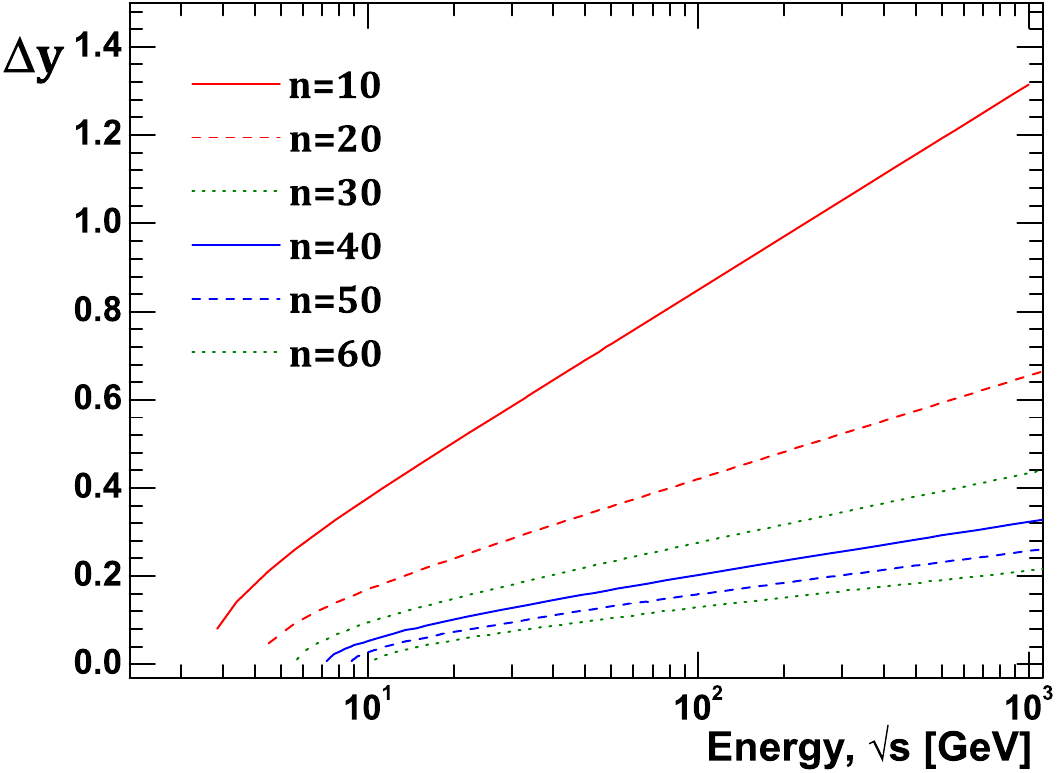}
    \label{fig:part1_fig05b}
  }
\end{center}
\caption{ The dependence of rapidity step $\Delta y $ of an arithmetic progression, which constrainedly maximizing the scattering amplitude, on energy $\sqrt s $ for the different numbers $n$ of particles on ``the comb". At high energies this dependence become logarithmic.}
\label{fig:part1_fig05}
\end{figure}

Now let us examine properties of the constrained maximum point following from the results of numerical computations. Some typical results are shown on Fig.\ref{fig:part1_fig05} and corresponding to it Table.\ref{fig:part1_table02}.

The column $y$ in Fig.\ref{fig:part1_fig05} contains the rapidities of particles
obtained using Maximize procedure (Mathcad) for which scattering amplitude $A_0$ defined by Eq.\ref{eq25} in the case of even number particles and by Eq.\ref{eq26} in the case of odd number particles has maximum reduction. Moreover, note that the order of numbers in the columns match to the order of arguments in the corresponding function.

\begin{figure}
\begin{center}
  \includegraphics[scale=0.43]{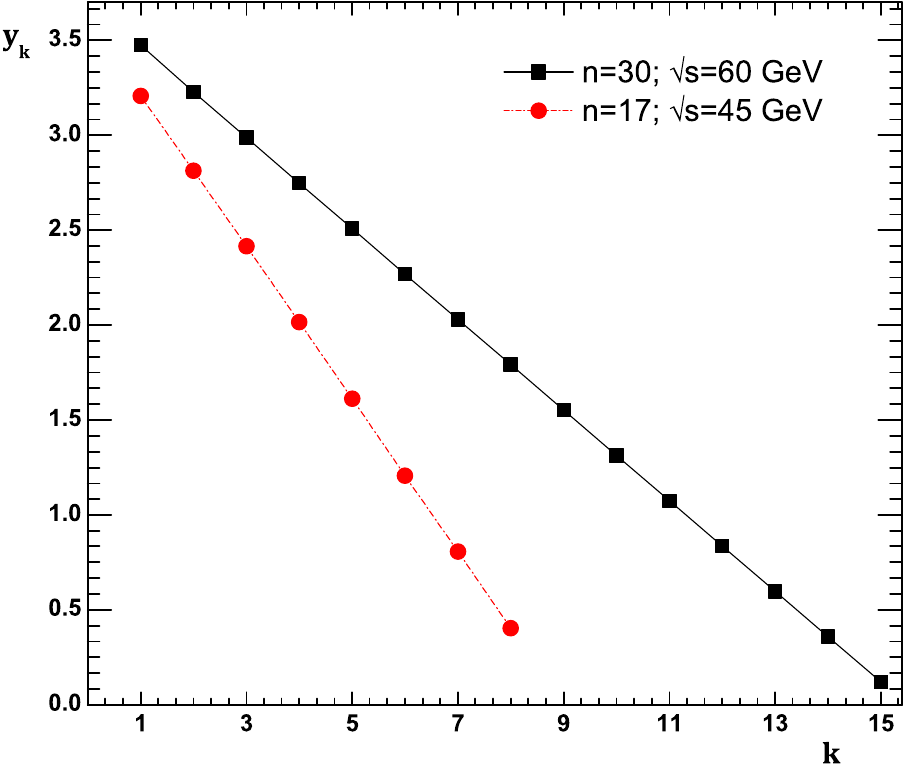}
\end{center}
 \caption{ Rapidity dependence on vertex number in the diagram Fig.\ref{fig:part1_fig02} at the constrained maximum point.}
\label{fig:part1_fig04}
\end{figure}

 For instance, in Fig.\ref{fig:part1_fig04} at $n=30$ the reduction determined by Eq.\ref{eq25} is the function of 15-en variables \\ ${A_0}\left( {n = 15, {y_1}, {y_2}, ..., {y_{15}}} \right)$ corresponding to the particles rapidities joined to the upper 15-en vertices of the diagram in Fig.\ref{fig:part1_fig02}. The column shown in Table.\ref{fig:part1_table02} contains fifteen numbers, in which the function ${A_0}\left( {n = 15,{y_1},{y_2},...,{y_{15}}} \right)$ has a maximum and at the same time the first number in the column is the value of $y_1$, the second number is the value of $y_2$ etc.

As it is apparent from Fig.\ref{fig:part1_fig04}, there is an interesting feature, which consists in the fact that the maximizing rapidities in the case of even $n$ and  in the case of odd $n$, are approximately equal to the numbers producing arithmetic progression. It is confirmed by the presented dependences in Fig.\ref{fig:part1_fig05}. $\Delta y$ values are approximately equal to each other (see Table.\ref{fig:part1_table02}).

\begin{table}
\begin{ruledtabular}
    \begin{tabular}{l|cccccc}
    \multicolumn{4}{c}{\footnotesize{Fig.\ref{fig:part1_fig04} n=30 $\sqrt{s}=60$ GeV}} \\
      \hline \hline
      index
& \multicolumn{1}{c}{$y_{k}$}
&\multicolumn{1}{c}{$\Delta y = y_{k}-y_{k+1}$}
&\multicolumn{1}{c}{$y_k/y_{15}$} \\
      \hline
1	&	3.475	& 0.2461	& 29.07	\\
2	&	3.229	& 0.2417	& 27.01	\\
3	&	2.987	& 0.2399	& 24.99	\\
4	&	2.747	& 0.2392	& 22.98	\\
5	&	2.508	& 0.2389	& 20.98    \\
6	&	2.269	& 0.2387	& 18.98	\\
7	&	2.03	& 0.2387	& 16.98	\\
8	&	1.792	& 0.2387	& 14.99	\\
9	&	1.553	& 0.2387	& 12.99	\\
10	&	1.314	& 0.2388	& 10.99	\\
11	&	1.075	& 0.2389	& 8.997	\\
12	&	0.8365	& 0.2389	& 6.998    \\
13	&	0.5976	& 0.239	& 4.999	\\
14	&	0.3586	& 0.2391	& 3	\\
15	&	0.1195	& -            & 1	\\
    \end{tabular}
\\
\vspace{0.5cm}
    \begin{tabular}{l|cccccc}
    \multicolumn{4}{c}{\footnotesize{Fig.\ref{fig:part1_fig04} n=17 $\sqrt{s}=45A$ GeV}} \\
      \hline \hline
      index
& \multicolumn{1}{c}{$y_{k}$}
&\multicolumn{1}{c}{$\Delta y = y_{k}-y_{k+1}$}
&\multicolumn{1}{c}{$y_k/y_{8}$} \\
      \hline
1		& 3.207    & 0.3929	& 7.951\\
2		& 2.814	& 0.3972	& 6.977\\
3		& 2.416	& 0.4016	& 5.992\\
4		& 2.015	& 0.4017	& 4.996\\
5		& 1.613	& 0.4061	& 4\\
6		& 1.207	& 0.3992	& 2.993\\
7		& 0.8079	& 0.4046	& 2.003\\
8		& 0.4033	& -             &1\\
    \end{tabular}
  \end{ruledtabular}
  \caption{ Resulting rapidities obtained using Maximize procedure (Mathcad) for which scattering amplitude $A_0$ defined by Eq.\ref{eq25} in the case of even number particles and by Eq.\ref{eq26} in the case of odd number particles has maximum reduction. The column $\Delta y$ contains a difference between every column element and the successor of this column. The last column contains ratios $\frac{{{y_k}}}{{{y_{15}}}},\;k = 1,\;2,...,\;15$ and $\frac{{{y_k}}}{{{y_8}}},\;k = 1,\;2,...,\;8$.}
\label{fig:part1_table02}
\end{table}

In addition, the proximity of the sequence of $y_k$ column elements to an arithmetic progression follows also from calculations of $\frac{y_k}{y_{8}}$ and $\frac{y_k}{y_{15}}$ columns, which are constructed by the following principle. If we assume that $y_k$ form an arithmetic progression with difference $\Delta y$, in the case of even $n$ we will obtain ${y_{\frac{n}{2} + 1}} = {y_{\frac{n}{2}}} - \Delta y$. On the other hand, from symmetry relations obtained in Section.\ref{SECTION_2} we have ${y_{\frac{n}{2} + 1}} =  - {y_{\frac{n}{2}}}$. From these two relations we obtain ${y_{\frac{n}{2}}} = \frac{{\Delta y}}{2}$. Then ${y_{\frac{n}{2} - 1}} = {y_{\frac{n}{2}}} + \Delta y = 3\frac{{\Delta y}}{2} = 3{y_{\frac{n}{2}}}$ and in the similar manner ${y_{\frac{n}{2} - 2}} = {y_{\frac{n}{2}}} + 2\Delta y = 5\frac{{\Delta y}}{2} = 5{y_{\frac{n}{2}}}$, etc. Hence, in case of the diagram with even $n$ the rapidity ratios $\frac{{{y_{\frac{n}{2} - 1}}}}{{{y_{\frac{n}{2}}}}},\frac{{{y_{\frac{n}{2} - 2}}}}{{{y_{\frac{n}{2}}}}},...,\frac{{{y_1}}}{{{y_{\frac{n}{2}}}}}$ must form the sequence of odd whole numbers. The columns $\frac{y_k}{y_{8}}$ and $\frac{y_k}{y_{15}}$ contains these ratios constructed by the column elements $y_k$ determined with the help of Maximize function \onlinecite{Brent2009161}. As seen from Table.\ref{fig:part1_table02}, the column elements $\frac{y_k}{y_{8}}$ and $\frac{y_k}{y_{15}}$  are really close to odd numbers.

In case of odd $n$ with allowance for ${y_{\frac{{n - 1}}{2} + 1}} = 0$ we have
\begin{eqnarray}%
 {y_{\frac{{n - 1}}{2}}} = {y_{\frac{{n - 1}}{2} + 1}} + \Delta y = \Delta y \cr
  {y_{\frac{{n - 1}}{2} - 1}} = {y_{\frac{{n - 1}}{2}}} + \Delta y = 2\Delta y
\end{eqnarray}

Then the ratios $\frac{{{y_{\frac{{n - 1}}{2}}}}}{{{y_{\frac{{n - 1}}{2}}}}},\frac{{{y_{\frac{{n - 1}}{2} - 1}}}}{{{y_{\frac{{n - 1}}{2}}}}},\frac{{{y_{\frac{{n - 1}}{2} - 2}}}}{{{y_{\frac{{n - 1}}{2}}}}}$, must produce the sequence of whole numbers $1,2,...$ . And Table.\ref{fig:part1_table01}, where columns $\frac{y_k}{y_{8}}$ and $\frac{y_k}{y_{15}}$ consist of such ratios, shows that these ratios are really close to the corresponding whole numbers. The similar results are obtained for different numbers of particles $n$ and energies $\sqrt s $.

The analytic form of arithmetic progression at any $n$ and $\sqrt s $ will be considered in more detail below the text, when we will given an analytical solution of the extremum problem for reductions Eq.\ref{eq25} and Eq.\ref{eq26}.

Moreover, the results of numerical computations confirm the well-known multi-peripheral model assumption about particle rapidity ordering, because at the maximum point the particle rapidities monotone increase at movement upward along the diagram in Fig.\ref{fig:part1_fig02}.

Besides of these results numerical computation allows to trace several other properties of the extremum point. In particular, if the rapidities in the maximum point form an arithmetic progression, the question arises, how the difference $\Delta y $ of this arithmetic progression depends on the energy $\sqrt s $ and the number of particles $n$? Result of numerical computation of the dependence of difference of an arithmetic progression $\Delta y $ on $\sqrt s $ at different numbers of particles on the ``comb" Fig.\ref{fig:part1_fig02} are shown in Fig.\ref{fig:part1_fig05}. At the same time the arithmetical average of column elements $\Delta y$ (the similar to shown in Table.\ref{fig:part1_table02}) was used as the value of $\Delta y $.

As its obvious from Fig.\ref{fig:part1_fig05}, the $\Delta y $ is a monotonically increasing function of energy $\sqrt s $ for a fixed number $n$ of particles. At the same time the dependence $\Delta y\left( {n = const,\sqrt s } \right)$ has some energy threshold. It is obvious, that at given $n$ such dependence makes sense only at
\begin{eqnarray}%
\sqrt s  \ge nm + 2M
\label{eq27}
\end{eqnarray}
(where $\sqrt s $ is not dimensionless by $m$), that corresponds to the total rest energy of particle in final state. Note that function $\Delta y\left( {\sqrt s } \right)$ reaches asymptotic quite quickly. Since Fig.\ref{fig:part1_fig05} has a logarithmic scale on the energy axis, its seen that this asymptotic behavior is characterized by the linear dependence on logarithm of energy normalized to 1 GeV. 
\begin{table}
   \begin{ruledtabular}
    \begin{tabular}{l|cccccc}
    & \multicolumn{3}{c}{$\sqrt{s}$} \\
       & \multicolumn{1}{c}{$10$ GeV} &\multicolumn{1}{c}{$100$ GeV}  &\multicolumn{1}{c}{$1$ TeV} \\
\hline
\large{$\frac{{\ln \left( {\Delta {y_4}} \right) - \ln \left( {\Delta {y_2}} \right)}}{{\ln (4) - \ln (2)}} $}  	& -1.009 & -1.005 & -1.002  \\  \\
\large{$\frac{{\ln \left( {\Delta {y_6}} \right) - \ln \left( {\Delta {y_2}} \right)}}{{\ln (6) - \ln (2)}}$} 	&-1.011  & -1.006 & -1.003  \\ \\
\large{$\frac{{\ln \left( {\Delta {y_{20}}} \right) - \ln \left( {\Delta {y_2}} \right)}}{{\ln (20) - \ln (2)}} $}    &-1.057 & -1.011 & -1.002  \\ \\
\large{$\frac{{\ln \left( {\Delta {y_{40}}} \right) - \ln \left( {\Delta {y_2}} \right)}}{{\ln (40) - \ln (2)}}$}	&-1.203 & -1.02 & -1.009  \\ \\
\large{$\frac{{\ln \left( {\Delta {y_{60}}} \right) - \ln \left( {\Delta {y_2}} \right)}}{{\ln (60) - \ln (2)}}$}	&-1.82   & -1.029& -1.012  \\
    \end{tabular}
  \end{ruledtabular}
  \caption{ The difference of rapidity $\Delta y $  from Fig.\ref{fig:part1_fig06}. From these results it is evident that the dependence is close to inverse proportion.}
\label{fig:part1_table03}
\end{table}
\begin{figure}
\begin{center}
\includegraphics[scale=0.41]{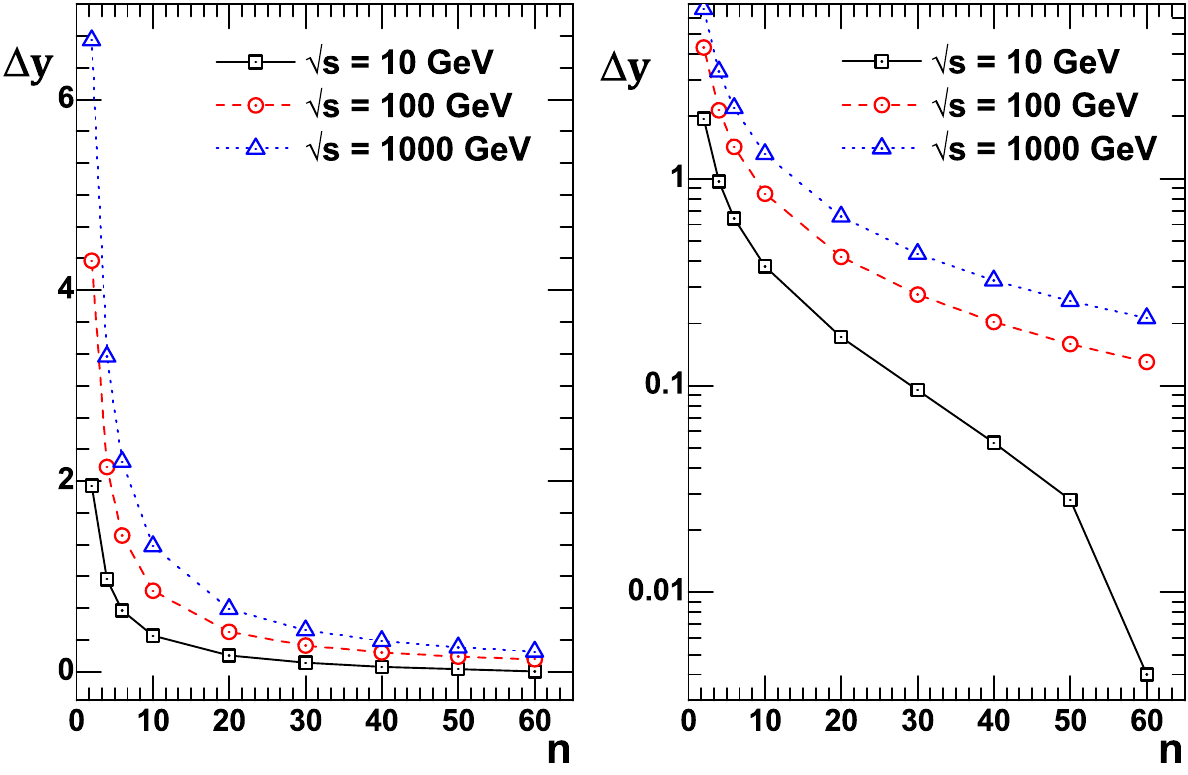}
\end{center}
\caption{ Dependence of difference $\Delta y $ in an arithmetic progression, which constrainedly maximizing the inelastic scattering amplitude for the different number of particles $n$ on the ``comb" at the fixed energies $\sqrt s $: 10 GeV, 100 GeV and 1000 GeV.}
\label{fig:part1_fig06}
\end{figure}

\begin{figure*}
\begin{center}
  \centering
  \subfigure[]{
  \includegraphics[scale=0.42]{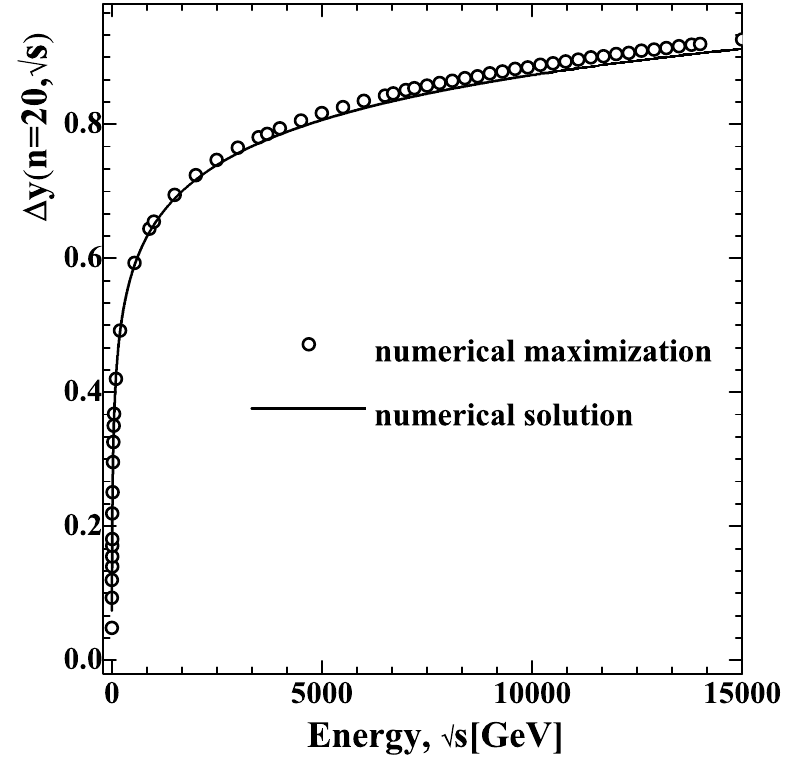}
  \label{fig:fig_part1_10a}
  }
  \subfigure[]{
  \includegraphics[scale=0.42]{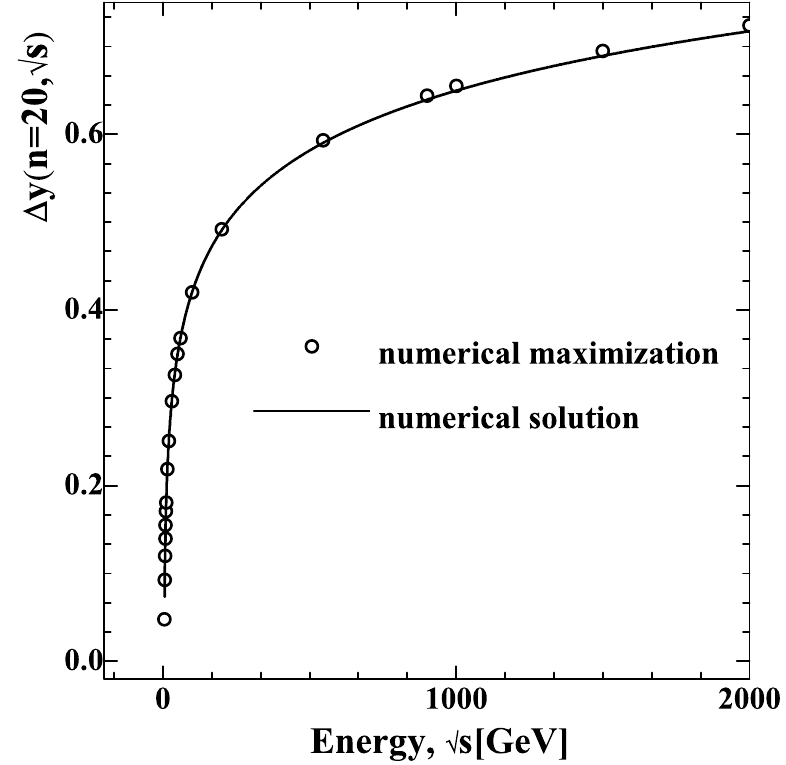}
    \label{fig:fig_part1_10b}
  }
  \subfigure[]{
  \includegraphics[scale=0.42]{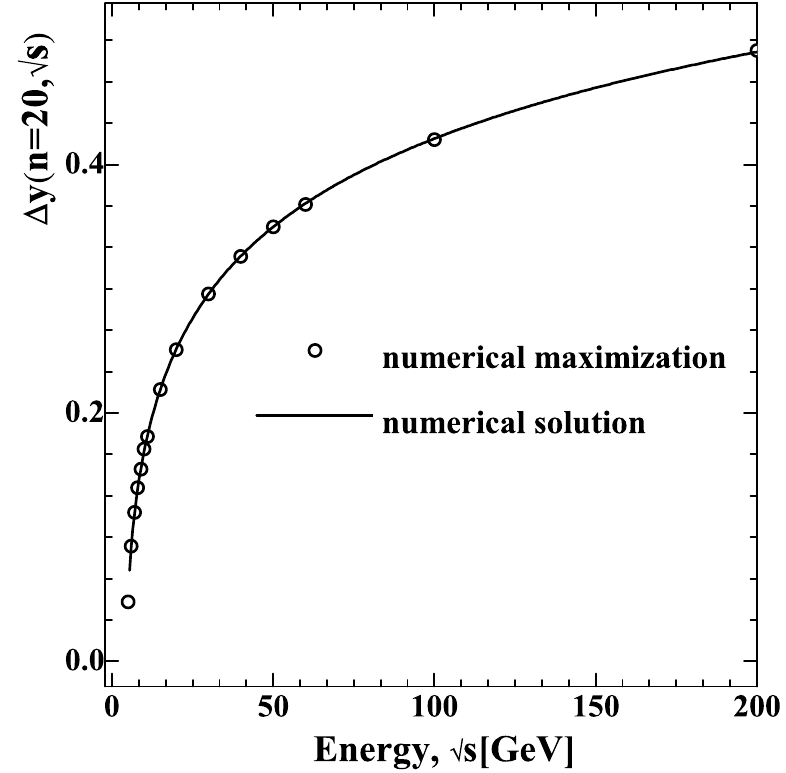}
  \label{fig:fig_part1_10c}
  }
\end{center}
 \caption{The results of the numerical solution of Eq.\ref{eq64} (solid line) with the results of numerical maximization (circles) of the magnitude $\Delta y (n,\sqrt{s})$ at $n=20$ for the different energy ranges, 
GeV: $5\div16000$ (\ref{fig:fig_part1_10a}); $5\div2000$ (\ref{fig:fig_part1_10b}); $5\div200$ (\ref{fig:fig_part1_10c}). Note, here we taken into account that $\Delta y (n,\sqrt{s})=2y_{n\over 2}$.}
 \label{fig:part1_fig08}
\end{figure*}

Output computation of the dependence of difference of an arithmetic progression $\Delta y $ on the number of particles for set of energies $\sqrt s = 10$ GeV, $100$ GeV, $1000$ GeV is presented in Fig.\ref{fig:part1_fig06}. From Fig.\ref{fig:part1_fig06} (where dependences $\Delta y(n) $ are given in linear and logarithmic scales) it is evident that, when $n$ is small in comparison with the boundary value determined by Eq.\ref{eq27}, this dependence is close to inversely.
\begin{figure}
\begin{center}
  \centering
  \subfigure[]{
  \includegraphics[scale=0.45]{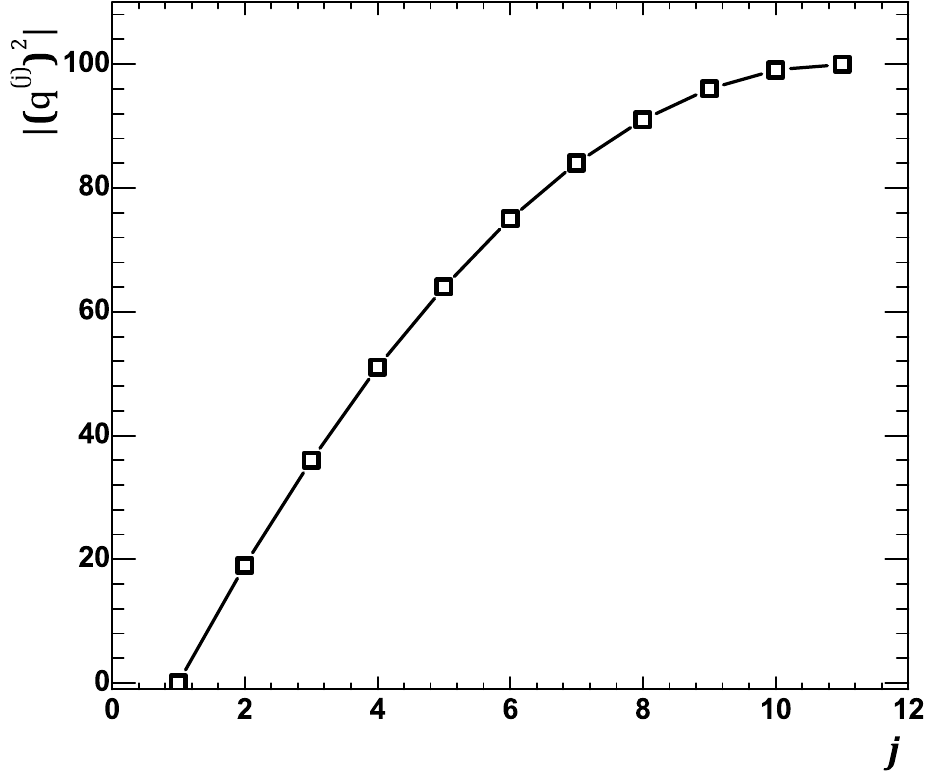}
  \label{fig:part1_fig07a}
  }
  \subfigure[]{
  \includegraphics[scale=0.45]{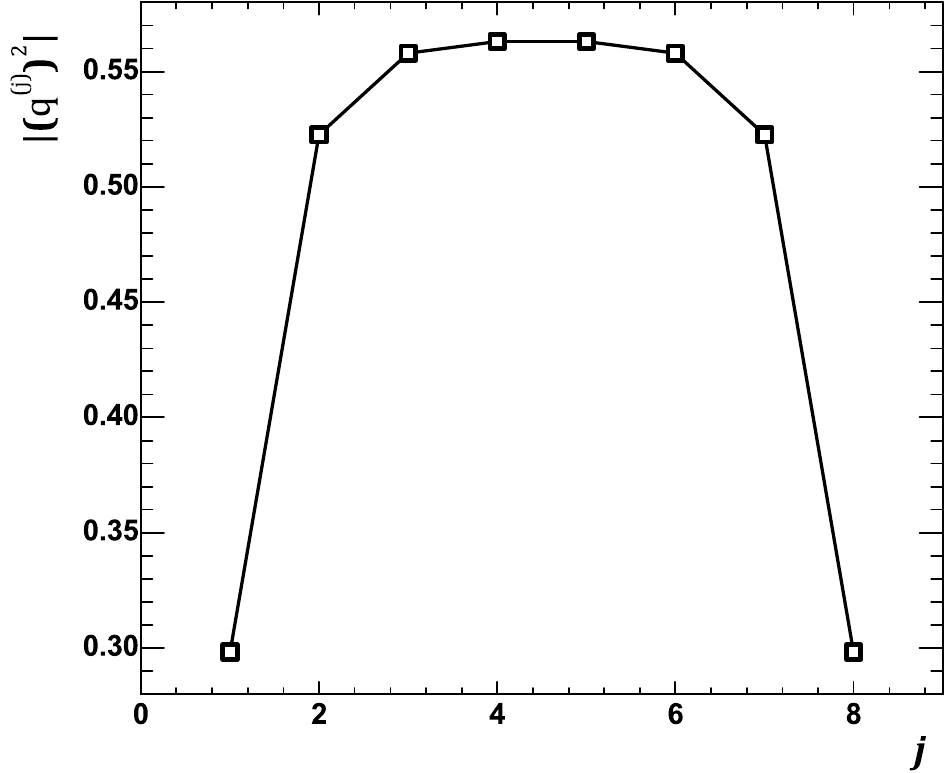}
    \label{fig:part1_fig07b}
  }
\end{center}
\caption{ The virtuality variation along ``the comb". \ref{fig:part1_fig07a} shows particles virtuality dependence on vertex number on diagram Fig.\ref{fig:part1_fig02}, for even number of particles ($n = 20$). Only the upper half of the diagram was considered, because as it follows from symmetry relations, discussed in Section.\ref{SECTION_2}, virtualities at the maximum point on the lines symmetrical to symmetry axis are equal. This also evident from \ref{fig:part1_fig07a}, which shows the virtuality variation along the whole ``the comb" for odd number of particles ($n = 7$).}
\label{fig:part1_fig07}
\end{figure}
In particular, if we examine $\ln \left( {\Delta y} \right)$ as a function of $\ln \left( n \right)$, it is possible to calculate corresponding difference of an arithmetic progressions for the given energy (they are marked as $\Delta {y_1},\Delta {y_2},...,\Delta {y_k}$ in  Fig.\ref{fig:part1_fig08}). As it follows from Table.\ref{fig:part1_table03}, ratios
\begin{eqnarray}%
  \frac{{\ln \left( {\Delta {y_j}} \right) - \ln \left( {\Delta {y_2}} \right)}}{{\ln \left( {{n_j}} \right) - \ln \left( {{2}} \right)}}
\end{eqnarray}
where $j = 1, 2, 3,...,k$ are close to $(-1)$,
which suggests that if we fix $\sqrt s $ and examine $n \ll \frac{{\sqrt s  - 2M}}{m}$, that is much smaller value than the maximum allowable by law of conservation of energy for a given value $\sqrt s $, then we have $\Delta y \sim {n^{ - 1}}$.

More specific information about the dependence $\Delta y\left( {n,\sqrt s } \right)$ can be obtained from analytical solution of the constrained extremum problem for scattering amplitude, which will be considered in the next section.

However, the greatest interest is the dependence of the absolute values of virtualities on energy (i.e. the virtual particle squared four-momentums corresponding to the diagram in Fig.\ref{fig:part1_fig02}, calculated at values of real particle four-momentums for which the scattering amplitude has a constrained maximum). Discussion of this problem leads us to possible mechanism of inelastic scattering cross-section growth with energy.

Let us designate the virtual particle four-momentums according to Fig.\ref{fig:part1_fig02}:
\begin{eqnarray}%
&& {k^{\left( 1 \right)}} = {P_1} - {P_3},{k^{\left( 2 \right)}} = {P_1} - {P_3} - {p_1},..., \nonumber\\
&& {k^{\left( j \right)}} = {P_1} - {P_3} - \sum\limits_{l = 1}^{j - 1} {{p_l}} ,...
\label{eq28}
\end{eqnarray}

The numbers of four-momentums are taken in brackets to distinguish them from the four-momentum components notation. For example, notation ${k^{\left( 0 \right)0}}$ denotes zero component of contravariant four-vector ${k^{\left( 0 \right)}}$.

As was noted above, all the quantities were made dimensionless by the mass $m$ before calculations, therefore we introduce the notation
\begin{eqnarray}%
  {q^{\left( j \right)}} = \frac{{{k^{\left( j \right)}}}}{m}
\label{eq29}
\end{eqnarray}

At the same time the magnitude $A$, which is determined by Eq.\ref{eq3} and coincident with scattering amplitude accurate within constant, can be written in dimensionless form like:
\begin{eqnarray}%
  A = \prod\limits_{j = 1}^{n + 1} {{{\left( {1 + \left| {{{\left( {{q^{\left( j \right)}}} \right)}^2}} \right|} \right)}^{ - 1}}}
\label{eq30}
\end{eqnarray}

\begin{figure}
\begin{center}
\includegraphics[scale=0.53]{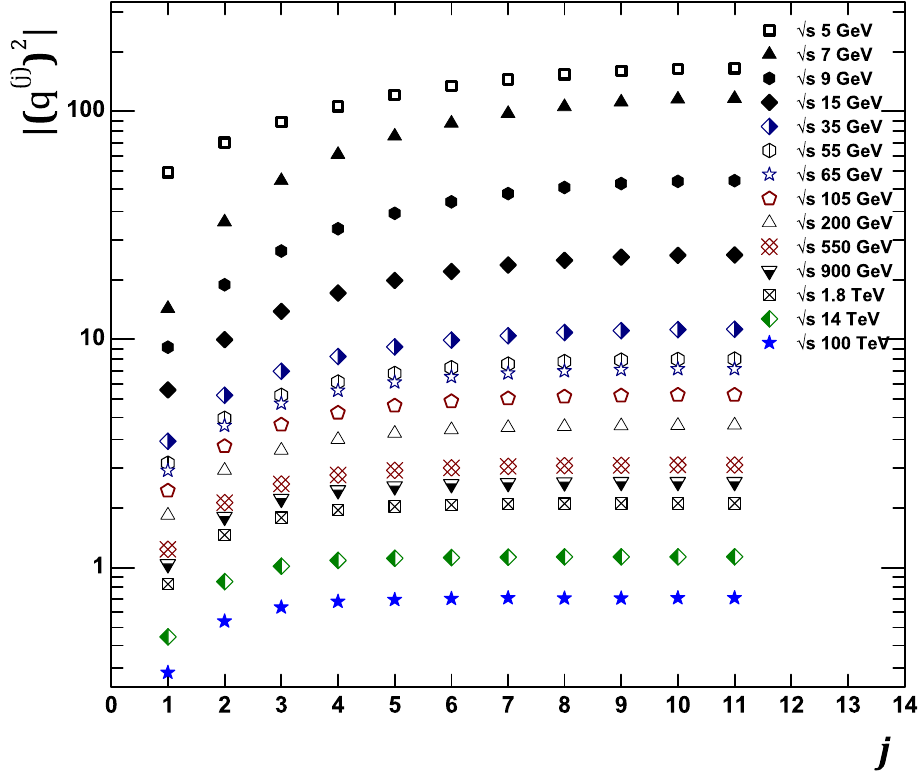}
\end{center}
\caption{Variation of the modulus of virtuality with energy for $n = 20$.}
\label{fig:part1_fig09}
\end{figure}

From numerical computation we know particles four-momentums in final state, for which function $A$ has constrained maximum. Therefore, with help of relations Eq.\ref{eq28} and Eq.\ref{eq29} we can calculate four-momentums ${q^{\left( j \right)}}$ and whereupon calculate the corresponding dimensionless virtualities. Results of such calculation presented on Fig.\ref{fig:part1_fig07} and Fig.\ref{fig:part1_fig09}.
As it well known, that there is reference frame where the particle energy vanishes [\onlinecite{DeBruijn:225131}] for particles with negative virtuality. Such a reference frame is called the standard reference system according to terminology of [\onlinecite{DeBruijn:225131}]. Obviously that the particle momentum in the standard reference system has the smallest possible value of all inertial frames of reference, and this smallest value is defined by particle`s virtuality.

Taking into account the Heisenberg uncertainty principle, we find that the virtuality characterizes a size of the domain, in which a particle can be detected with high accuracy, if the measurements were made in the standard reference system of this particle. At the same time, the more particle`s virtuality, the smaller this size. Thus, the virtuality makes it possible to judge the spatial extension of those domains, where inelastic processes take place described by the diagrams of type Fig.\ref{fig:part1_fig02}. These space domains form the virtual ``coats" of colliding particles, therefore these spatial extensions define typical sizes of colliding particles $P_1$ and $P_2$, see Fig.\ref{fig:part1_fig02}. From Fig.\ref{fig:part1_fig07} it is evident that virtualities increase with movement from the diagram edges to its center. This is easy to explain, because all the virtualities are negative and therefore
\begin{eqnarray}%
  \mbox{\fontsize{10}{10}\selectfont $ {
\left( {{k^{(1)}}} \right)^2} = {\left( {P_1^0-P_3^0} \right)^2}-{\left( {{P_{1\parallel }}-{P_{3\parallel }}} \right)^2}-{({\overrightarrow P _{3 \bot }})^2} < 0
$}\nonumber\\
\end{eqnarray}

Taking into account that transversal components of momentum are equal to zero at the maximum point, we have:
\begin{eqnarray}%
  {\left( {{P_{1\parallel }} - {P_{3\parallel }}} \right)^2} > \left( {P_1^0 - P_3^0} \right)
\label{eq31}
\end{eqnarray}

According to the law of conservation of energy we have $P_3^0 < P_1^0$. At the same time\\
$P_1^0 = \sqrt {{M^2} + {{({P_{1\parallel }})}^2}} $\\
and\\
$P_3^0 = \sqrt {{M^2} + {{({P_{3\parallel }})}^2}} $,\\
therefore ${({P_{3\parallel }})^2} < {({P_{1\parallel }})^2}$. Since both these expressions are positive, we have ${P_{3\parallel }} < {P_{1\parallel }}$. Therefore due to positivity of expressions in brackets, we obtain from Eq.\ref{eq31}:
\begin{eqnarray}%
  {P_{1\parallel }} - {P_{3\parallel }} > P_1^0 - P_3^0
\end{eqnarray}

However, when we move from ${k^{\left( 1 \right)}}$ to ${k^{\left( 2 \right)}}$, we subtract $ch\left( {{y_1}} \right)$ from the lower side $P_1^0 - P_3^0$ and $sh\left( {{y_1}} \right)$ from the higher side ${P_{1\parallel }} - {P_{3\parallel }}$. Consequently, the difference of these quantities increases, and so naturally to expect that the difference of squares, i.e., virtuality $\left( {{k^{(2)}}} \right)^2$, increases too Fig.\ref{fig:part1_fig07}.
Turning to each subsequent virtuality, we will subtract a hyperbolic cosine of corresponding rapidity from the energy and subtract a hyperbolic sine of the same rapidity from the longitudinal momentum. The difference between the longitudinal momentum and the energy will increase, which explains the increase in differences of their squares, i.e. virtualities.

Behavior of virtuality at the maximum point for different energies is shown in Fig.\ref{fig:part1_fig09}. From the results in Fig.\ref{fig:part1_fig09} it follows that at the maximum point the virtuality monotone decreases with the energy growth. Note from Eq.\ref{eq30}, that decrease of the virtuality should lead to amplitude increasing with energy at the maximum point, and for reasons outlined in Section.\ref{INTRO} this should lead to increase of partial cross sections with energy $\sqrt s$ growth.
Such an effect is a consequence of ​rapidity increasing at the maximum point, and therefore in principle can not be taken into account within framework of multi-Regge kinematics [\onlinecite{levin_2, KozlovNSU_2007, Lipatov:2008, Lipatov:2004}] due to the fact that this dependence is neglected in the integrand of Eq.\ref{eq1}.

Further, in the second part of our paper we will discuss the question of whether the amplitude growth can lead to growth in cross-section ${\sigma _n}$ calculated by Laplace method and in total scattering cross-section ${\sigma_{total}}$.
\section{An analytical solution for the constrained extremum problem in the approximation of equal-denominators}
\label{SECTION_4}
We now consider the case of even number of particles $n$ in the diagram of Fig.\ref{fig:part1_fig02}. The scattering amplitude reduction $A_0(n, y_1,y_2,\ldots,y_{n/2})$, defined by Eq.\ref{eq25} and nondimensionalized with the mass $m$, can be expressed as:
\begin{eqnarray}%
&& {A_0}\left( {n,{y_1},{y_2},...,{y_{\frac{n}{2}}}} \right) =  \nonumber\\
&& = {\left( {1 - {{\left( {\sum\limits_{k = 1}^{\frac{n}{2}} {ch\left( {{y_k}} \right)} } \right)}^2} + {{\left( {{P_{1\parallel }}}-{{P_{3\parallel }}} \right)}^2}} \right)^{ - 2}} \nonumber\\
&& \times {\left( {1 + {{\left( { {P_{1\parallel }}-{P_{3\parallel }} - \sum\limits_{k = 1}^{\frac{n}{2}} {sh\left( {{y_k}} \right)} } \right)}^2}} \right)^{ - 1}} \nonumber\\
&& \mbox{\fontsize{9}{10}\selectfont $ \times \prod\limits_{j = 2}^{\frac{n}{2}} {{{\left( {1 - {{\left( {\sum\limits_{k = j}^{\frac{n}{2}} {ch\left( {{y_k}} \right)} } \right)}^2} + {{\left( { {P_{1\parallel }}-{P_{3\parallel }} - \sum\limits_{k = 1}^{j - 1} {sh\left( {{y_k}} \right)} } \right)}^2}} \right)}^{ - 2}}} $} \nonumber\\
\label{eq33}
\end{eqnarray}
Here $P_{1\parallel}=\sqrt{{s/4}-M^2}$, and instead of $\sqrt{s}$ and $M$ we use their 
nondimensionalized (with the mass $m$) values. Because we search for the constrained extremum upon the condition of energy-momentum conservation, it is assumed that $P_3$ is described by Eq.\ref{eq23}. Again all values are nondimensionalized with the mass $m$. Taking into account the symmetry relation, normalization and the introduction of rapidity (see Eq.\ref{eq12}), we obtain 
an alternative expression for $E_p$ then Eq.\ref{eq6}:
\begin{eqnarray}%
  {E_p} = \sqrt s  - 2\sum\limits_{k = 1}^{\frac{n}{2}} {ch\left( {{y_k}} \right)}
\label{eq34}
\end{eqnarray}
For further calculations we use the following definition:
\begin{eqnarray}%
 E = \sum\limits_{k = 1}^{\frac{n}{2}} {ch\left( {{y_k}} \right)} \quad {\rm and} \quad \Delta P = {P_{1\parallel }} - {P_{3\parallel }}
\label{eq35}
\end{eqnarray}
Note, that due to Eqs.\ref{eq23}, \ref{eq34} and Eq.\ref{eq35} quantity $\Delta P$ depends on rapidity as the composite function of $E$, which is denoted as $\Delta P(E)$.  The first term of Eq.\ref{eq33} depends upon rapidity only via $E$.

Rather than looking for the maximum of function $A_0(n, y_1,y_2,\ldots,y_{n/2})$ we can instead look for the maximum of its logarithm. This shall be defined as $L$, where:
\begin{eqnarray}%
&&  L =  - 2\ln \left( {1 - {{\left( E \right)}^2} + {{\left( {\Delta P\left( E \right)} \right)}^2}} \right) \nonumber\\
&&  \mbox{\fontsize{9}{10}\selectfont $ - 2\sum\limits_{j = 2}^{\frac{n}{2}} {\ln \left( {1 - {{\left( {\sum\limits_{k = j}^{\frac{n}{2}} {ch\left( {{y_k}} \right)} } \right)}^2} + {{\left( {\Delta P\left( E \right) - \sum\limits_{k = 1}^{j - 1} {sh\left( {{y_k}} \right)} } \right)}^2}} \right)}$ } \nonumber\\
&&   - \ln \left( {1 + {{\left( {\Delta P\left( E \right) - \sum\limits_{k = 1}^{\frac{n}{2}} {sh\left( {{y_k}} \right)} } \right)}^2}} \right)
\end{eqnarray}
We now make the following definitions:
\begin{eqnarray}%
&&{Z_1} = 1 - {\left( E \right)^2} + {\left( {\Delta P\left( E \right)} \right)^2} \cr
&&{Z_j} = 1 - {\left( {\sum\limits_{k = j}^{\frac{n}{2}} {ch\left( {{y_k}} \right)} } \right)^2} + {\left( {\Delta P\left( E \right) - \sum\limits_{k = 1}^{j - 1} {sh\left( {{y_k}} \right)} } \right)^2} \cr
&&{Z_{\frac{n}{2} + 1}} = 1 + {\left( {\Delta P\left( E \right) - \sum\limits_{k = 1}^{\frac{n}{2}} {sh\left( {{y_k}} \right)} } \right)^2} \nonumber\\
\label{eq4.5}
\end{eqnarray}
where $j = 1,2,...,\frac{n}{2}$

After taking into account Eq.\ref{eq6}, all variables of function $A_0(n, y_1,y_2,\ldots,y_{n/2})$ and therefore of its logarithm become independent. Therefore, the point of maximum can be found under the condition that partial derivatives with respect to
all variables are zero. The equations for the extreme point can be written down in a form
\begin{eqnarray}%
&&  \mbox{\fontsize{13}{14}\selectfont $ \frac{{\partial L}}{{\partial {y_1}}} = \frac{{\partial L}}{{\partial E}}sh\left( {{y_1}} \right) + 4ch\left( {{y_1}} \right)\sum\limits_{j = 2}^{\frac{n}{2}} {\frac{{\Delta P\left( E \right) - \sum\limits_{k = 1}^{j - 1} {sh\left( {{y_k}} \right)} }}{{{Z_j}}}}$ }  \nonumber\\
&&  \mbox{\fontsize{13}{14}\selectfont $ + 2ch\left( {{y_1}} \right)\frac{{\Delta P\left( E \right) - \sum\limits_{k = 1}^{\frac{n}{2}} {sh\left( {{y_k}} \right)} }}{{{Z_{\frac{n}{2} + 1}}}} = 0$ } 
\label{eq4.6}
\end{eqnarray}
\\
\begin{eqnarray}%
&&  \mbox{\fontsize{13}{14}\selectfont $ \frac{{\partial L}}{{\partial {y_l}}} = \frac{{\partial L}}{{\partial E}}sh\left( {{y_l}} \right) + 4sh\left( {{y_l}} \right)\sum\limits_{j = 2}^l {\frac{{\sum\limits_{k = j}^{\frac{n}{2}} {ch\left( {{y_k}} \right)} }}{{{Z_j}}}}$ } \nonumber\\
&&  \mbox{\fontsize{13}{14}\selectfont $ + 4ch\left( {{y_l}} \right)\sum\limits_{j = l + 1}^{\frac{n}{2}} {\frac{{\Delta P\left( E \right) - \sum\limits_{k = 1}^{j - 1} {sh\left( {{y_k}} \right)} }}{{{Z_j}}}}$ }  \nonumber\\
&&  \mbox{\fontsize{13}{14}\selectfont $ + 2ch({y_l})\frac{{\Delta P(E) - \sum\limits_{k = 1}^{\frac{n}{2}} {sh({y_k})} }}{{{Z_{\frac{n}{2} + 1}}}} = 0$ }
\end{eqnarray}
where $l = 2,\;3,\;...\;\frac{n}{2} - 1$. \\
\begin{eqnarray}%
&&  \mbox{\fontsize{12}{12}\selectfont $ \frac{{\partial L}}{{\partial {y_{\frac{n}{2}}}}} = \frac{{\partial L}}{{\partial E}}sh\left( {{y_{\frac{n}{2}}}} \right) + 4sh\left( {{y_{\frac{n}{2}}}} \right)\sum\limits_{j = 2}^{\frac{n}{2}} {\frac{{\sum\limits_{k = j}^{\frac{n}{2}} {ch\left( {{y_k}} \right)} }}{{{Z_j}}}}$ } \nonumber\\
&&  \mbox{\fontsize{13}{14}\selectfont $ + 2ch\left( {{y_{\frac{n}{2}}}} \right)\frac{{\Delta P\left( E \right) - \sum\limits_{k = 1}^{\frac{n}{2}} {sh\left( {{y_k}} \right)} }}{{{Z_{\frac{n}{2} + 1}}}} = 0$ }
\label{eq4.8}
\end{eqnarray}
Eqs.\ref{eq4.6} - \ref{eq4.8} form the system of equations to be solved to find the extremum point.
An approximation can be made to this system of equations. This we will call ``the equal-denominators approximation”.

As seen from Eq.\ref{eq30}, the amplitude is a product of fractions whose denominators contain the expression greater than unit. However, function $f(x) = 1/x$ slowly varying at $x>1$, that follows from its derivative. In addition, as discussed in Section.\ref{SECTION_3}, virtualities increase, and hence, as it evident from  Eq.\ref{eq31} and arguments made after this relation, the denominators increase with movement from the edges of the ``comb" to its center. However, as we are looking for the maximum point, virtualities at this point should be as small as possible.
Therefore, they should increase as we move from the edges of the ``comb" to the center as slowly as possible.
Hence, we can expect that the denominators lightly differs from each other in the desired maximum point, then for the further analysis of this system at the maximum point we have adopted an approximation, in which all the denominators are equal between themselves. Their approximate common value will be denoted as $Z$,
\begin{eqnarray}%
  {Z_j} \approx Z, \quad j = 1,2, \cdots ,\frac{n}{2} + 1
\label{eq41}
\end{eqnarray}
In this case, the system of equations for the maximum point takes the form:
\begin{eqnarray}%
&&  \mbox{\fontsize{13}{14}\selectfont $ \frac{Z}{2}\frac{{\partial L}}{{\partial E}} + 2\frac{{ch\left( {{y_1}} \right)}}{{sh\left( {{y_1}} \right)}}\sum\limits_{j = 2}^{\frac{n}{2}} {\left( {\Delta P\left( E \right) - \sum\limits_{k = 1}^{j - 1} {sh\left( {{y_k}} \right)} } \right)}$ }  \nonumber\\
&&  \mbox{\fontsize{13}{14}\selectfont $  + \frac{{ch\left( {{y_1}} \right)}}{{sh\left( {{y_1}} \right)}}\left( {\Delta P\left( E \right) - \sum\limits_{k = 1}^{\frac{n}{2}} {sh\left( {{y_k}} \right)} } \right) = 0 $ } 
\label{eq42}
\end{eqnarray}
\begin{eqnarray}%
&& \frac{Z}{2}\frac{{\partial L}}{{\partial E}} + 2\sum\limits_{j = 2}^l {\left( {\sum\limits_{k = j}^{\frac{n}{2}} {ch\left( {{y_k}} \right)} } \right)}  \nonumber\\
&&+ 2\frac{{ch\left( {{y_l}} \right)}}{{sh\left( {{y_l}} \right)}}\sum\limits_{j = l + 1}^{\frac{n}{2}} {\left( {\Delta P\left( E \right) - \sum\limits_{k = 1}^{j - 1} {sh\left( {{y_k}} \right)} } \right)} \nonumber\\
&& + \frac{{ch\left( {{y_l}} \right)}}{{sh\left( {{y_l}} \right)}}\left( {\Delta P\left( E \right) - \sum\limits_{k = 1}^{\frac{n}{2}} {sh\left( {{y_k}} \right)} } \right) = 0 
\label{eq43}
\end{eqnarray}
here $l = 2,\;3,...,\;\frac{n}{2} - 1$.
\begin{eqnarray}%
&& \frac{Z}{2}\frac{{\partial L}}{{\partial E}} + 2\sum\limits_{j = 2}^{\frac{n}{2}} {\left( {\sum\limits_{k = j}^{\frac{n}{2}} {ch\left( {{y_k}} \right)} } \right)}  \nonumber\\
&&+ \frac{{ch\left( {{y_{\frac{n}{2}}}} \right)}}{{sh\left( {{y_{\frac{n}{2}}}} \right)}}\left( {\Delta P\left( E \right) - \sum\limits_{k = 1}^{\frac{n}{2}} {sh\left( {{y_k}} \right)} } \right) = 0 
\label{eq44}
\end{eqnarray}

From the approximation Eq.\ref{eq41}, we obtain ${Z_{\frac{n}{2}}} \approx {Z_{\frac{n}{2} + 1}}$. Taking into account Eq.\ref{eq4.5} results in the following identity:
\begin{eqnarray}%
  \Delta P\left( E \right) - \sum\limits_{k = 1}^{{n \mathord{\left/
 {\vphantom {n 2}} \right.
 \kern-\nulldelimiterspace} 2}} {sh\left( {{y_k}} \right)}  = \frac{1}{{2sh\left( {{y_{{n \mathord{\left/
 {\vphantom {n 2}} \right.
 \kern-\nulldelimiterspace} 2}}}} \right)}}
\label{eq45}
\end{eqnarray}

Substituting Eq.\ref{eq45} into the system of Eq.\ref{eq42} - \ref{eq44} gives
\begin{eqnarray}%
&&\frac{Z}{2}\frac{{\partial L}}{{\partial E}} +  2\frac{{ch\left( {{y_1}} \right)}}{{sh\left( {{y_1}} \right)}}\sum\limits_{j = 2}^{\frac{n}{2}} {\left( {\frac{1}{{2sh\left( {{y_{\frac{n}{2}}}} \right)}} + \sum\limits_{k = j}^{\frac{n}{2}} {sh\left( {{y_k}} \right)} } \right)}  \nonumber\\
&&+ \frac{{ch\left( {{y_1}} \right)}}{{2sh\left( {{y_{\frac{n}{2}}}} \right)sh\left( {{y_1}} \right)}} = 0
\label{eq46}
\end{eqnarray}
\begin{eqnarray}
&& \frac{Z}{2}\frac{{\partial L}}{{\partial E}} + 2\sum\limits_{j = 2}^l {\left( {\sum\limits_{k = j}^{\frac{n}{2}} {ch\left( {{y_k}} \right)} } \right)}   \nonumber\\
&&  + 2\frac{{ch\left( {{y_l}} \right)}}{{sh\left( {{y_l}} \right)}}\sum\limits_{j = l + 1}^{\frac{n}{2}} {\left( {\frac{1}{{2sh\left( {{y_{\frac{n}{2}}}} \right)}} + \sum\limits_{k = j}^{\frac{n}{2}} {sh\left( {{y_k}} \right)} } \right)}  \nonumber\\
&&  + \frac{{ch\left( {{y_l}} \right)}}{{2sh\left( {{y_{\frac{n}{2}}}} \right)sh\left( {{y_l}} \right)}} = 0
\label{eq47}
\end{eqnarray}
here $l = 2,\;3,...,\;\frac{n}{2} - 1$.
\begin{eqnarray}
  \mbox{\fontsize{13}{10}\selectfont $\frac{Z}{2}\frac{{\partial L}}{{\partial E}} + 2\sum\limits_{j = 2}^{\frac{n}{2}} {\left( {\sum\limits_{k = j}^{\frac{n}{2}} {ch\left( {{y_k}} \right)} } \right)}  + \frac{{ch\left( {{y_{\frac{n}{2}}}} \right)}}{{2{{\left( {sh\left( {{y_{\frac{n}{2}}}} \right)} \right)}^2}}} = 0$ }\nonumber\\
\label{eq48}
\end{eqnarray}
Eqs.\ref{eq46} - \ref{eq48} form the system of equations for the search of constrained maximum point of inelastic scattering amplitude in the  approximation of equal denominators Eq.\ref{eq41}. To evaluate this system we consider Eq.\ref{eq47} with $l=\frac{n}{2}-1$:
\begin{eqnarray}
&&  \frac{Z}{2}\frac{{\partial L}}{{\partial E}} + 2\sum\limits_{j = 2}^{\frac{n}{2} - 1} {\left( {\sum\limits_{k = j}^{\frac{n}{2}} {ch\left( {{y_k}} \right)} } \right)}\nonumber\\
&& + \frac{{ch\left( {{y_{\frac{n}{2} - 1}}} \right)}}{{sh\left( {{y_{\frac{n}{2} - 1}}} \right)}}\left( {\frac{3}{{2sh\left( {{y_{\frac{n}{2}}}} \right)}} + 2sh\left( {{y_{\frac{n}{2}}}} \right)} \right) = 0
\label{eq49}
\end{eqnarray}
Subtracting Eq.\ref{eq49} from Eq.\ref{eq48} we have
\begin{eqnarray}
&&    2ch\left( {{y_{\frac{n}{2}}}} \right) + \frac{{ch\left( {{y_{\frac{n}{2}}}} \right)}}{{2{{\left( {sh\left( {{y_{\frac{n}{2}}}} \right)} \right)}^2}}} \nonumber\\
&&  - \frac{{ch\left( {{y_{\frac{n}{2} - 1}}} \right)}}{{sh\left( {{y_{\frac{n}{2} - 1}}} \right)}}\left( {\frac{3}{{2sh\left( {{y_{\frac{n}{2}}}} \right)}} + 2sh\left( {{y_{\frac{n}{2}}}} \right)} \right) = 0 
\end{eqnarray}
It follows from this equation that
\begin{eqnarray}
  \mbox{\fontsize{13}{10}\selectfont $ {\rm th}\left( {{y_{\frac{n}{2} - 1}}} \right) = \frac{{\frac{3}{{2sh\left( {{y_{\frac{n}{2}}}} \right)}} + 2sh\left( {{y_{\frac{n}{2}}}} \right)}}{{2ch\left( {{y_{\frac{n}{2}}}} \right) + \frac{{ch\left( {{y_{\frac{n}{2}}}} \right)}}{{2{{\left( {sh\left( {{y_{\frac{n}{2}}}} \right)} \right)}^2}}}}}
  = {\rm th}\left( {3{y_{\frac{n}{2}}}} \right) $ } \nonumber\\
\label{eq51}
\end{eqnarray}
Taking into account that the hyperbolic tangent is a monotonous function along the whole real axis, it follows from Eq.\ref{eq51}:
\begin{eqnarray}
  {y_{\frac{n}{2} - 1}} = 3{y_{\frac{n}{2}}}
\label{eq52}
\end{eqnarray}

Note, that this result agrees with the numerical results shown in Fig.\ref{fig:part1_fig04}, and related to it Table.\ref{fig:part1_table02}(see column $y_k/y_{15}$ for $n=30$).
We now proceed to prove by induction that
\begin{eqnarray}
  \mbox{\fontsize{11}{10}\selectfont $ {y_{\frac{n}{2} - k}} = \left( {2k + 1} \right){y_{\frac{n}{2}}},\;\;k = 1,\;2,...,\;\frac{n}{2} - 1$ }
\label{eq53}
\end{eqnarray}
Eq.\ref{eq53} is already proved for $k=1$, since, it coincides with Eq.\ref{eq52}. 
Suppose that this equation is true for $k = 1, 2, ..., \frac{n}{2} - l - 1$ (i.e., at ${y_{{n \mathord{\left/
 {\vphantom {n 2}} \right.
 \kern-\nulldelimiterspace} 2} - 1}},{y_{{n \mathord{\left/
 {\vphantom {n 2}} \right.
 \kern-\nulldelimiterspace} 2} - 2}},...,{y_{l + 1}}$) and we wish to prove that it is true for $k = \frac{n}{2} - l$ (i.e., at ${y_l}$).
Subtracting Eq.\ref{eq47} from Eq.\ref{eq48} we obtain:
\begin{widetext}
\begin{equation}
  2\sum\limits_{j = l + 1}^{\frac{n}{2}} {\left( {\sum\limits_{k = j}^{\frac{n}{2}} {ch\left( {{y_k}} \right)} } \right)}  + \frac{{ch\left( {{y_{\frac{n}{2}}}} \right)}}{{2{{\left( {sh\left( {{y_{\frac{n}{2}}}} \right)} \right)}^2}}} 
   - 2\frac{{ch\left( {{y_l}} \right)}}{{sh\left( {{y_l}} \right)}}\sum\limits_{j = l + 1}^{\frac{n}{2}} {\left( {\frac{1}{{2sh\left( {{y_{\frac{n}{2}}}} \right)}} + \sum\limits_{k = j}^{\frac{n}{2}} {sh\left( {{y_k}} \right)} } \right)} 
   - \frac{{ch\left( {{y_l}} \right)}}{{2sh\left( {{y_{\frac{n}{2}}}} \right)sh\left( {{y_l}} \right)}} = 0
\label{eq54}
\end{equation}
\end{widetext}

Note, that sums $\sum\limits_{k = j}^{n/2} {ch\left( {{y_k}} \right)}$ and $\sum\limits_{k = j}^{\frac{n}{2}} {sh\left( {{y_k}} \right)}$ include only those $y_k$ with respect $l + 1 \le j \le \frac{n}{2}$, which are covered by the assumption of induction.
Hence, from Eq.\ref{eq53}, we get $\sum\limits_{k = j}^{\frac{n}{2}} {sh\left( {{y_k}} \right)}$ and this makes it 
possible to calculate the sums included in Eq.\ref{eq54}. After transformations this has a form:
\begin{eqnarray}
  {\rm th}\left( {{y_l}} \right) = {\rm th}\left( {\left( {2\left( {\frac{n}{2} - l} \right) + 1} \right){y_{\frac{n}{2}}}} \right)
\end{eqnarray}
so that
\begin{eqnarray}
&& {y_l} = \left( {2\left( {\frac{n}{2} - l} \right) + 1} \right){y_{\frac{n}{2}}} \cr
&& {y_{\frac{n}{2} - l}} = \left( {2l + 1} \right){y_{\frac{n}{2}}}
\end{eqnarray}
i.e., it coincides with the proved expression in Eq.\ref{eq53}.

For the diagrams with even number of particles we have shown that in the approximation of equal denominators (see Eq.\ref{eq41}), the rapidities produce an arithmetic progression at the maximum point and the ratios of all the rapidities to the minimum rapidity produce a sequence of odd integers.

To determine the values ​​of these rapidities which maximize the scattering amplitude, we also have to calculate ${y_{{n \mathord{\left/
 {\vphantom {n 2}} \right.
\kern-\nulldelimiterspace} 2}}}$, in term of all rapidities. This can be done using the same equal-denominators approximation as in Eq.\ref{eq41}. In view of Eq.\ref{eq48}, computing the sums in Eq.\ref{eq53} we have:
\begin{eqnarray}
  \frac{Z}{2}\frac{{\partial L}}{{\partial E}} + \frac{{ch\left( {\left( {n - 1} \right){y_{\frac{n}{2}}}} \right)}}{{2{{\left( {sh\left( {{y_{\frac{n}{2}}}} \right)} \right)}^2}}} = 0
\label{eq57}
\end{eqnarray}
Now we can calculate the derivative $\frac{{\partial L}}{{\partial E}}$, taking into account Eq.\ref{eq53} for the magnitude $L$: 
\begin{eqnarray}
  \mbox{\fontsize{12}{14}\selectfont $ \frac{{\partial L}}{{\partial E}} = \frac{{4E}}{{{Z_1}}} - 4{\wp}\frac{{\partial \Delta P\left( E \right)}}{{\partial E}} $}
\label{eq58_1}
\end{eqnarray}
where
\begin{eqnarray}
  \mbox{\fontsize{11}{14}\selectfont $ {\wp} = \frac{{\Delta P\left( E \right)}}{{{Z_1}}} + \sum\limits_{j = 2}^{\frac{n}{2}} {\frac{{\Delta P\left( E \right) - \sum\limits_{k = 1}^{j - 1} {sh\left( {{y_k}} \right)} }}{{{Z_j}}}}   + \frac{1}{2}\frac{{\Delta P\left( E \right) - \sum\limits_{k = 1}^{\frac{n}{2}} {sh\left( {{y_k}} \right)} }}{{{Z_{\frac{n}{2} + 1}}}} $}\nonumber\\
\label{eq58_2}
\end{eqnarray}

Using the equal-denominators approximation Eq.\ref{eq41}, we obtain:
\begin{eqnarray}
\frac{Z}{2}\frac{{\partial L}}{{\partial E}} = 2E - 2\wp '\frac{{\partial \Delta P\left( E \right)}}{{\partial E}} 
\label{eq59_1}
\end{eqnarray}
where
\begin{eqnarray}
&& \wp ' = \Delta P\left( E \right) + \sum\limits_{j = 2}^{\frac{n}{2}} {\left( {\Delta P\left( E \right) - \sum\limits_{k = 1}^{j - 1} {{\mathop{\rm sh}\nolimits} \left( {{y_k}} \right)} } \right)} \nonumber\\
&& + \frac{1}{2}\left( {\Delta P\left( E \right) - \sum\limits_{k = 1}^{\frac{n}{2}} {{\mathop{\rm sh}\nolimits} \left( {{y_k}} \right)} } \right) 
\label{eq59_2}
 \end{eqnarray}

After transformations with respect to Eq.\ref{eq45} we get:
\begin{eqnarray}
  \mbox{\fontsize{14}{14}\selectfont $ \frac{Z}{2}\frac{{\partial L}}{{\partial E}} = \frac{{sh\left( {n{y_{\frac{n}{2}}}} \right)}}{{sh\left( {{y_{\frac{n}{2}}}} \right)}} + \frac{{sh\left( {\left( {n + 1} \right){y_{\frac{n}{2}}}} \right)}}{{2s{h^2}\left( {{y_{\frac{n}{2}}}} \right)}}\frac{{\partial {P_{3\parallel }}\left( E \right)}}{{\partial E}} = 0$ } \nonumber\\
\label{eq60}
\end{eqnarray}
Substituting Eq.\ref{eq60} into Eq.\ref{eq57}, we can reduce the resulting equation to the form:
\begin{eqnarray}
  \mbox{\fontsize{10}{10}\selectfont $ sh\left( {\left( {n + 1} \right){y_{\frac{n}{2}}}} \right)\frac{{\partial {P_{3\parallel }}\left( E \right)}}{{\partial E}} + ch\left( {\left( {n + 1} \right){y_{\frac{n}{2}}}} \right) = 0$ }
\label{eq61}
\end{eqnarray}

The derivative $\frac{{\partial {P_{3\parallel }}\left( E \right)}}{{\partial E}}$ can be computed from Eq.\ref{eq23} with allowance for Eq.\ref{eq34} and Eq.\ref{eq35}. 
Then the expression for ${{P_{3\parallel }}}$, which is non-dimensionalised by mass $m$, can be written as:
\begin{eqnarray}
  {P_{3\parallel }} = \sqrt {{{\left( {{\sqrt s }/2 - E} \right)}^2} - {M^2}}
\label{eq62}
\end{eqnarray}
where it is assumed that $\sqrt s $ and particle masses $M$ at the ends of the``comb'' are non-dimensionlized with the mass $m$.
The pion mass was set to m and the proton mass to $M$.

Substituting the derivative Eq.\ref{eq62} into Eq.\ref{eq61}, and after performing simple transformations, we obtain the equation
\begin{eqnarray}
  \frac{{\sqrt s }}{2} - E = M \cdot ch\left( {\left( {n + 1} \right){y_{\frac{n}{2}}}} \right)
\label{eq63}
\end{eqnarray}

Note, that the rapidity corresponding to momentum ${P_{3\parallel }}$ is equal to $\left( {n + 1} \right){y_{{n \mathord{\left/
 {\vphantom {n 2}} \right.
 \kern-\nulldelimiterspace} 2}}}$, as it follows from Eq.\ref{eq62} and Eq.\ref{eq63}.

This expression could be obtained from Eq.\ref{eq53} if one includes $k = \frac{n}{2}$. That is, the arithmetic progression Eq.\ref{eq53} lengthens by one term. One might consider that rapidities of particles on the ``comb” edges at the maximum point ``continues” an arithmetic progression formed by the internal particles of the ``comb”.

This fact once again indicates the close relation between the equal denominators approximation Eq.\ref{eq41} and the production of arithmetic progression by the rapidities at the maximum point. In other words, this arithmetic progression is the consequence of equal-denominators approximation.

It is also possible to verify permissibility of this approximation in the following way. Taking into account that
\begin{eqnarray}
  E = \sum\limits_{k = 1}^{\frac{n}{2}} {ch\left( {{y_k}} \right)}  = \frac{{sh\left( {n{y_{\frac{n}{2}}}} \right)}}{{2sh\left( {{y_{\frac{n}{2}}}} \right)}}
\label{eq63_a}
\end{eqnarray}
 (as it follows from Eq.\ref{eq53} and Eq.\ref{eq35}) we get instead of Eq.\ref{eq63}:
\begin{eqnarray}
  \frac{{\sqrt s }}{2} - \frac{{sh\left( {n{y_{\frac{n}{2}}}} \right)}}{{2sh\left( {{y_{\frac{n}{2}}}} \right)}} = M \cdot ch\left( {\left( {n + 1} \right){y_{\frac{n}{2}}}} \right)
\label{eq64}
\end{eqnarray}

\begin{figure}
\begin{center}
  \subfigure[]{
  \includegraphics[scale=0.29]{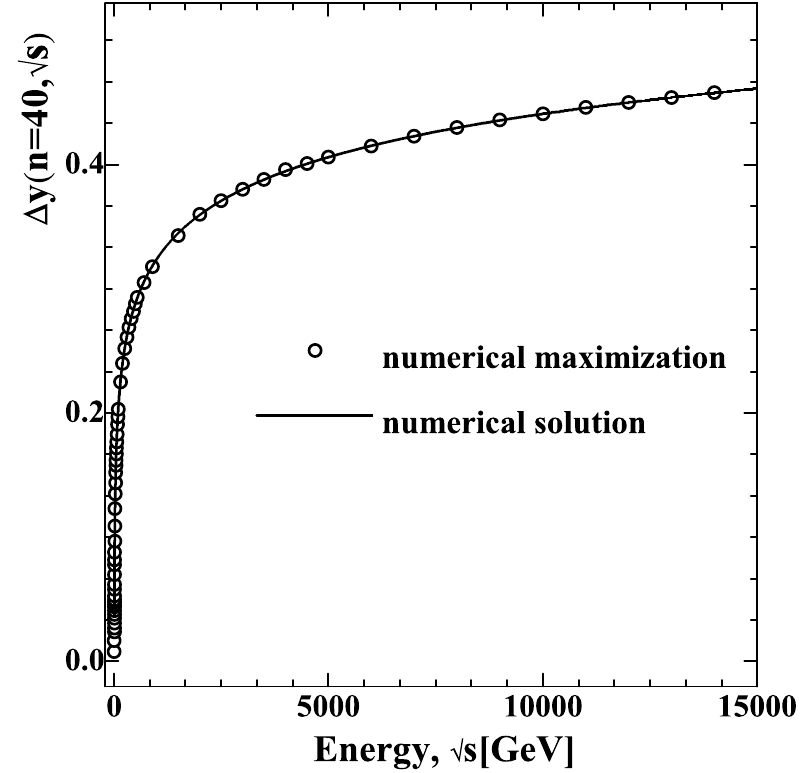}
  \label{fig:fig_part1_11a}
  }
  \subfigure[]{
  \includegraphics[scale=0.29]{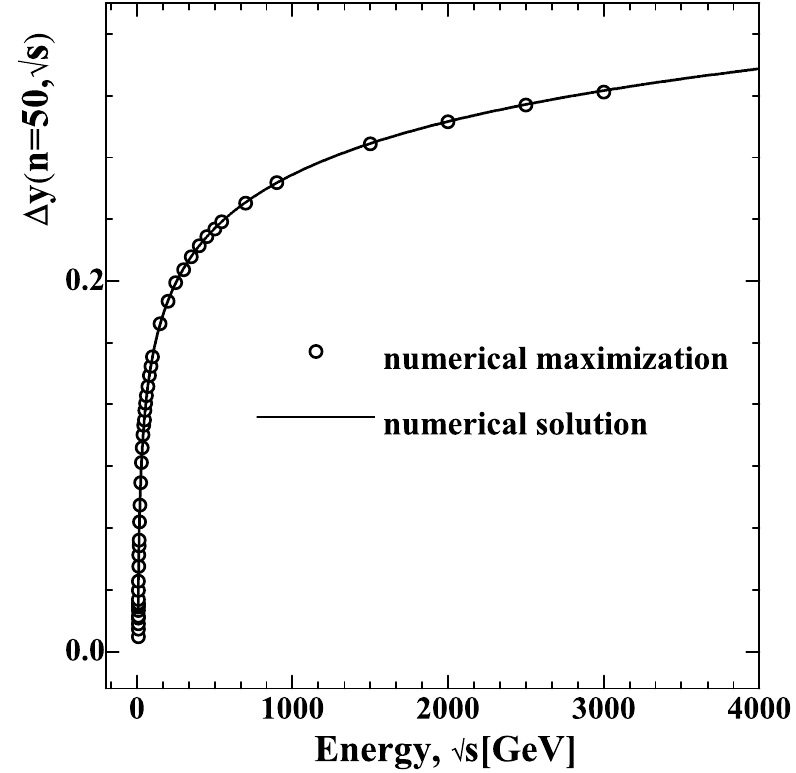}
    \label{fig:fig_part1_11b}
  }
\end{center}
\caption{ The numerical solution of Eq.\ref{eq64}(solid line) and numerical maximization (circles) of the magnitude $\Delta y(n, \sqrt s )$ at $n=40$ (\ref{fig:fig_part1_11a}) and $n=50$ (\ref{fig:fig_part1_11b}).} 
  \label{fig:fig_part1_11}
\end{figure}

\begin{figure*}
\begin{center}
\subfigure[]{
    \includegraphics[scale=0.42]{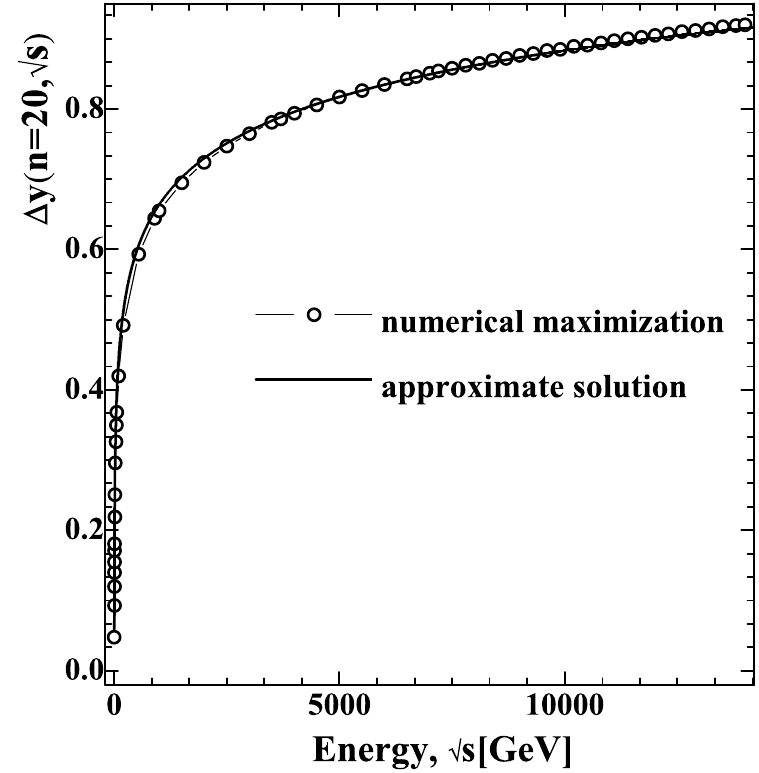}
  \label{fig:fig_part1_12a}
}
\subfigure[]{
  \includegraphics[scale=0.42]{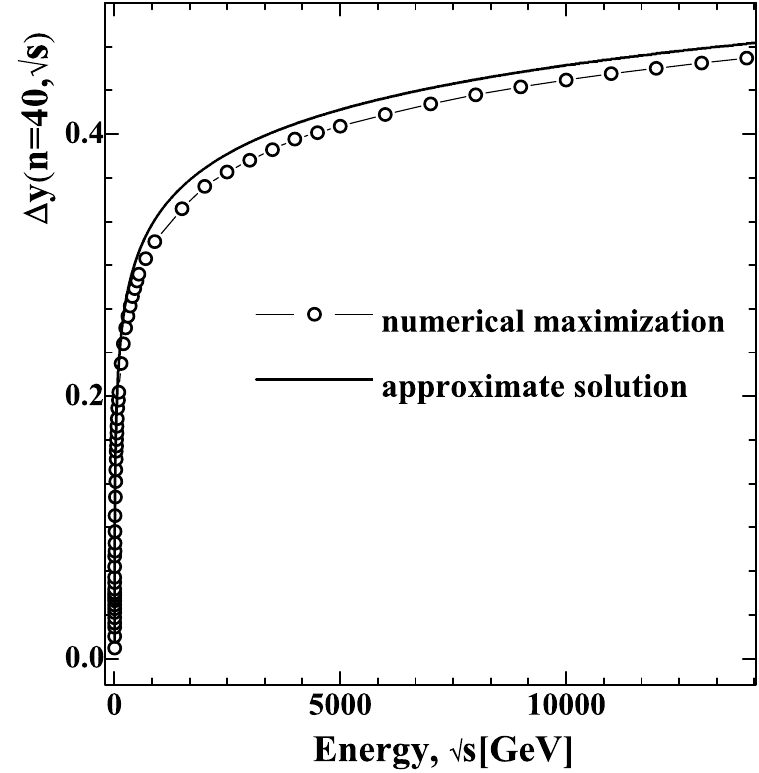}
  \label{fig:fig_part1_12c}
}
\subfigure[]{
  \includegraphics[scale=0.42]{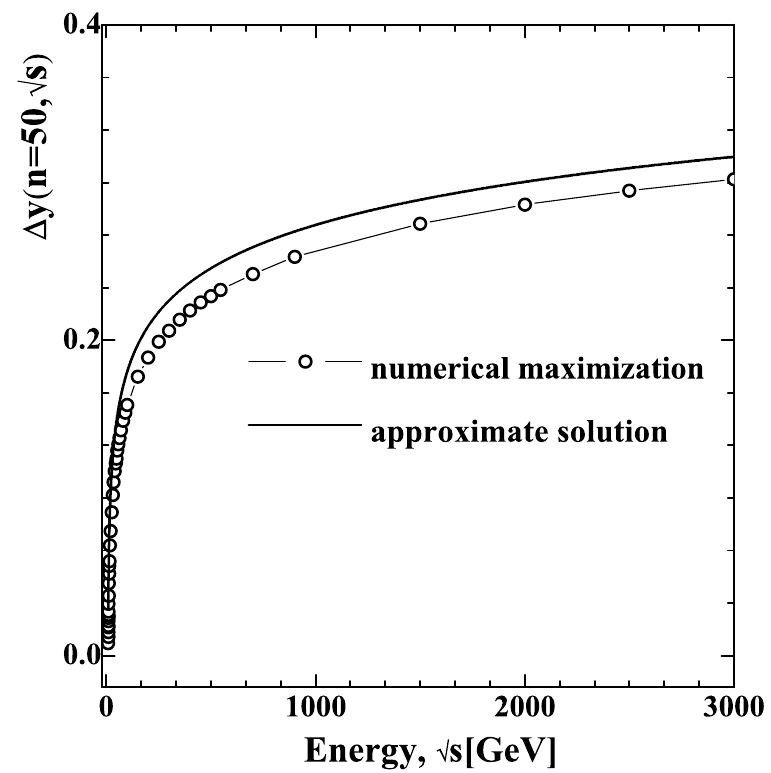}
  \label{fig:fig_part1_12e}
}
\subfigure[]{
  \includegraphics[scale=0.42]{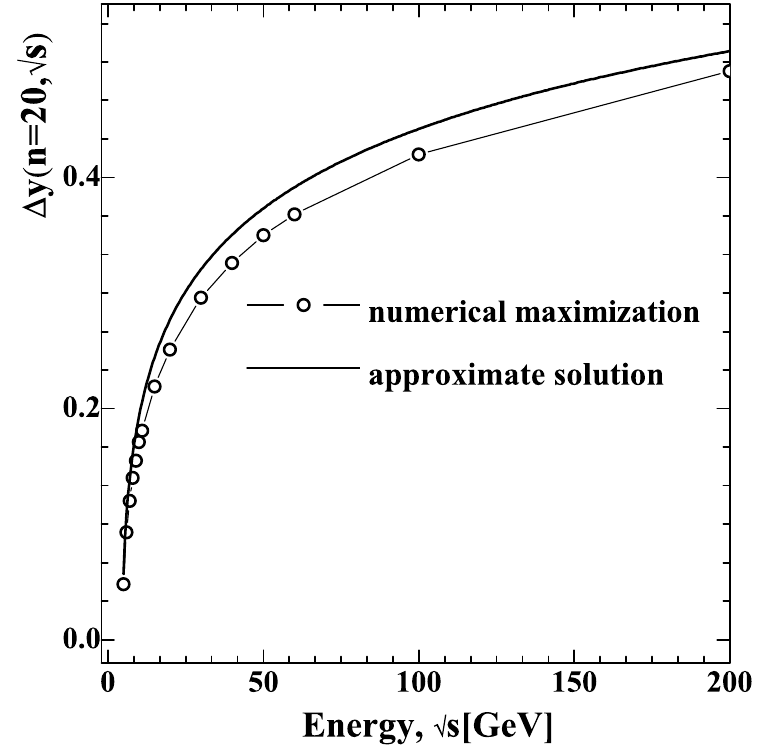}
  \label{fig:fig_part1_12b}
}
\subfigure[]{
  \includegraphics[scale=0.42]{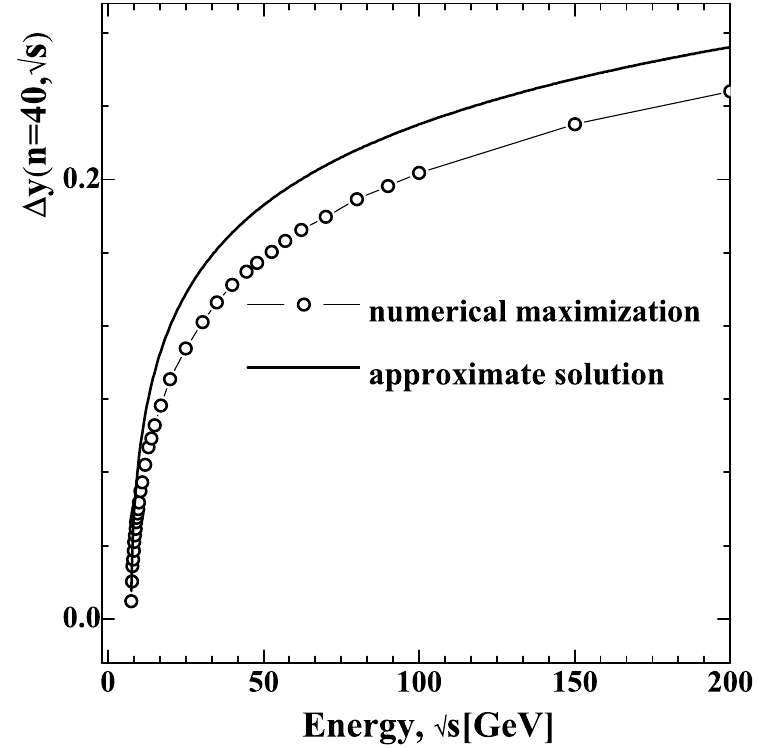}
    \label{fig:fig_part1_12d}
}
\subfigure[]{
  \includegraphics[scale=0.42]{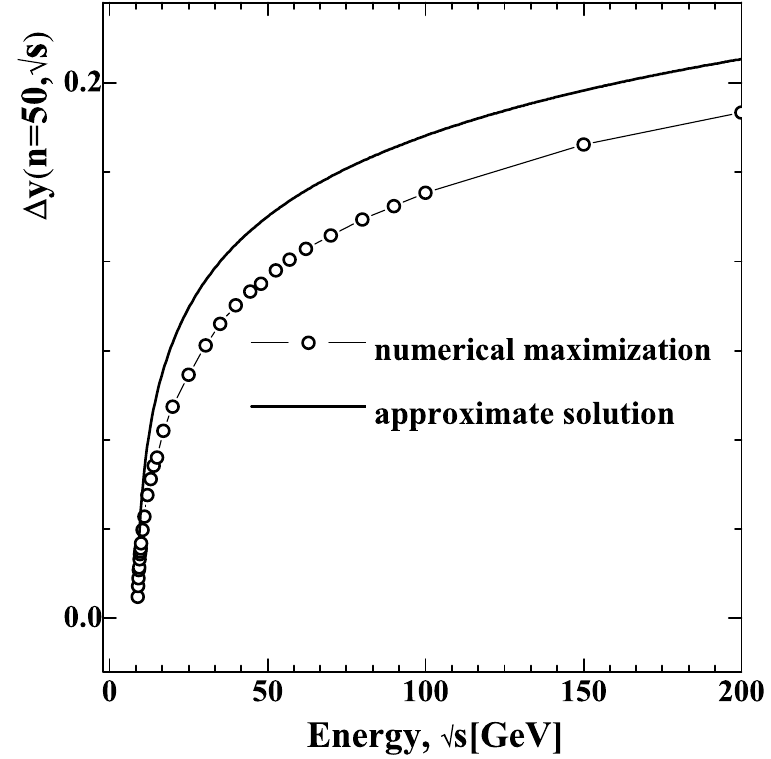}
  \label{fig:fig_part1_12f}
}
\end{center}
\caption{
Comparison the approximate solution of Eq.\ref{eq64} (solid line) with the results of numerical computation (circles) for $n=20$ (\ref{fig:fig_part1_12a}), (\ref{fig:fig_part1_12b}); $n=40$ (\ref{fig:fig_part1_12c}), (\ref{fig:fig_part1_12d}); $n=50$ (\ref{fig:fig_part1_12e}), (\ref{fig:fig_part1_12f}). The range of low energies close to the threshold branch point (in dimensionless form) is shown on (\ref{fig:fig_part1_12b}), (\ref{fig:fig_part1_12d}) and (\ref{fig:fig_part1_12f}). The good agreement of results shows the applicability of the approximation of equal denominators Eq.\ref{eq41} resulting in the Eq.\ref{eq63}.}
\label{fig:fig_part1_12}
\end{figure*}

This equation does not admit an exact analytical solution, and therefore we will consider an approximate solution. However, we can verify admissibility of the approximations, which resulted in Eq.\ref{eq64}, using Mathcad 2001 (or any other computer software for engineering and scientific calculations) to solve this equation numerically for different energies
$\sqrt s$ and comparing it with the result of numerical determination of the maximum point and shown in Fig.\ref{fig:part1_fig05}. The results of such comparison are shown in Fig.\ref{fig:fig_part1_12} and Fig.\ref{fig:fig_part1_11}.

As seen from Fig.\ref{fig:fig_part1_11} and Fig.\ref{fig:fig_part1_12}, the ``exact'' numerical solution of Eq.\ref{eq64} practically 
indistinguishable from the results of numerical computation.
This is the evidence of the fact that the equal-denominators approximation Eq.\ref{eq41} in
Eq.\ref{eq64} is admissible approximation.

Now let us turn to the approximate analytical solution of
Eq.\ref{eq64}. Note, that function ${\rm sh}\left( {n{y_{{n \mathord{\left/
 {\vphantom {n 2}} \right.
 \kern-\nulldelimiterspace} 2}}}} \right)/2 {\rm sh} \left( {{y_{{n \mathord{\left/
 {\vphantom {n 2}} \right.
 \kern-\nulldelimiterspace} 2}}}} \right)$ in Eq.\ref{eq64} changes slowly at small values of ${{y_{{n \mathord{\left/
 {\vphantom {n 2}} \right.
 \kern-\nulldelimiterspace} 2}}}}$ and can be approximately replaced by $n/2$ at ${y_{{n \mathord{\left/
 {\vphantom {n 2}} \right.
 \kern-\nulldelimiterspace} 2}}} \to 0$. This approximation leads to the following solution:
\begin{eqnarray}
  {y_{\frac{n}{2}}} = \frac{1}{{n + 1}} {\rm arccosh}\left( {\frac{{\sqrt s  - n}}{{2M}}} \right)
\label{eq65}
\end{eqnarray}

The approximate solution of Eq.\ref{eq65} and the results of numerical computation are presented in Fig.\ref{fig:fig_part1_12}, where it is evident that Eq.\ref{eq65} gives a somewhat overstated value in comparison with numerical computation. It is naturally, because using approximation $\frac{{sh\left( {n{y_{{n \mathord{\left/
 {\vphantom {n 2}} \right.
 \kern-\nulldelimiterspace} 2}}}} \right)}}{{2sh\left( {{y_{{n \mathord{\left/
 {\vphantom {n 2}} \right.
 \kern-\nulldelimiterspace} 2}}}} \right)}} \approx \frac{n}{2}$ in Eq.\ref{eq64}, we underestimate value of function $\frac{{sh\left( {n{y_{{n \mathord{\left/
 {\vphantom {n 2}} \right.
 \kern-\nulldelimiterspace} 2}}}} \right)}}{{2sh\left( {{y_{{n \mathord{\left/
 {\vphantom {n 2}} \right.
 \kern-\nulldelimiterspace} 2}}}} \right)}}$ and thereby overstated the value of the hyperbolic cosine in the right-hand side of Eq.\ref{eq64}.

Nevertheless, as seen in Fig.\ref{fig:fig_part1_12}, the absolute uncertainty of approximation Eq.\ref{eq65} does not increase with the energy growth, and since ${y_{{n \mathord{\left/
 {\vphantom {n 2}} \right.
 \kern-\nulldelimiterspace} 2}}}$ increases, the relative uncertainty decreases. This can also be explained with reasoning from Eq.\ref{eq64}. Since $\frac{{sh\left( {n{y_{{n \mathord{\left/
 {\vphantom {n 2}} \right.
 \kern-\nulldelimiterspace} 2}}}} \right)}}{{2sh\left( {{y_{{n \mathord{\left/
 {\vphantom {n 2}} \right.
 \kern-\nulldelimiterspace} 2}}}} \right)}}$ becomes small in comparison with $Mch\left( {\left( {n + 1} \right){y_{{n \mathord{\left/
 {\vphantom {n 2}} \right.
 \kern-\nulldelimiterspace} 2}}}} \right)$ at sufficiently high energies $\sqrt s $ (and ${{y_{{n \mathord{\left/
 {\vphantom {n 2}} \right.
 \kern-\nulldelimiterspace} 2}}}}$ accordingly) therefore, the accuracy of approximation for function $\frac{{sh\left( {n{y_{{n \mathord{\left/
 {\vphantom {n 2}} \right.
 \kern-\nulldelimiterspace} 2}}}} \right)}}{{2sh\left( {{y_{{n \mathord{\left/
 {\vphantom {n 2}} \right.
 \kern-\nulldelimiterspace} 2}}}} \right)}}$ has no importance. By neglecting $\frac{{sh\left( {n{y_{{n \mathord{\left/
 {\vphantom {n 2}} \right.
 \kern-\nulldelimiterspace} 2}}}} \right)}}{{2sh\left( {{y_{{n \mathord{\left/
 {\vphantom {n 2}} \right.
 \kern-\nulldelimiterspace} 2}}}} \right)}}$ in Eq.\ref{eq64} in comparison with $Mch\left( {\left( {n + 1} \right){y_{{n \mathord{\left/
 {\vphantom {n 2}} \right.
 \kern-\nulldelimiterspace} 2}}}} \right)$ and by neglecting $n$ in Eq.\ref{eq65} with respect to $\sqrt s $, we obtain the same result. This means that approximation Eq.\ref{eq65} ensures the ``correct" asymptotic of value ${{y_{{n \mathord{\left/
 {\vphantom {n 2}} \right.
 \kern-\nulldelimiterspace} 2}}}}$ at high $\sqrt s $.

Let us note some features of Eq.\ref{eq65}. First, the approximate solution of Eq.\ref{eq65}, nondimensionalized with the
mass $m$, has a threshold branch-point when $\sqrt s = n + 2M $. This means that the difference of the rapidity arithmetic progression, which maximizes the amplitude of inelastic process, also has this singularity. The contribution of inelastic processes to the imaginary part of the elastic scattering amplitude, after calculation with the help of Laplace`s method, will be expressed in terms of the difference of the arithmetic progression $\Delta y \left( \sqrt s, n\right)$. Therefore it is possible to expect that the threshold feature, 
via the difference of arithmetic progression, will be incorporated into the imaginary part of the elastic scattering amplitude. And this feature is required by unitary condition. Note, that Eq.\ref{eq65} has logarithmic asymptotic behavior at the energies 
substantially exceeding the threshold value and at $\Delta y = 2{y_{\frac{n}{2}}}{n^{ - 1}}$ from Eq.\ref{eq65}, coinciding with the results of numerical computation (see Section.\ref{SECTION_3}).

For the diagrams in Fig.\ref{fig:part1_fig02} with an odd number of particles, the whole procedure is similar to the one
described above for the diagrams with even number of particles.
First, we will take derivative of the logarithm of amplitude restriction Eq.\ref{eq26} with respect to all rapidities ${y_1},\;{y_2},\;...,\;{y_{\frac{{n - 1}}{2}}}$. After that, it is possible to use the equal-denominators approximation:
\begin{eqnarray}
  {Z_1} \approx ... \approx {Z_{\frac{{n - 1}}{2}}} \approx {Z_{\frac{{n - 1}}{2} + 1}} = Z
\end{eqnarray}
From the equality ${Z_{\frac{{n - 1}}{2} - 1}} \approx {Z_{\frac{{n - 1}}{2}}}$ we will obtain a relation similar to Eq.\ref{eq45}:
\begin{eqnarray}
  {P_{1\parallel }} - {P_{3\parallel }} - \sum\limits_{j = 1}^{\frac{{n - 1}}{2}} {sh\left( {{y_j}} \right)}  = \frac{{ch\left( {\frac{1}{2}{y_{\frac{{n - 1}}{2}}}} \right)}}{{2sh\left( {\frac{1}{2}{y_{\frac{{n - 1}}{2}}}} \right)}}
\label{eq67}
\end{eqnarray}
Now, using the same recipe as in Eq.\ref{eq51}, for the derivatives of the logarithm of the scattering amplitude restriction with respect to ${y_{\frac{{n - 1}}{2}}}$ and ${y_{\frac{{n - 1}}{2} - 1}}$, and taking into account
Eq.\ref{eq67}, we have
\begin{eqnarray}
  {y_{\frac{{n - 1}}{2} - 1}} = 2{y_{\frac{{n - 1}}{2}}}
\end{eqnarray}
This result agrees with results of the numerical calculation presented in Table.\ref{fig:part1_table02}(see column $y_k/y_{8}$ for $n=17$).
Hence it can be shown by induction that
\begin{eqnarray}
  \mbox{\fontsize{10}{10}\selectfont $ {y_{\frac{{n - 1}}{2} - (k - 1)}} = k{y_{\frac{{n - 1}}{2}}},\;\;k = 1,\;2,...,\;\frac{{n - 1}}{2} - 1$ }
\label{eq79a}
\end{eqnarray}
Thus, all rapidities can be expressed in terms of ${y_{\frac{{n - 1}}{2}}}$.

Repeating the same calculations as in Eqs.\ref{eq57}-\ref{eq64}, we obtain the following expression from the approximation of equal-denominators:
\begin{eqnarray}
  \mbox{\fontsize{10}{10}\selectfont $ \frac{{\sqrt s }}{2} - \frac{{sh\left( {\frac{n}{2}{y_{\frac{{n - 1}}{2}}}} \right)}}{{2sh\left( {\frac{{{y_{\frac{{n - 1}}{2}}}}}{2}} \right)}} = M \cdot ch\left( {\left( {\frac{{n - 1}}{2} + 1} \right){y_{\frac{{n - 1}}{2}}}} \right)$ }
\end{eqnarray}
Using an approximation similar to the one which results in Eq.\ref{eq65}, we get
\begin{eqnarray}
  {y_{\frac{{n - 1}}{2}}} = \frac{2}{{n + 1}} {\rm arccosh} \left( {\frac{{\sqrt s  - n}}{{2M}}} \right)
\label{eq71}
\end{eqnarray}

From Eq.\ref{eq71} it is evident that in the case of an odd number $n$, the common rapidity difference, which constrainedly maximizes the scattering amplitude, also has a threshold branch point. Note, that the difference of arithmetic progression is equal to
${y_{\frac{{n - 1}}{2}}}$ in the case of odd $n$ and to $2{y_{\frac{n}{2}}}$ in the case of even $n$, i.e., as agreed, the approximation for the common difference of arithmetic progression
$\Delta y\left( {\sqrt s,\;  n} \right)$ is expressed by the same formula for both even and odd $n$

The obtained analytical results enables us to see how the mechanism of virtuality reduction ``works” with the energy growth.
In approximation of equal-denominators  Eq.\ref{eq41} with allowance for Eq.\ref{eq45} the amplitude value at the maximum point (in case of even $n$) can be written as:
\begin{eqnarray}
  {A^{(0),n}} = {\left( {1 + \frac{1}{{4s{h^2}\left( {{y_{\frac{n}{2}}}} \right)}}} \right)^{ - \left( {n + 1} \right)}}
\end{eqnarray}
The quantity $\frac{1}{{4s{h^2}\left( {{y_{\frac{n}{2}}}} \right)}}$ included in this expression sets the characteristic value of virtuality at the maximum point of amplitude at the approximation of equal-denominators.

Taking into account increase of ${y_{\frac{n}{2}}}$ with energy $\sqrt s $ growth, which is approximately described by Eq.\ref{eq65}, note that virtuality at the maximum point really decreases and the maximum value of amplitude grows with energy $\sqrt s $ growth.

The demonstrated results were obtained in the approximation of equal denominators Eq.\ref{eq41}. This approximation is acceptable in order to demonstrate that the scattering amplitude has a constrained maximum point and in order to find that maximum. However, to calculate the value of the amplitude at this point, this approximation provides insufficiently accurate result, so for the next section will be calculated  amplitude at the maximum point in a more accurate approximation.

\section{The approximate calculation of the sum of logarithms at the point of constrained maximum}
\label{SECTION_AndriiAddon_01}
Taking into account the aforementioned results (see Fig.\ref{fig:part1_fig04}  and Eqs.\ref{eq53},\ref{eq79a}) for the calculation of constrained maximum point of multi-peripheral scattering amplitude, one gets the following representation of $A^{(0),n}$:
\begin{eqnarray}
&& {A^{\left( 0 \right),n}} = \exp \left( { - \sum\limits_{k = 0}^n {\ln \left( {{Z_k}} \right)} } \right) = \exp \left( { - 2\sum\limits_{k = 0}^{\frac{n}{2}} {\ln \left( {{Z_k}} \right)} } \right)  \nonumber\\
\label{eq_AndriiAddon_11}
\end{eqnarray}
where
\begin{widetext}
\begin{equation}
{Z_k} = 1 - {\left( {\frac{{{\mathop{\rm sh}\nolimits} \left( {\left( {n - 2k} \right)\frac{{\Delta y}}{2}} \right)}}{{2{\mathop{\rm sh}\nolimits} \left( {\frac{{\Delta y}}{2}} \right)}}} \right)^2} + {\left( {\sqrt {\frac{s}{4} - M{\,^2}}  - M{\mathop{\rm sh}\nolimits} \left( {\left( {n + 1} \right)\frac{{\Delta y}}{2}} \right) - \frac{{{\mathop{\rm ch}\nolimits} \left( {n\frac{{\Delta y}}{2}} \right) - {\mathop{\rm ch}\nolimits} \left( {\left( {n - 2k} \right)\frac{{\Delta y}}{2}} \right)}}{{2{\mathop{\rm sh}\nolimits} \left( {\frac{{\Delta y}}{2}} \right)}}} \right)^2}
\label{eq_AndriiAddon_12}
\end{equation}
\end{widetext}

Through $\Delta y$ we denote (as before) the difference of arithmetic progression at the point of constrained maximum. The aim of this chapter is to obtain the approximate solution for the sum $\sum\limits_{k = 0}^{\frac{n}{2}} {\ln \left( {{Z_k}} \right)}$, which enters the exponent of Eq.\ref{eq_AndriiAddon_11}.

In the previous sections (see Figs.\ref{fig:part1_fig07}, \ref{fig:part1_fig09}) has been shown that Feynman denominators grow if one moves from comb's edges to its center. In other words, the maximum of $Z_k$ values Eq.\ref{eq_AndriiAddon_12} is $Z_{n/2}$. Furthermore, since Feynman denominators are obliged to grow while moving towards the comb's center, the maximum of scattering amplitude is attained when the aforementioned growth is minimal. This means that the denominators at the point of maximum differ little from each other.

The results of numerical calculations enable to claim that the higher is the energy, the smaller is the difference of each Feynman denominator from the central one on the comb, with the exception of two outermost denominators (see Section.\ref{SECTION_3} and Figs.\ref{fig:part1_fig07}, \ref{fig:part1_fig09}). This can be also shown analytically (see Eq.\ref{eq_AndriiAddon_29})

Thus, one can try to express all denominators except two outermost ones in terms of $Z_{n/2}$.
\begin{eqnarray}%
&&  \mbox{\fontsize{12}{14}\selectfont $  \ln \left( {{Z_k}} \right) = \ln \left( {{Z_{\frac{n}{2}}}} \right) $ } \nonumber\\
&&  \mbox{\fontsize{9}{14}\selectfont $  + \ln \left( {1 - \frac{{\left( {P - {P_{3\left\| {} \right.}} - \frac{{ch\left( {n\frac{{\Delta y}}{2}} \right)}}{{2sh\left( {\frac{{\Delta y}}{2}} \right)}}} \right)}}{{{Z_{\frac{n}{2}}}sh\left( {\frac{{\Delta y}}{2}} \right)}}\left( {1 - ch\left( {\left( {n - 2k} \right)\frac{{\Delta y}}{2}} \right)} \right)} \right) $ }\nonumber\\
\label{eq_AndriiAddon_13}
\end{eqnarray}%

Taking into account that  $Z_k$ differs little from $Z_{n/2}$, we get
\begin{eqnarray}%
&&  \mbox{\fontsize{12}{14}\selectfont $ \ln \left( {{Z_k}} \right) = \ln \left( {{Z_{\frac{n}{2}}}} \right) $ }\nonumber\\
&&  \mbox{\fontsize{10}{14}\selectfont $  - \frac{{\left( {P - {P_{3\left\| {} \right.}} - \frac{{ch\left( {n\frac{{\Delta y}}{2}} \right)}}{{2sh\left( {\frac{{\Delta y}}{2}} \right)}}} \right)}}{{{Z_{\frac{n}{2}}}sh\left( {\frac{{\Delta y}}{2}} \right)}}\left( {1 - ch\left( {\left( {n - 2k} \right)\frac{{\Delta y}}{2}} \right)} \right) $ }
\label{eq_AndriiAddon_14}
\end{eqnarray}%
Substituting approximation Eq.\ref{eq_AndriiAddon_14} into expression for scattering amplitude leads to:
\begin{eqnarray}%
&&  \mbox{\fontsize{12}{14}\selectfont $  {A^{\left( 0 \right),n}} = \frac{1}{{{{\left( {{Z_0}} \right)}^2}{{\left( {{Z_{\frac{n}{2}}}} \right)}^{n - 1}}}} $ }\nonumber\\
&&  \mbox{\fontsize{9}{14}\selectfont $   \times \exp \left( {\frac{{\left( {P - {P_{3\left\| {} \right.}} - \frac{{ch\left( {n{y_{\frac{n}{2}}}} \right)}}{{2sh\left( {{y_{\frac{n}{2}}}} \right)}}} \right)}}{{{Z_{\frac{n}{2}}}sh\left( {{y_{\frac{n}{2}}}} \right)}}\left( {n - 1 - \frac{{sh\left( {\left( {n - 1} \right){y_{\frac{n}{2}}}} \right)}}{{sh\left( {{y_{\frac{n}{2}}}} \right)}}} \right)} \right) $ }\nonumber\\
\label{eq_AndriiAddon_15}
\end{eqnarray}%
where
\begin{eqnarray}%
&&  {Z_0} = 1 - {\left( {\sqrt s/2 - Mch\left( {\left( {n + 1} \right)\frac{{\Delta y}}{2}} \right)} \right)^2}\nonumber\\
&&  + {\left( {\sqrt {s/4 - {M^2}}  - Msh\left( {\left( {n + 1} \right)\frac{{\Delta y}}{2}} \right)} \right)^2}
\label{eq_AndriiAddon_16}
\end{eqnarray}%
\begin{eqnarray}%
  \mbox{\fontsize{12}{14}\selectfont $ {Z_{\frac{n}{2}}} = 1 + \left( {\sqrt {s/4 - {M^2}} } \right. $ }\nonumber\\
  \mbox{\fontsize{12}{14}\selectfont $ {\left. { - Msh\left( {\left( {n + 1} \right)\frac{{\Delta y}}{2}} \right) - \frac{{ch\left( {n\frac{{\Delta y}}{2}} \right) - 1}}{{2sh\left( {\frac{{\Delta y}}{2}} \right)}}} \right)^2} $ }
\label{eq_AndriiAddon_17}
\end{eqnarray}%
Relations Eqs.\ref{eq_AndriiAddon_15}-\ref{eq_AndriiAddon_17} are expressing scattering amplitude in terms of solution $\Delta y$  of transcendental equation Eq.\ref{eq64}. Our next goal is to express scattering amplitude $A^{(0),n}$ in terms of energy $\sqrt s$, the parameter which is characterizes the scattering process. Furthermore, we can restrict ourselves to considering energies far from threshold ${\left( {\sqrt s } \right)_T} = 2M + n$. This is due to the fact that near the threshold the behavior of partial cross-section is determined mostly by the volume of final-state particles phase space, which vanishes in the limit of $\sqrt s  \to {\left( {\sqrt s } \right)_T} + 0$, while $A^{(0),n}$  remains restricted with some non-zero value. Therefore, the exact value of magnitude  $A^{(0),n}$ at such energies is insufficient.

\section{The approximate solution of the transcendental equation expressing the difference of rapidity arithmetic progression at the point of constrained maximum of the scattering amplitude}
\label{SECTION_AndriiAddon_02}

Consider the transcendental equation Eq.\ref{eq64} taking into account the fact that
$\Delta y(n,\sqrt s)=2y_{\frac{n}{2}}$.  At energies sufficiently higher the threshold value, when is $\Delta y/2$  not small anymore, the energy of final-state protons
\begin{eqnarray}%
  {P_{30}} + {P_{40}} = 2M{\mathop{\rm ch}\nolimits} \left( {\left( {n + 1} \right)\frac{{\Delta y}}{2}} \right)
\end{eqnarray}%
is much higher than the pions energy $\frac{{{\mathop{\rm sh}\nolimits} \left( {\frac{n}{2}\Delta y} \right)}}{{{\mathop{\rm sh}\nolimits} \left( {\frac{1}{2}\Delta y} \right)}}$
which is caused with the greatness of proton mass $M$ (in the units of pion mass) and with the fact that at not small $\Delta y/2$  one gets
\begin{eqnarray}%
  \mbox{\fontsize{12}{14}\selectfont $  2M{\mathop{\rm ch}\nolimits} \left( {\left( {n + 1} \right)\frac{{\Delta y}}{2}} \right)\sim \exp \left( {\left( {n + 1} \right)\frac{{\Delta y}}{2}} \right) $}
\label{eq_AndriiAddon_17a}
\end{eqnarray}%
while
\begin{eqnarray}%
  \mbox{\fontsize{12}{14}\selectfont $  \frac{{{\mathop{\rm sh}\nolimits} \left( {\frac{n}{2}\Delta y} \right)}}{{{\mathop{\rm sh}\nolimits} \left( {\frac{1}{2}\Delta y} \right)}}\sim \exp \left( {\left( {n - 1} \right)\frac{{\Delta y}}{2}} \right) $}
\label{eq_AndriiAddon_17b}
\end{eqnarray}%

\begin{figure}
\begin{center}
  \includegraphics[scale=0.4]{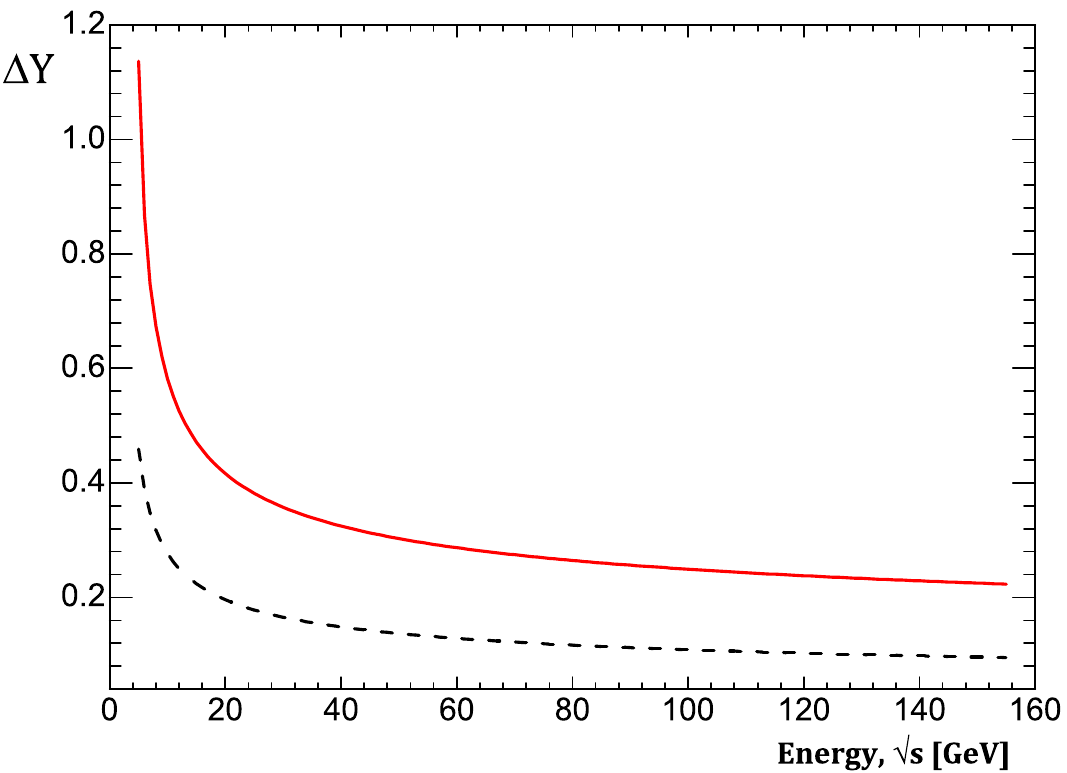}
\end{center}
\caption{
The dependence of difference between initial and final-state protons rapidities on energy $\sqrt s$, calculated in the point of constrained maximum of scattering amplitude for $n=10$ (solid line), and $n=20$  (dash line).}
\label{fig:AnDiplom_fig_03}
\end{figure}

\begin{figure}
\begin{center}
  \includegraphics[scale=0.45]{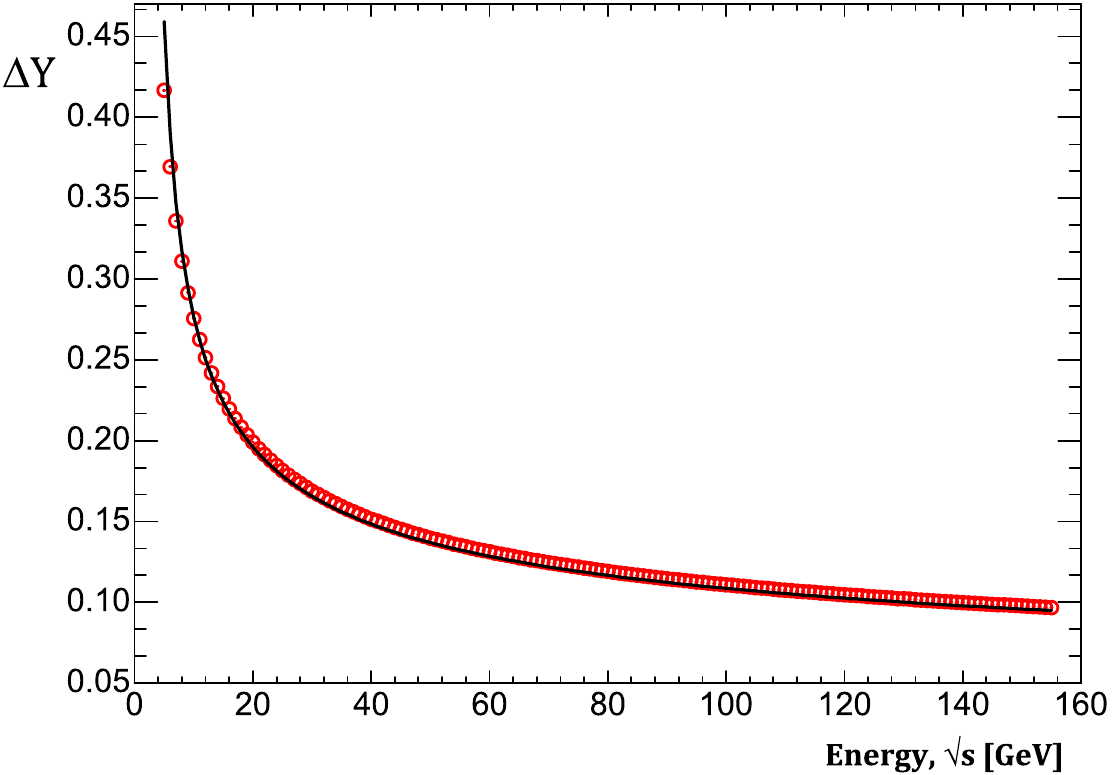}
\end{center}
\caption{
The comparison of approximate solution of Eq.\ref{eq_AndriiAddon_23} (circle) with results of numerical solution of Eq.\ref{eq64} with respect to magnitude $\Delta Y$  (solid line) at $n=10$.}
\label{fig:AnDiplom_fig_04}
\end{figure}

The fact that most of energy in c.m.s. framework is carried by secondary protons in its turn means that the rapidity of each of these protons is close on absolute value to initial proton's rapidity (let's denote it  $Y^*$). Namely,  ${Y^*} - (n + 1)\frac{{\Delta y}}{2} \ll 1$ (see Fig.\ref{fig:AnDiplom_fig_03}).  Let's enter the new variable instead of  $\Delta y$
\begin{eqnarray}%
  \mbox{\fontsize{12}{14}\selectfont $ \Delta Y = {Y^*} - (n + 1)\frac{{\Delta y}}{2} $}
\label{eq_AndriiAddon_18}
\end{eqnarray}%
Taking into account $\sqrt s /2 = M{\mathop{\rm ch}\nolimits} \left( {{Y^ * }} \right)$  we'll represent Eq.\ref{eq64} as
\begin{eqnarray}%
  \mbox{\fontsize{13}{14}\selectfont $ 2M sh\left( {{Y^*} - \frac{{\Delta Y}}{2}} \right)sh\left( {\frac{{\Delta Y}}{2}} \right) = \frac{{sh\left( {\frac{n}{{n + 1}}\left( {{Y^*} - \Delta Y} \right)} \right)}}{{2sh\left( {\frac{1}{{n + 1}}\left( {{Y^*} - \Delta Y} \right)} \right)}} $} \nonumber\\
\label{eq_AndriiAddon_19}
\end{eqnarray}%
Neglecting the small magnitude $\Delta Y$  with respect to large magnitude $Y^*$ one gets
\begin{eqnarray}%
  \mbox{\fontsize{14}{14}\selectfont $ sh\left( {\frac{{\Delta Y}}{2}} \right) = \frac{{sh\left( {\frac{{n{Y^*}}}{{n + 1}}} \right)}}{{4Msh\left( {{Y^*}} \right)sh\left( {\frac{{{Y^*}}}{{n + 1}}} \right)}} $}
\label{eq_AndriiAddon_20}
\end{eqnarray}%
Taking into account the smallness of $\Delta Y$
\begin{eqnarray}%
  \mbox{\fontsize{14}{14}\selectfont $ \Delta Y = \frac{{sh\left( {\frac{{n{Y^*}}}{{n + 1}}} \right)}}{{2Msh\left( {{Y^*}} \right)sh\left( {\frac{{{Y^*}}}{{n + 1}}} \right)}} $}
\label{eq_AndriiAddon_21}
\end{eqnarray}%
Since the arguments of hyperbolic sine and cosine functions are large
\begin{eqnarray}%
  \mbox{\fontsize{13}{14}\selectfont $ sh\left( {\frac{{\Delta Y}}{2}} \right) \approx \frac{1}{{2M\left( {\exp \left( {\frac{2}{{n + 1}}{Y^*}} \right) - 1} \right)}} \approx \frac{1}{{2M\left( {{{\left( {\frac{{\sqrt s }}{M}} \right)}^{\frac{2}{{n + 1}}}} - 1} \right)}} $}\nonumber\\
\label{eq_AndriiAddon_22}
\end{eqnarray}%
or
\begin{eqnarray}%
  \mbox{\fontsize{12}{14}\selectfont $ \Delta Y \approx \frac{1}{{M\left( {{{\left( {\frac{{\sqrt s }}{M}} \right)}^{\frac{2}{{n + 1}}}} - 1} \right)}} $}
\label{eq_AndriiAddon_23}
\end{eqnarray}%
In order to control the applicability of made above approximations let's compare the solution of Eq.\ref{eq_AndriiAddon_23} with the one, which can be obtained from Eq.\ref{eq_AndriiAddon_18} substituting the ``exact" solution of transcendental Eq.\ref{eq64} (where $\Delta y(n,\sqrt s)=2y_{\frac{n}{2}}$). The result of such a comparison for $n=10$  is depicted on Fig.\ref{fig:AnDiplom_fig_04}. These results enable to conclude that the entered approximations are applicable at least for rather high multiplicity of final-state particles.

Now we can pass to the approximations for other magnitudes, which enter to Eq.\ref{eq_AndriiAddon_15}, expressing the aforementioned scattering a multitude at the point of constrained maximum.

\section{The analytical representation of Feynman denominators at high energies}
\label{SECTION_AndriiAddon_03}
\begin{figure}
\begin{center}
  \includegraphics[scale=0.45]{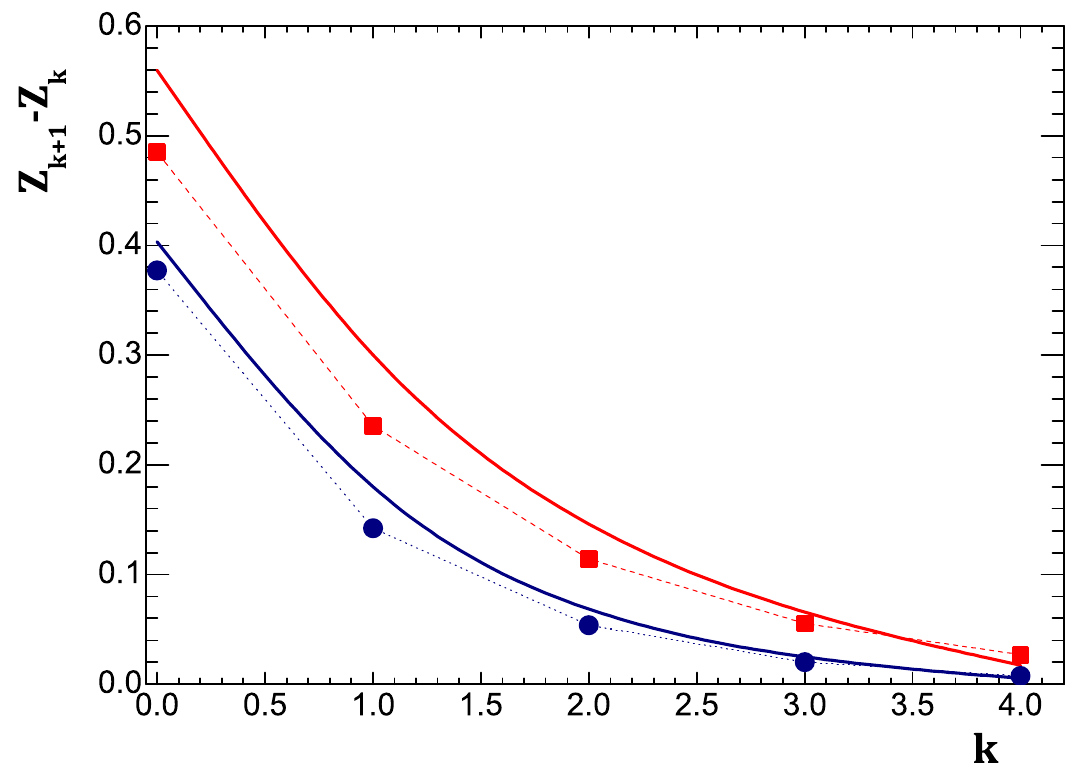}
\end{center}
\caption{
The comparison for the magnitude of Feynman denominator jump on the comb at the point of constrained maximum obtained in the approximation Eq.\ref{eq_AndriiAddon_29} (dashed lines with boxes and circles) with its exact value (red line) for $n=10$ at $\sqrt s =50$ GeV and at $\sqrt s =200$ GeV (blue line).}
\label{fig:AnDiplom_fig_05}
\end{figure}

Now, it is possible to analytically show the applicability of transformation Eq.\ref{eq_AndriiAddon_14}. The Feynman denominators on the comb may be represented as follows
\begin{widetext}
\begin{equation}
{Z_i} = 1 
+ \left( {\frac{{{e^{\left( {n - 2i} \right)\frac{{\Delta y}}{2}}} - {e^{ - n\frac{{\Delta y}}{2}}}}}{{{e^{\frac{{\Delta y}}{2}}} - {e^{ - \frac{{\Delta y}}{2}}}}} + M\left( {{e^{ - {Y_3}}} - {e^{ - \left( {{Y_3} + dY} \right)}}} \right)} \right) 
 \times \left( {\sum\limits_{l = 1}^i {{e^{ - \left( {n + 1 - 2l} \right)\frac{{\Delta y}}{2}}}}  + M\left( {{e^{ - {Y_3}}} - {e^{ - \left( {{Y_3} + dY} \right)}}} \right)} \right) 
\label{eq_AndriiAddon_24}
\end{equation}
\end{widetext}
where
\begin{eqnarray}%
  {Y_3} = \rm {arcsh}\left( P_{3\parallel }/{M} \right)
\end{eqnarray}%
If we consider the energies higher than threshold one  $\sqrt s \ll \left[ {{{\left( {\sqrt s } \right)}_T} = 2M + n} \right]$ we can neglect the product of two small quantities in this expression. Then we get
\begin{eqnarray}%
&&{Z_i} \approx 1 + \frac{{{e^{\left( {n - 2i} \right)\frac{{\Delta y}}{2}}}}}{{{e^{\frac{{\Delta y}}{2}}} - {e^{ - \frac{{\Delta y}}{2}}}}}\nonumber\\
&&  \times \left( {\sum\limits_{l = 1}^i {{e^{ - \left( {n + 1 - 2l} \right)\frac{{\Delta y}}{2}}}}  + M\left( {{e^{ - {Y_3}}} - {e^{ - \left( {{Y_3} + dY} \right)}}} \right)} \right)\nonumber\\
\label{eq_AndriiAddon_25}
\end{eqnarray}%
Then, let's consider the expression for $Z_0$. Taking into account Eq.\ref{eq_AndriiAddon_18} one can represent Eq.\ref{eq_AndriiAddon_16} as follows
\begin{eqnarray}%
  \mbox{\fontsize{12}{14}\selectfont ${Z_0} = 1 + 4{M^2}s{h^2}\left( {\frac{{\Delta Y}}{2}} \right)$}
\label{eq_AndriiAddon_26}
\end{eqnarray}%
Taking into account Eq.\ref{eq_AndriiAddon_22} we get
\begin{eqnarray}%
  \mbox{\fontsize{12}{14}\selectfont ${Z_0} = 1 + {\left( {{{\left( {\frac{{\sqrt s }}{M}} \right)}^{\frac{2}{{n + 1}}}} - 1} \right)^{ - 2}}$}
\label{eq_AndriiAddon_27}
\end{eqnarray}%
Substituting  Eq.\ref{eq_AndriiAddon_27} into  Eq.\ref{eq_AndriiAddon_25} will result in
\begin{eqnarray}%
  \mbox{\fontsize{12}{14}\selectfont ${Z_i} = 1 + {\left( {{{\left( {\frac{{\sqrt s }}{M}} \right)}^{\frac{2}{{n + 1}}}} - 1} \right)^{ - 2}} + \sum\limits_{k = 1}^i {{{\left( {\frac{M}{{\sqrt s }}} \right)}^{2\frac{i}{{n + 1}}}}} $}\nonumber\\
\label{eq_AndriiAddon_28}
\end{eqnarray}%
or
\begin{eqnarray}%
  \mbox{\fontsize{12}{14}\selectfont ${Z_{i + 1}} - {Z_i} = {e^{ - \left( {i + 1} \right)\Delta y}} = {\left( {\frac{M}{{\sqrt s }}} \right)^{2\frac{{i + 1}}{{n + 1}}}} $}
\label{eq_AndriiAddon_29}
\end{eqnarray}%
Namely, one can see that at high energies the difference between Feynman denominators falls exponentially. In other words, at high energies, the largest change of Feynman denominator occurs at the movement from zero (1st on the comb) to 1st (2nd on the comb) denominator. The result Eq.\ref{eq_AndriiAddon_29} is illustrated on a Fig.\ref{fig:AnDiplom_fig_05}
\begin{figure}
\begin{center}
  \includegraphics[scale=0.45]{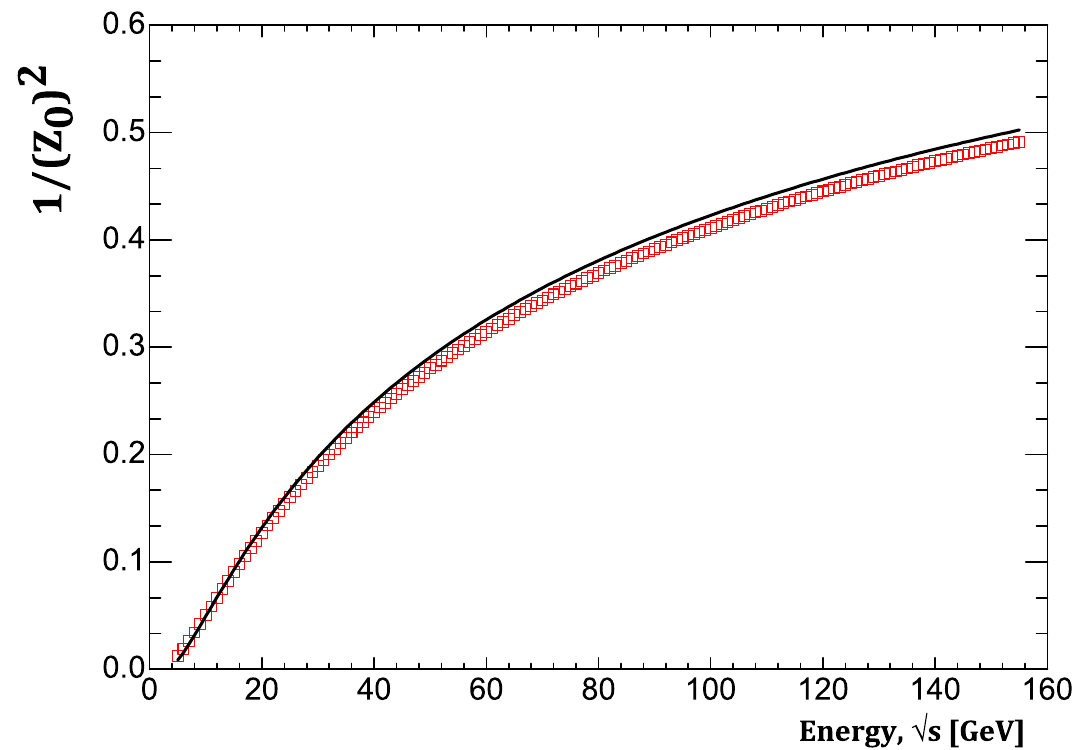}
\end{center}
\caption{
The dependence of $\frac{1}{{{{\left( {{Z_0}} \right)}^2}}}$ on energy $\sqrt s$ at $n=10$ obtained from Eq.\ref{eq_AndriiAddon_16} by substituting the solution of transcendental equation Eq.\ref{eq64} $\textendash$ solid line; with the approximation Eq.\ref{eq_AndriiAddon_27} $\textendash$ boxes.}
\label{fig:AnDiplom_fig_06}
\end{figure}

\section{The analytical expression for scattering amplitude dependence on energy $\sqrt s$ at the point of constrained maximum}
\label{SECTION_AndriiAddon_04}

In the previous analysis among the other results we derived an expression for $Z_0$  Eq.\ref{eq_AndriiAddon_27}. In order to verify a precision of this approximation we'll compare it with an exact solution which can be obtained from Eq.\ref{eq_AndriiAddon_16} by substituting the solution of transcendental equation Eq.\ref{eq_AndriiAddon_19}. As one can see from Fig.\ref{fig:AnDiplom_fig_06} this is an adequate approximation.

\begin{figure}
\begin{center}
  \includegraphics[scale=0.42]{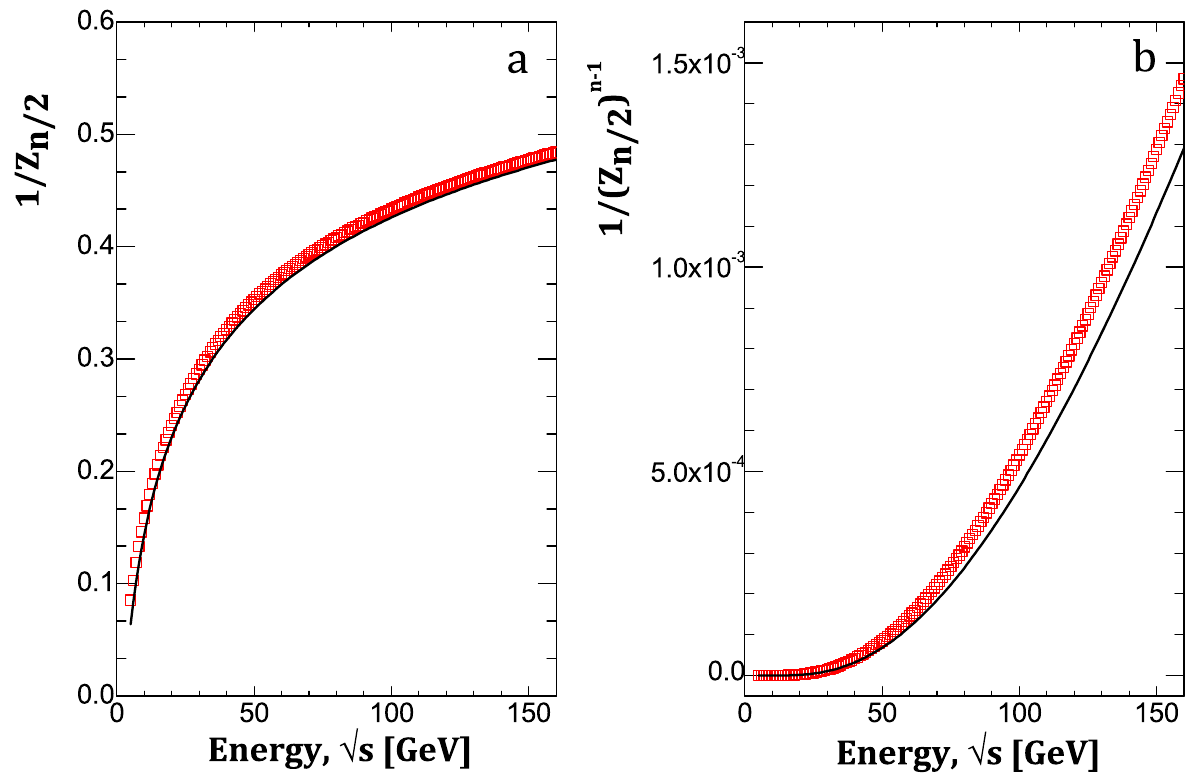}
\end{center}
\caption{
The dependence of ${\left( {{Z_{\frac{n}{2}}}} \right)^{ - 1}}$  (a) and   ${\left( {{Z_{\frac{n}{2}}}} \right)^{ - (n-1)}}$ (b) on energy $\sqrt s$ at $n=10$. Solid line corresponds to Eq.\ref{eq_AndriiAddon_17} with substituting the numerical solution of Eq.\ref{eq_AndriiAddon_19}, while dashed line refers to approximation  Eq.\ref{eq_AndriiAddon_32}.}
\label{fig:AnDiplom_fig_07}
\end{figure}

Thus, we got $Z_0$ described analytically. Now let's get an analytical expression for two other multipliers entering to Eq.\ref{eq_AndriiAddon_15}. First we'll rewrite $Z_{n/2}$ Eq.\ref{eq_AndriiAddon_17} in form
\begin{eqnarray}%
&&{Z_{\frac{n}{2}}} = 1 + \left( {2Msh\left( {\frac{{\Delta Y}}{2}} \right)ch\left( {{Y^*} - \frac{{\Delta Y}}{2}} \right)} \right. \nonumber\\
&& {\left. { - \frac{{ch\left( {\frac{n}{{n + 1}}\left( {{Y^*} - \Delta Y} \right)} \right) - 1}}{{2sh\left( {\frac{{{Y^*} - \Delta Y}}{{n + 1}}} \right)}}} \right)^2} 
\label{eq_AndriiAddon_30}
\end{eqnarray}%
Neglecting as before $\Delta Y$ with respect to $Y^*$ and taking the common factor $ch(Y^*)$ out of the brackets we get:
\begin{eqnarray}%
  \mbox{\fontsize{11}{14}\selectfont ${Z_{\frac{n}{2}}} = 1 + s{\left( {sh\left( {\frac{{\Delta Y}}{2}} \right) - \frac{{ch\left( {\frac{n}{{n + 1}}\left( {{Y^*}} \right)} \right) - 1}}{{2M\left( {sh\left( {\frac{{n + 2}}{{n + 1}}{Y^*}} \right) - sh\left( {\frac{n}{{n + 1}}{Y^*}} \right)} \right)}}} \right)^2} $}\nonumber\\
\label{eq_AndriiAddon_31}
\end{eqnarray}%
In further, we would get rid of a small exponential summands, which enters to hyperbolic sine and cosine, and that leads us to:
\begin{eqnarray}%
  \mbox{\fontsize{12}{14}\selectfont ${Z_{\frac{n}{2}}} \approx 1 + {\left( {\frac{{{{\left( {\frac{{\sqrt s }}{M}} \right)}^{\frac{1}{{n + 1}}}}}}{{{{\left( {\frac{{\sqrt s }}{M}} \right)}^{\frac{2}{{n + 1}}}} - 1}}} \right)^2}$}
\label{eq_AndriiAddon_32}
\end{eqnarray}%
Again, the obtained approximation is compared with the results of numerical calculations Fig.\ref{fig:AnDiplom_fig_07}
\begin{figure}
\begin{center}
  \includegraphics[scale=0.4]{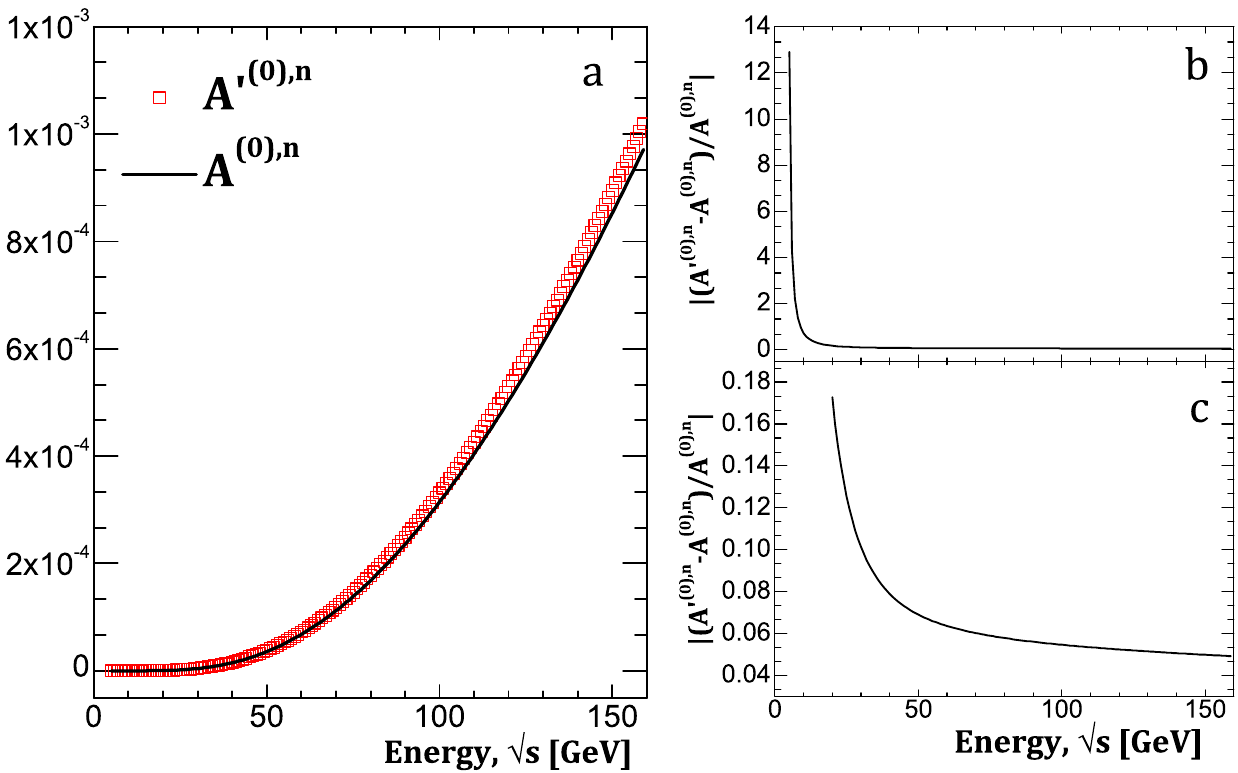}
\end{center}
\caption{The dependence of scattering amplitude in the point of constrained maximum on energy $\sqrt s$ (a). Solid line corresponds to exact solution obtained from Eqs.\ref{eq_AndriiAddon_11}, \ref{eq_AndriiAddon_12} with a substitution of numerical solution of transcendental equation Eq.\ref{eq64}, while dashed line represents the approximation Eqs.\ref{eq_AndriiAddon_36};
(b) the relative deviation in energy range $\sqrt s =5 \div155$ GeV; (c) the relative deviation in energy range $\sqrt s =20 \div155$ GeV.}
\label{fig:AnDiplom_fig_08}
\end{figure}
Now the only thing left is to represent
\begin{eqnarray}%
  \mbox{\fontsize{10}{14}\selectfont $\sqrt {\frac{s}{4} - M{\,^2}}  - M{\mathop{\rm sh}\nolimits} \left( {\left( {n + 1} \right)\frac{{\Delta y}}{2}} \right) - \frac{{{\mathop{\rm ch}\nolimits} \left( {n\frac{{\Delta y}}{2}} \right)}}{{2{\mathop{\rm sh}\nolimits} \left( {\frac{{\Delta y}}{2}} \right)}}$} \nonumber
\end{eqnarray}%
in a more convenient form. First let's rewrite it as follows:
\begin{eqnarray}%
&&  \mbox{\fontsize{12}{14}\selectfont $ \sqrt {\frac{s}{4} - M{\,^2}}  - M{\mathop{\rm sh}\nolimits} \left( {\left( {n + 1} \right)\frac{{\Delta y}}{2}} \right) - \frac{{{\mathop{\rm ch}\nolimits} \left( {n\frac{{\Delta y}}{2}} \right)}}{{2{\mathop{\rm sh}\nolimits} \left( {\frac{{\Delta y}}{2}} \right)}} =  $} \nonumber\\
&&  \mbox{\fontsize{12}{14}\selectfont $  =  - 2M\frac{{{\mathop{\rm sh}\nolimits} \left( {{Y^ * } + \frac{{n - 1}}{{n + 1}}\frac{{\Delta {y_1}}}{2}} \right) - {\mathop{\rm sh}\nolimits} \left( {\frac{{n - 1}}{{n + 1}}\left( {{Y^ * } - \frac{{\Delta {y_1}}}{2}} \right)} \right)}}{{{\mathop{\rm sh}\nolimits} \left( {\frac{{2n}}{{n + 1}}{Y^ * }} \right)}}{\mathop{\rm sh}\nolimits} \left( {\frac{{\Delta {y_1}}}{2}} \right) $} \nonumber\\
\label{eq_AndriiAddon_33}
\end{eqnarray}%
Performing the same trick as before, namely, neglecting  $\Delta Y$ with respect to  $Y^*$ and getting rid of small exponential summands we get:
\begin{eqnarray}%
  \mbox{\fontsize{11}{14}\selectfont $\sqrt {\frac{s}{4} - M{\,^2}}  - M{\mathop{\rm sh}\nolimits} \left( {\left( {n + 1} \right)\frac{{\Delta y}}{2}} \right) - \frac{{{\mathop{\rm ch}\nolimits} \left( {n\frac{{\Delta y}}{2}} \right)}}{{2{\mathop{\rm sh}\nolimits} \left( {\frac{{\Delta y}}{2}} \right)}} \approx  - \frac{1}{{\left( {\frac{{\sqrt s }}{M}} \right)}} $} \nonumber\\
\label{eq_AndriiAddon_34}
\end{eqnarray}%
Similarly
\begin{eqnarray}%
  \mbox{\fontsize{14}{14}\selectfont $\frac{{{\mathop{\rm sh}\nolimits} \left( {\left( {n - 1} \right)\frac{{\Delta y}}{2}} \right)}}{{{\mathop{\rm sh}\nolimits} \left( {\frac{{\Delta y}}{2}} \right)}} \approx \frac{{{{\left( {\frac{{\sqrt s }}{M}} \right)}^{\frac{n}{{n + 1}}}}}}{{{{\left( {\frac{{\sqrt s }}{M}} \right)}^{\frac{2}{{n + 1}}}} - 1}} $}
\label{eq_AndriiAddon_35}
\end{eqnarray}%
Finally, let's gather all the results. Substituting Eq.\ref{eq_AndriiAddon_27}, Eq.\ref{eq_AndriiAddon_32}, Eq.\ref{eq_AndriiAddon_34} and Eq.\ref{eq_AndriiAddon_35} into Eq.\ref{eq_AndriiAddon_15} we get an analytic representation of scattering amplitude at the point of constrained maximum:
\begin{eqnarray}%
&& A^{(0),n}=\left( 1+a(\sqrt s,n)\right)^{-2}   \nonumber\\
&& \times \left( 1+  b(\sqrt s,n) \right)^{-(n-1)} \exp\left( c(\sqrt s,n) \right)  
\label{eq_AndriiAddon_36}
\end{eqnarray}%
where
\begin{eqnarray}%
&&   a(\sqrt s,n)=\left( \frac{1}{ \left( {\sqrt s}/M \right)^{\frac{2}{n+1}}-1 }\right)^2  \nonumber\\
&&  b(\sqrt s,n)=\left( \frac{\left( {\sqrt s}/M \right)^{\frac{1}{n+1}}}{\left( {\sqrt s}/M \right)^{\frac{2}{n+1}}-1} \right)^2  \nonumber\\
&&   c\left( {\sqrt s ,n} \right) = 2\left( {1 - \left( {n - 1} \right){{\left( {\frac{{\sqrt s }}{M}} \right)}^{ - \frac{n}{{n + 1}}}}{a^{ - \frac{1}{2}}}\left( {\sqrt s ,n} \right)} \right)  \nonumber\\
&&   \times {\left( {{a^{ - 1}}\left( {\sqrt s ,n} \right) + {{\left( {\frac{{\sqrt s }}{M}} \right)}^{\frac{2}{{n + 1}}}}} \right)^{ - 1}}  
\label{eq_AndriiAddon_36a}
\end{eqnarray}%
$a(\sqrt s,n)$ and $b(\sqrt s,n)$ determine the characteristic value of virtuality at the maximum point of scattering amplitude, and
$c(\sqrt s,n)$  determines the variation of virtuality along the ``comb". In other words, the following estimate takes place
\begin{eqnarray}%
  \mbox{\fontsize{12}{14}\selectfont $
a(\sqrt s, n) \le \left| {{{\left( {{q^{(j)}}} \right)}^2}} \right| \le  b(\sqrt s, n)
$}
\label{eq:eq_part_virtuality1}
\end{eqnarray}%
where $\left| \left( q^{(j)} \right)^2 \right|$ is the absolute value of virtuality corresponding to $j$-th internal line on the ``comb" in the point of constrained maximum.
The comparison of scattering amplitude dependence on energy $\sqrt s$ obtained in analytical way with the one obtained by numerically solving Eqs.\ref{eq_AndriiAddon_11}, \ref{eq_AndriiAddon_12} and substituting numerical solution of Eq.\ref{eq64} is presented on Fig.\ref{fig:AnDiplom_fig_08}.

Notation $A^{(0),n}$ was introduced to distinguish the exact value of amplitude defined by the relations Eq.\ref{eq_AndriiAddon_11} and Eq.\ref{eq_AndriiAddon_12} from its approximate value Eq.\ref{eq_AndriiAddon_36}. To characterize the accuracy of approximation Eq.\ref{eq_AndriiAddon_36} was introduced a relative deviation (see Fig.\ref{fig:AnDiplom_fig_08}b and Fig.\ref{fig:AnDiplom_fig_08}c):
\begin{equation}
  \mbox{\fontsize{14}{14}\selectfont $  \varepsilon  = \left| {\frac{{{{A'}^{\left( 0 \right),n}} - {A^{\left( 0 \right),n}}}}{{{A^{\left( 0 \right),n}}}}} \right| $}
\label{eq:epsilon}
\end{equation}

\section{Discussion and conclusions}
\label{SECTION_Discussion_and_Conclusions}

The main results and conclusion of this paper is that the multi-peripheral scattering amplitude indeed has a point of constrained maximum under conditions of the energy-momentum conservation law.
This leads to the fact that the principal contribution to the multi-dimensional integrals, which representing the inelastic scattering cross-section with production of a given number of particles, makes a small neighborhood of a maximum point.

Analyzing properties of the maximum point leads to some differences of the physical picture of the multi-peripheral processes from ones, which leads to the basic formulas of Reggeon theory [\onlinecite{Baker19761, Nikitin:113716, Ter-Martirosyan}]. In particular, the rapidities of secondary particles at the maximum point are arranged and are equidistant from each other, as assumed in the justification of Reggeon formulas. At the same time in the presented model the distance between the adjacent rapidities (i.e., difference of an arithmetic progression) depends on energy $\sqrt s $ and on number of particles $n$ (see Eq.\ref{eq65} and Eq.\ref{eq71}), but not a constant value close to unity, as it accepted, for instance, in [\onlinecite{Nikitin:113716}] and [\onlinecite{Ter-Martirosyan}]. The assumption, that the interval between adjacent rapidities does not depend on energy plays an important role for the ground of power dependence of the imaginary part of elastic scattering amplitude on Lorentz-invariant $s$, that results in the Regge pole.

It is useful to rewrite the energy term entering both sides of
Eq.\ref{eq:eq_part_virtuality1} 
in the following form
\begin{equation}
  \mbox{\fontsize{13}{14}\selectfont $ \left(  \frac{\sqrt s}{M}\right)^{\frac{1}{n+1}} = \exp\left(\frac{1}{n+1} \ln\left( \frac{\sqrt s}{M} \right) \right) $}
\label{eq:eq_part_virtuality2}
\end{equation}

The growth of exponent with energy $\sqrt s$ is much weaker than the corresponding decrease, related to an increase in the number of particles $n$.  Thus, one can see that for large $n$, $\left( \sqrt s /M \right)^{\frac{1}{n+1}} \sim 1$ even at high energies ($\sqrt s>>M$). 

As a result, the difference of energy and longitudinal momentum squares is not negligible compared to the transverse momentum for each virtuality on the ``comb”. This conclusion comes in contradiction with the statement that virtualities can be reduced to the transverse momentum squared. This statement is usually claimed in the standard approach
[\onlinecite{Byckling:100542, Collins:111502, bfkl_1976, Nikitin:113716, Ter-Martirosyan, levin_2, KozlovNSU_2007, Lipatov:2008}].

Taking into account the growth of $\left( \sqrt s /M \right)^{\frac{2}{n+1}}$ with energy $\sqrt s$ growth, we see that virtuality at the maximum point do decreases, and the maximum value of amplitude grows with the growth of energy $\sqrt s$. 
Note also, that for large $n$, the magnitude $\left( \sqrt s /M \right)^{\frac{1}{n+1}}$ is close to unity 
for a wide energy range. This means that the growth of cross-section is indeed much steeper than the one attained in the Regge-based theories [\onlinecite{springerlink:10.1007/BF02781901, Collins:111502, Kaidalov:2003, levin_2, levin}], which is described by a factor of $ln^{n-2}\left( \sqrt s/ M \right)$. 
Moreover, this energy range is wider for larger $n$
The asymptotic behavior for different $n$ is reached at different $\sqrt s$, which casts doubt on the validity of the asymptotic formulas of multi-Regge kinematics.

In addition, as it follows from an examination of multi-peripheral diagrams in [\onlinecite{Collins:111502}] and [\onlinecite{Nikitin:113716}],  Reggeon formulas are derived in case of total disregard to the dependence of the integrand for the cross section of production a certain number of particles on the rapidity of these particles.  In such an approach a cross-section is defined by the value of rapidity phase volume, which is also well noticeable from the comparison of dependence of a set number particle production on $\sqrt {s}$
with the calculation of so-called volume of ``transversal-truncated phase space'' [\onlinecite{Byckling:100542}].

Obviously, that at the described above approach, the dependence of scattering amplitude on rapidity is substantial,
because just this dependence determines the value of scattering amplitude in the neighborhood of the most probable configuration of momentums.

Existence of the constrained maximum point in multi-peripheral scattering amplitude and properties of this maximum allow us to apply this information for the calculation of inelastic scattering cross-section. This is the main purpose of our next paper.

\appendix
\section{The proof of spatial similarity of virtual particle four-momentums in the diagram of Fig.\ref{fig:part1_fig02}}
\label{Appendix_A}
We carry out the proof by ``contradiction". Suppose that any of the four-momentums of virtual particles, for example, ${P_1} - {P_3} - {p_1} - {p_2} - ... - {p_k}$ is time-like. Note, that this four-momentum must be equal to ${P_4} - {P_2} + {p_n} + {p_{n - 1}} + ... + {p_{k + 1}}$ by virtue of the law of energy-momentum conservation.

We assume, that time component of the time-like four-momentum (energy) can not vanish in any inertial system, which means that the sign of the time component of such four-momentum is the Lorentz-invariant quantity, i.e., can not be changed when we move from one inertial system to another.

If we assume that the sign of the time components of four-momentum ${P_1} - {P_3} - {p_1} - {p_2} - ... - {p_k}$ (or four-vector ${P_4} - {P_2} + {p_n} + {p_{n - 1}} + ... + {p_{k + 1}}$) is positive, the following inequalities must be simultaneously satisfied in all the inertial systems:
\begin{eqnarray}
&&  {\left( {{P_1}} \right)^0} - {\left( {{P_3}} \right)^0} - {\left( {{p_1}} \right)^0} - ... - {\left( {{p_{k - 1}}} \right)^0} - {\left( {{p_k}} \right)^0} > 0 \nonumber\\
\\
&&  {\left( {{P_4}} \right)^0} - {\left( {{P_2}} \right)^0} + {\left( {{p_n}} \right)^0} + {\left( {{p_{n - 1}}} \right)^0} + ... + {\left( {{p_{k + 1}}} \right)^0} > 0  \nonumber\\
\end{eqnarray}

Given that both inequalities include the energy components of the energy-momentum four-vectors of real particles, satisfying to the mass shell conditions, we can rewrite them as
\begin{eqnarray}
&& \sqrt {{M^2} + {{\left( {{{\vec P}_1}} \right)}^2}}  - \sqrt {{M^2} + {{\left( {{{\vec P}_3}} \right)}^2}} -\sqrt {{m^2} + {{\left( {{{\vec p}_1}} \right)}^2}}  \nonumber\\
&& ...- \sqrt {{m^2} + {{\left( {{{\vec p}_{k-1}}} \right)}^2}}  - \sqrt {{m^2} + {{\left( {{{\vec p}_k}} \right)}^2}}  > 0 \\ \nonumber\\
&&\sqrt {{M^2} + {{\left( {{{\vec P}_4}} \right)}^2}}  - \sqrt {{M^2} + {{\left( {{{\vec P}_2}} \right)}^2}}+ \sqrt {{m^2} + {{\left( {{{\vec p}_n}} \right)}^2}} \nonumber\\
&& + \sqrt {{m^2} + {{\left( {{{\vec p}_{n-1}}} \right)}^2}} + ...+ \sqrt {{m^2} + {{\left( {{{\vec p}_{k + 1}}} \right)}^2}}  > 0
\end{eqnarray}
Note, that first of these inequalities does not hold in the rest frame of particle $P_1$ because
\begin{eqnarray}
&& \sqrt {{M^2}}  - \sqrt {{M^2} + {{\left( {{{\vec P}_3}} \right)}^2}}- \sqrt {{m^2} + {{\left( {{{\vec p}_1}} \right)}^2}}  \nonumber\\
&& ... - \sqrt {{m^2} + {{\left( {{{\vec p}_{k-1}}} \right)}^2}} - \sqrt {{m^2} + {{\left( {{{\vec p}_k}} \right)}^2}} \; < 0
\end{eqnarray}
If we assume that time component of considered four-vector has a negative sign, then we obtain, that
in all inertial systems the following inequalities must be simultaneously satisfied:
\begin{eqnarray}
&& \sqrt {{M^2} + {{\left( {{{\vec P}_1}} \right)}^2}}  - \sqrt {{M^2} + {{\left( {{{\vec P}_3}} \right)}^2}}  - \sqrt {{m^2} + {{\left( {{{\vec p}_1}} \right)}^2}}  \nonumber\\
&&... - \sqrt {{m^2} + {{\left( {{{\vec p}_{k-1}}} \right)}^2}} - \sqrt {{m^2} + {{\left( {{{\vec p}_k}} \right)}^2}}  < 0 \\ \nonumber\\
&&\sqrt {{M^2} + {{\left( {{{\vec P}_4}} \right)}^2}}  - \sqrt {{M^2} + {{\left( {{{\vec P}_2}} \right)}^2}} + \sqrt {{m^2} + {{\left( {{{\vec p}_n}} \right)}^2}}  \nonumber\\
&& + \sqrt {{m^2} + {{\left( {{{\vec p}_{n-1}}} \right)}^2}}  + ... + \sqrt {{m^2} + {{\left( {{{\vec p}_{k + 1}}} \right)}^2}}  < 0
\end{eqnarray}

However, second of these inequalities can not take place in the rest frame of particle $P_2$.

Thus, we can satisfy of energy-momentum conservation law in all inertial systems only in case, if the time component of considered four-vector changes a sign when moving from one reference frame to other. However, in this case, the four-vector can not be time-like.  Because presented argumentation is applicable to any of virtual four-momentums in the diagram Fig.\ref{fig:part1_fig02}, they must all be space-like. Quod erat demonstrandum. 

\section{The proof of Eq.\ref{eq14} from Section.\ref{SECTION_2}}
\label{Appendix_B}
Using function ${A_\parallel }\left( {n,{y_1},{y_2},...,{y_n}} \right)$ defined by Eq.\ref{eq:eq_10}, let us examine function ${{A'}_\parallel }\left( {n,{y_1},{y_2},...,{y_n}} \right)$, which defined by relation:
\begin{eqnarray}
&& {{A'}_\parallel }\left( {n,{y_1},{y_2}, \cdots ,{y_{\frac{n}{2} - 1}},{y_{\frac{n}{2}}},{y_{\frac{n}{2} + 1}}, \cdots ,{y_{n - 1}},{y_n}} \right)  \nonumber\\
&&   = {A_\parallel }\left( {n, - {y_n}, - {y_{n - 1}}, \cdots , - {y_{\frac{n}{2}}}, \cdots , - {y_2}, - {y_1}} \right) 
\label{eq2.1}
\end{eqnarray}
The expression
\begin{eqnarray}
  {{A'}_\parallel }\left( {n,{y_1},{y_2},...,{y_n}} \right)
\label{eq2.1a}
\end{eqnarray}
we would obtain if we were choose the direction of longitudinal momentum ${P_{2\parallel }}$ as the positive direction of collision axis and interchange the four-momentums of particles in the diagram, arranged symmetrically about the axis in Fig.\ref{fig:part1_fig03} (numeration of the diagram vertices corresponds to Fig.\ref{fig:part1_fig02}).

In order to write function Eq.\ref{eq2.1a} in explicit form, at first we will rewrite it in following way:
\begin{eqnarray}
&& {A_\parallel } = {\left( {m - {{\left( {\frac{{\sqrt s }}{2} - {P_{30}}} \right)}^2} + {{\left( {{S_k}} \right)}^2}} \right)^{ - 1}} \nonumber\\
&& \times \prod\limits_{k = 1}^{n - 1} {{{\left( {m - {{\left( {{E_k}} \right)}^2} + {{\left( {{S_k} - m\sum\limits_{l = 1}^k {sh\left( {{y_l}} \right)} } \right)}^2}} \right)}^{ - 1}}}  \nonumber\\
&&\times {\left( {m - {{\left( {{E_n}} \right)}^2} + {{\left( {{S_k} - m\sum\limits_{l = 1}^n {sh\left( {{y_l}} \right)} } \right)}^2}} \right)^{ - 1}}
\label{eq2.2}
\end{eqnarray}
where
\begin{eqnarray}
&&  {S_k} = \sqrt {s/4 - {M^2}}  - {P_{3\parallel }} \\
&&  {E_k} = {\sqrt s }/2 - \left( {{P_{30}} + m\sum\limits_{l = 1}^k {ch\left( {{y_l}} \right)} } \right) \\
&&  {E_n} = {\sqrt s }/2 - \left( {{P_{30}} + m\sum\limits_{l = 1}^n {ch\left( {{y_l}} \right)} } \right)
\end{eqnarray}

Moreover, as
\begin{eqnarray}
  \mbox{\fontsize{11}{10}\selectfont $ {P_{3\parallel }}\left( {{y_1},{y_2},...,{y_n}} \right) = \frac{1}{2}{P_{\parallel p}}   + \frac{1}{2}{E_p} \sqrt {1 - \frac{{4{M^2}}}{{{E_p}^2 - {P_{\parallel p}}^2}}} $}\nonumber\\
\end{eqnarray}
with
\begin{eqnarray}
&&  \mbox{\fontsize{11}{10}\selectfont $ {E_p}={E_p}\left( {{y_1},{y_2},...,{y_n}} \right) = \frac{{\sqrt s }}{2} - m\sum\limits_{k = 1}^n {ch\left( {{y_k}} \right)}$ }   \\
&&  \mbox{\fontsize{11}{10}\selectfont $ {P_{\parallel p}}={P_{\parallel p}}\left( {{y_1},{y_2},...,{y_n}} \right) =  - m\sum\limits_{k = 1}^n {sh\left( {{y_k}} \right)}$ }
\end{eqnarray}
then we obtain:
\begin{eqnarray}
&&  \mbox{\fontsize{10}{10}\selectfont $ {P_{3\parallel }}\left( { - {y_n},..., - {y_2}, - {y_1}} \right) =  $ }  \nonumber\\
&&  \mbox{\fontsize{10}{10}\selectfont $ - \frac{1}{2}{P_{\parallel p}}  + \frac{1}{2}{E_p}\sqrt {1 - \frac{{4{M^2}}}{{{E_p}^2 - {P_{\parallel p}}^2}}}  =  - {P_{4\parallel }}\left( {{y_1},{y_2},...,{y_n}} \right)  $}  \nonumber\\
\end{eqnarray}
and accordingly
\begin{eqnarray}
&&   {P_{30}}\left( { - {y_n},..., - {y_2}, - {y_1}} \right) \nonumber\\
&&  = \sqrt {{m^2} + {{\left( { - {P_{4\parallel }}\left( {{y_1},{y_2},...,{y_n}} \right)} \right)}^2}} 
   = {P_{40}}\left( {{y_1},{y_2},...,{y_n}} \right) \nonumber\\
\end{eqnarray}
With this we get
\begin{eqnarray}
&&{{A'}_\parallel } = {\left( {m - {{\left( {\frac{{\sqrt s }}{2} - {P_{40}}} \right)}^2} + {{\left( {S'_k} \right)}^2}} \right)^{ - 1}} \nonumber\\
&& \times \prod\limits_{k = 1}^{n - 1} {{{\left( {m - {{\left( {{E'_k}} \right)}^2} + {{\left( {S'_k + m\sum\limits_{i = n}^{n - k + 1} {sh\left( {{y_i}} \right)} } \right)}^2}} \right)}^{ - 1}}}  \nonumber\\
&& \times {\left( {m - {{\left( {{E'_n}} \right)}^2} + {{\left( {S'_k + m\sum\limits_{i = 1}^n {sh\left( {{y_i}} \right)} } \right)}^2}} \right)^{ - 1}}
\label{eq2.5}
\end{eqnarray}
where
\begin{eqnarray}
&& S'_k = \sqrt {s/4 - {M^2}}  + {P_{4\parallel }} \\
&& {E'_k} = {\sqrt s }/2 - \left( {{P_{40}} + m\sum\limits_{i = 1}^{n - k + 1} {ch\left( {{y_i}} \right)} } \right)  \\
&& {E'_n} = {\sqrt s }/2 - \left( {{P_{40}} + m\sum\limits_{i = 1}^n {ch\left( {{y_i}} \right)} } \right) 
\end{eqnarray}
Or in more convenient form expression Eq.\ref{eq2.5} looks like:
\begin{eqnarray}
&& {{A'}_\parallel } = {\left( {m - {{\left( {\frac{{\sqrt s }}{2} - {P_{40}}} \right)}^2} + {{\left( {S'_k} \right)}^2}} \right)^{ - 1}} \nonumber\\
&& \times \prod\limits_{k = 1}^{n - 1} {{{\left( {m - {{\left( {{E''_k}} \right)}^2} + {{\left( {S'_k + m\sum\limits_{i = n - k + 1}^n {sh\left( {{y_i}} \right)} } \right)}^2}} \right)}^{ - 1}}}  \nonumber\\
&& \times {\left( {m - {{\left( {{E''_n}} \right)}^2} + {{\left( {S'_k + m\sum\limits_{i = 1}^n {sh\left( {{y_i}} \right)} } \right)}^2}} \right)^{ - 1}}
\label{eq2.6}
\end{eqnarray}
where
\begin{eqnarray}
&& {E''_k} = {\sqrt s }/2 - \left( {{P_{40}} + m\sum\limits_{i = n - k + 1}^n {ch\left( {{y_i}} \right)} } \right) \\
&& {E''_n} = {\sqrt s }/2 - \left( {{P_{40}} + m\sum\limits_{i = 1}^n {ch\left( {{y_i}} \right)} } \right) 
\end{eqnarray}

We take into account that amplitude depends on the rapidity, which satisfy the conservation laws of energy and longitudinal momentum components:
\begin{eqnarray}
&& \sqrt s  - {P_{30}} - {P_{40}} - m\sum\limits_{i = 1}^n {ch\left( {{y_i}} \right)}  = 0 \cr
&&  {P_{3\parallel }} + {P_{4\parallel }} + m\sum\limits_{i = 1}^n {sh\left( {{y_i}} \right)}  = 0
\label{eq2.7}
\end{eqnarray}
From the conservation of the longitudinal component of momentum,  we get
\begin{eqnarray}
&& {P_{3\parallel }} + m\sum\limits_{i = 1}^{n - k} {sh\left( {{y_i}} \right)}  =  - {P_{4\parallel }} - m\sum\limits_{i = n - k + 1}^n {sh\left( {{y_i}} \right)}; \quad k < n \nonumber\\
&& {P_{3\parallel }} + m\sum\limits_{i = 1}^n {sh\left( {{y_i}} \right)}  =  - {P_{4\parallel }}; \quad k = n  
\label{eq2.8}
\end{eqnarray}

Furthermore, from the energy conservation law we have
\begin{eqnarray}
  \mbox{\fontsize{9}{10}\selectfont $  {\sqrt s }/2 - {P_{40}} =  - \left( {{\sqrt s }/2 - {P_{30}} - m\sum\limits_{i = 1}^n {ch\left( {{y_i}} \right)} } \right) $}
\end{eqnarray}
and
\begin{eqnarray}
  \mbox{\fontsize{10}{10}\selectfont $ {\sqrt s }/2 - \left( {{P_{40}} + m\sum\limits_{i = n - k + 1}^n {ch\left( {{y_i}} \right)} } \right) $}\nonumber\\
  \mbox{\fontsize{10}{10}\selectfont $  =  - \left( {{\sqrt s }/2 - {P_{30}} - m\sum\limits_{i = 1}^{n - k} {ch\left( {{y_i}} \right)} } \right) $}
\label{eq2.9}
\end{eqnarray}
Substituting Eq.\ref{eq2.9} to Eq.\ref{eq2.6} we obtain:
\begin{eqnarray}
&&{{A'}_\parallel } = {\left( {m - {{\left( {\frac{{\sqrt s }}{2} - {P_{30}}} \right)}^2} + {{\left( {S''_k} \right)}^2}} \right)^{ - 1}} \nonumber\\
&&\times \prod\limits_{k = 1}^{n - 1} {{{\left( {m - {{\left( {{E'''_k}} \right)}^2} + {{\left( {S''_k - m\sum\limits_{i = 1}^{n - k} {sh\left( {{y_i}} \right)} } \right)}^2}} \right)}^{ - 1}}}  \nonumber\\
&& \times {\left( {m - {{\left( {{E'''_n}} \right)}^2} + {{\left( {S''_k - m\sum\limits_{i = 1}^n {sh\left( {{y_i}} \right)} } \right)}^2}} \right)^{ - 1}} 
\label{eq2.10}
\end{eqnarray}
where
\begin{eqnarray}
&& S''_k = \sqrt {s/4 - {M^2}}  - {P_{3\parallel }}  \\
&& {E'''_n} = {\sqrt s }/2 - \left( {P_{30}} - m\sum\limits_{i = 1}^n {ch\left( {{y_i}} \right)} \right ) \\
&& {E'''_k} = {\sqrt s }/2 - \left( {P_{30}} + m\sum\limits_{i = 1}^{n - k} {ch\left( {{y_i}} \right)} \right) 
\end{eqnarray}
Replacing index $k=n-j$ in product, which is included in the expression for amplitude, we obtain the relation coincident with Eq.\ref{eq:eq_10} (taking into account Eq.\ref{eq12})
\begin{eqnarray}
  \mbox{\fontsize{10}{10}\selectfont $ {A_\parallel }\left( {n,{y_1},{y_2},...,{y_n}} \right) = {A'_\parallel }\left( {n,{y_1},{y_2},...,{y_n}} \right)$ }
\label{eq2.11}
\end{eqnarray}
Taking into account Eq.\ref{eq2.1}, we obtain the required relation Eq.\ref{eq14}:
\begin{eqnarray}%
&&  {A_\parallel }\left( {n,{y_1},{y_2},...,{y_{\frac{n}{2}}},{y_{\frac{n}{2} + 1}},...,{y_n}} \right) \nonumber\\
&& = {A_\parallel }\left( {n, - {y_n}, - {y_{n - 1}},..., - {y_{\frac{n}{2} + 1}}, - {y_{\frac{n}{2}}},..., - {y_1}} \right)\nonumber\\
\end{eqnarray}

\section{Calculation of the energies of the virtual lines that intersect with the axis of symmetry of the diagrams in Fig.\ref{fig:part1_fig03}}
\label{Appendix_C}
Let us examine at first the case of an even number of particles. The law of conservation of energy in c.m.s has a form:
\begin{eqnarray}
  \mbox{\fontsize{10}{10}\selectfont $  \sqrt {s}  = \sum\limits_{k = 1}^n {m \cdot ch\left( {{y_k}} \right)} + \sqrt {{M^2} + {\left( {{P_{3\parallel }}} \right)}^2}  + \sqrt {{M^2} + {{\left( {{P_{4\parallel }}} \right)}^2}} $}\nonumber\\
\end{eqnarray}
Since, constrained maximum of the scattering amplitude gives a symmetric configuration of ${y_{n - k + 1}} =  - {y_k},\;k = 1,\;2,...,\frac{n}{2}$ and ${P_{4\parallel }} =  - {P_{3\parallel }}$ (that follows from Eq.\ref{eq23}), then at the maximum point we get:
\begin{eqnarray}
  \mbox{\fontsize{12}{10}\selectfont $ 2\sum\limits_{k = 1}^{\frac{n}{2}} {m \cdot ch\left( {{y_k}} \right)}  + 2\sqrt {{M^2} + {{\left( {{P_{3\parallel }}} \right)}^2}}  = \sqrt s $ }\nonumber\\
\end{eqnarray}
hence
\begin{eqnarray}
  \mbox{\fontsize{12}{10}\selectfont $ \frac{{\sqrt s }}{2} - \sqrt {{M^2} + {{\left( {{P_{3\parallel }}} \right)}^2}}  - \sum\limits_{k = 1}^{\frac{n}{2}} {m \cdot ch\left( {{y_k}} \right)}  = 0$ }\nonumber\\
\label{eq3.2}
\end{eqnarray}
However expression Eq.\ref{eq3.2} corresponds to the energy transferred through the central link of the diagram between vertices with numbers of $\frac{n}{2}$ and $\frac{n}{2} + 1$ (Fig.\ref{fig:part1_fig02}, Fig.\ref{fig:part1_fig03}), therefore this energy is equal to zero at the maximum point.

Now let ladder has odd number of particles $n$. Writing the law of conservation energy, we select in the sum of pion energies the term corresponding to the central particle in the diagram:
\begin{eqnarray}
&&  \mbox{\fontsize{10}{10}\selectfont $ \sum\limits_{k = 1}^{\frac{{n - 1}}{2}} {m \cdot ch\left( {{y_k}} \right)}  + m \cdot ch\left({{y_{\frac{{n - 1}}{2} + 1}}} \right) + \sum\limits_{k = \frac{{n - 1}}{2} + 2}^n {m \cdot ch\left( {{y_k}} \right)} $}  \nonumber\\
&&  \mbox{\fontsize{10}{10}\selectfont $  + \sqrt {{M^2} + {{\left( {{P_{3\parallel }}} \right)}^2}} \sqrt {{M^2} + {{\left( {{P_{4\parallel }}} \right)}^2}}  = \sqrt s   $}
\label{eq3.3}
\end{eqnarray}

The maximum of inelastic scattering amplitude is reached at the symmetric configuration. It is characterized by the fact that the central particle has zero rapidity and the particles located symmetrically about the center particle have mutually oppositely rapidities, therefore
\begin{eqnarray}
  \mbox{\fontsize{12}{10}\selectfont $ 2\sum\limits_{k = 1}^{\frac{{n - 1}}{2}} {m \cdot ch\left( {{y_k}} \right)}  + m + 2\sqrt {{M^2} + {{\left( {{P_{3\parallel }}} \right)}^2}}  = \sqrt s $}\nonumber\\
\end{eqnarray}
It follows that
\begin{eqnarray}
  \mbox{\fontsize{12}{10}\selectfont $ \frac{{\sqrt s }}{2} - \sqrt {{M^2} + {{\left( {{P_{3\parallel }}} \right)}^2}}  - \sum\limits_{k = 1}^{\frac{{n - 1}}{2}} {m \cdot ch\left( {{y_k}} \right)}  = \frac{m}{2} $}\nonumber\\
\label{eq3.4}
\end{eqnarray}

Thus, the energy, which flows between $\frac{{n - 1}}{2}$-th and $\frac{{n - 1}}{2} + 1$-th particles at the most probable configuration, is equal to $\frac{m}{2}$. As $\frac{{n - 1}}{2} + 1$-th particle takes away energy $m$ (which has zero rapidity at the maximum point), then the following link in the diagram will be transfered energy $\frac{-m}{2}$.

\vspace{2cm} 
\normalsize {\textbf{REFERENCES}}
\bibliography{References_JMPh}

\end{document}